\documentclass[a4paper,fleqn,usenatbib,useAMS]{mnras}

\usepackage{graphicx}	
\usepackage{amsmath}	
\usepackage{amssymb}	
\usepackage{multicol}        
\usepackage{bm}		
\usepackage{pdflscape}	
\usepackage{color}
\usepackage[para,online,flushleft]{threeparttable}

\let\oldAA\AA
\renewcommand{\AA}{\text{\normalfont\oldAA}}

\usepackage[T1]{fontenc}
\usepackage{ae,aecompl}

\usepackage{amsmath}

\title[Structural decomposition of barred galaxies]{Galaxy Zoo: Secular evolution of barred galaxies from structural decomposition of multi-band images}

\author[Sandor Kruk et al.]{
Sandor J. Kruk,$^{1}$\thanks{E-mail: \href{mailto:sandor.kruk@physics.ox.ac.uk}{sandor.kruk@physics.ox.ac.uk}}
Chris J. Lintott,$^{1}$ 
Steven P. Bamford,$^{2}$
Karen L. Masters,$^{3}$
\newauthor
Brooke D. Simmons,$^{1,4}$\thanks{Einstein Fellow}
Boris H{\"a}u{\ss}ler,$^{5}$
Carolin N. Cardamone,$^{6}$
Ross E. Hart,$^{2}$
\newauthor 
Lee Kelvin,$^{7}$
Kevin Schawinski,$^{8}$
Rebecca J. Smethurst,$^{1,2}$
Marina Vika$^{9}$
\thanks{This investigation has been made possible by the participation of over 350,000 users in the Galaxy Zoo project. Their contributions are acknowledged at \href{http://authors.galaxyzoo.org}{http://authors.galaxyzoo.org}}\\
$^{1}$Oxford Astrophysics, Department of Physics, University of Oxford, Denys Wilkinson Building, Keble Road, Oxford, OX1 3RH, UK\\
$^{2}$School of Physics and Astronomy, The University of Nottingham, University Park, Nottingham NG7 2RD, UK\\
$^{3}$Institute of Cosmology and Gravitation, University of Portsmouth, Dennis Sciama Building, Barnaby Road, Portsmouth, PO1 3FX, UK\\
$^{4}$Center for Astrophysics and Space Sciences (CASS), Department of Physics, University of California, San Diego, CA 92093, USA\\
$^{5}$ESO - European Southern Observatory, Alonso de Cordova 3107, Vitacura, Casilla 19001, Santiago, Chile\\
$^{6}$Math and Science Department, Wheelock College, 200 The Riverway, Boston, MA 02215, USA\\
$^{7}$Astrophysics Research Institute, Liverpool John Moores University, IC2, Liverpool Science Park, 146 Brownlow Hill, Liverpool L3 5RF, UK\\
$^{8}$Institute for Astronomy, Department of Physics, ETH Z{\"u}rich, Wolfgang-Pauli Strasse 27, CH-8093 Z{\"u}rich, Switzerland\\
$^{9}$Institute for Astronomy, Astrophysics, Space Applications and Remote Sensing, National Observatory of Athens, Penteli, 15236, Athens, Greece}

\date{Last updated 21 September 2017 }

\pubyear{2017}

\begin{document}
\label{firstpage}
\pagerange{\pageref{firstpage}--\pageref{lastpage}}
\maketitle

\begin{abstract}

We present the results of two-component (disc+bar) and three-component (disc+bar+bulge) multiwavelength 2D photometric decompositions of barred galaxies in five SDSS bands (\textit{ugriz}). This sample of $\sim$3,500 nearby ($z<0.06$) galaxies with strong bars selected from the Galaxy Zoo citizen science project is the largest sample of barred galaxies to be studied using photometric decompositions which include a bar component. With detailed structural analysis we obtain physical quantities such as the bar- and bulge-to-total luminosity ratios, effective radii, Sérsic indices and colours of the individual components. We observe a clear difference in the colours of the components, the discs being bluer than the bars and bulges. An overwhelming fraction of bulge components have S\'ersic indices consistent with being pseudobulges. By comparing the barred galaxies with a mass-matched and volume-limited sample of unbarred galaxies, we examine the connection between the presence of a large-scale galactic bar and the properties of discs and bulges. We find that the discs of unbarred galaxies are significantly bluer compared to the discs of barred galaxies, while there is no significant difference in the colours of the bulges. We find possible evidence of secular evolution via bars that leads to the build-up of pseudobulges and to the quenching of star formation in the discs. We identify a subsample of unbarred galaxies with an inner lens/oval and find that their properties are similar to barred galaxies, consistent with an evolutionary scenario in which bars dissolve into lenses. This scenario deserves further investigation through both theoretical and observational work. 


\end{abstract}

\begin{keywords}
galaxies: general, galaxies: evolution, galaxies: structure, galaxies: bulges, galaxies: star formation, galaxies: stellar content
\end{keywords}




\section{Introduction}

\vspace{3mm}

Galactic bars have been known to exist ever since the discovery of the first galaxies, and their abundance in the local Universe led Edwin Hubble to dedicate a major part of his classification scheme to barred spiral galaxies \citep{Hubble1936}. Observational studies confirmed that stellar bars are common in disc components, with a fraction of 30\% showing bars at optical wavelengths \citep{Sellwood1993,Masters2011}, rising to 70\% in the infrared, if weaker bars are included \citep{Sheth2008}.

Simulations show that galactic bars arise because of instabilities in the disc and that they can develop over a large range of disc masses and can persist for a long time \citep{Combes1981,Shen2004,Debattista2006}. Considerable theoretical work on the formation of bars has been carried out by \citet{Athanassoula2013} who found that the gas fraction of a galaxy plays a major role in the formation and evolution of a bar: large-scale bars are harder to form in gas-rich discs than in gas poor ones. In an earlier study, \citet{Athanassoula2002} also found that bars can redistribute the angular momentum in the interstellar medium and are efficient at funneling gas to the centre of the galaxy. This can cause an increase in central star formation \citep{Hawarden1986} and can lead to the formation of so-called ``pseudobulges" \citep{Kormendy2004} which have properties (such as ordered stellar orbits) similar to disc galaxies, rather than ellipticals. The gas that falls in the central parts of the galaxy might trigger AGN (e.g. \citealt{Noguchi1988,Wada1992}); different authors have investigated if the presence of an AGN is correlated with the presence of a bar, finding contradictory answers (e.g. \citealt{Galloway2015,Cheung2015,Cisternas2015,Goulding2017}). This is not surprising given that timescales vary considerably for bar driven motions ($\sim$ Gyr, \citealt{Athanassoula2000}) and AGN activity ($\sim$ Myr, \citealt{Hickox2014}).

\begin{figure*}
 \includegraphics[width=\textwidth]{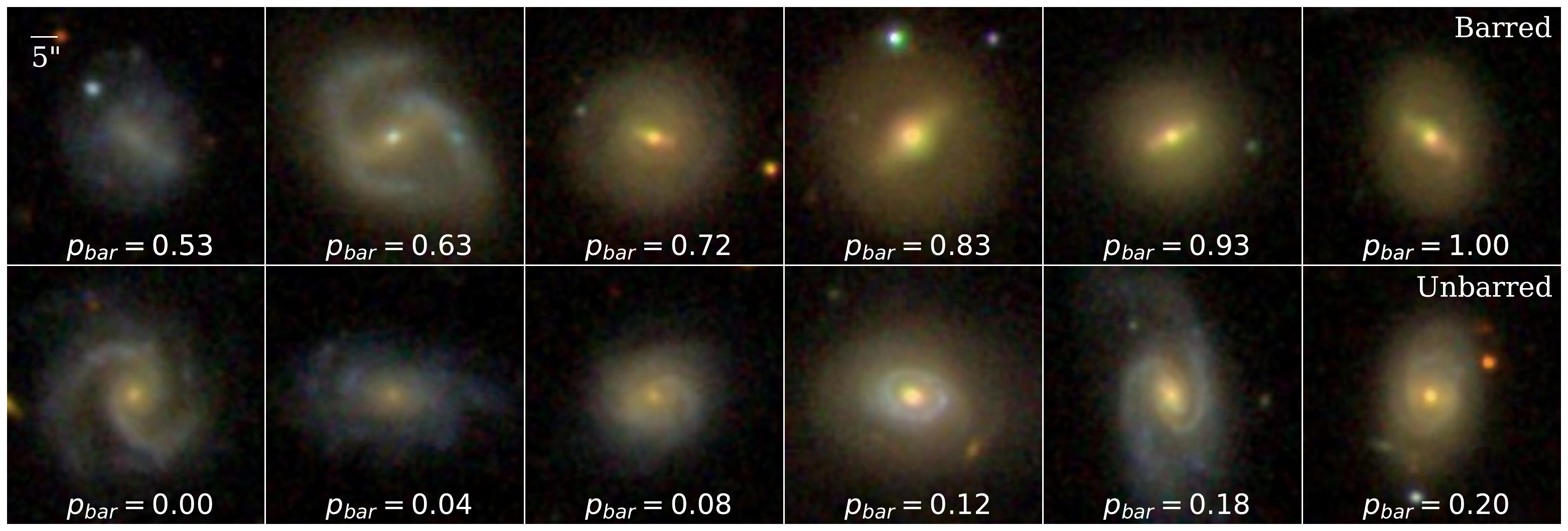}
 \caption{Examples of barred and unbarred galaxies in the volume-limited samples used in our study. The $p_{\mathrm{bar}}$ shows the Galaxy Zoo debiased likelihood of a disc galaxy being barred, based on the volunteers' inspection and classifications. The galaxies were randomly selected, approximately equally spaced in $p_{\mathrm{bar}}$ values.}
 \label{mosaic}
\end{figure*}

Recent observational studies involving large samples show that the fraction of disc galaxies which have a bar increases with redder, gas-poor galaxies \citep{Masters2012}, with over half of the red disc galaxies being barred (see Fig. 3 in \citealt{Masters2011}). Using a similarly large sample of barred galaxies, \citet{Cheung2013} finds that the likelihood of a galaxy hosting a bar is anti-correlated with the specific star formation rate, regardless of stellar mass or bulge prominence. They suggest that the observed trends are driven by the gas fraction in the discs, pointing towards a scenario in which the secular evolution of barred galaxies is driven by bars. 

Nevertheless, the role of bars in quenching the star formation, making a disc galaxy become `red and dead' and the details of this process are still unclear. To investigate these, one has to study the stellar populations of the individual components (bars, discs and bulges) separately, in detail. One way to achieve this is with Integral Field Spectroscopy (IFS), whereby spectra of various parts of galaxies are obtained simultaneously. However, until now there have been few IFS studies to observationally characterize the influence of bars on nearby galaxies, and they were limited to small samples (e.g. BaLROG - Bars in Low Redshift Optical Galaxies with SAURON IFS, \citealt{Seidel2015}, NGC 4371 with MUSE, \citealt{Gadotti2015}). Only with the advent of large IFS surveys such as CALIFA \citep{Califa2012}, SAMI \citep{Sami2012}, MaNGA \citep{Manga2016} it is now becoming possible to study the internal stellar populations for large samples. 

The alternative is to study the stellar populations of bars, discs and bulges by photometric decomposition of galaxy images \citep{Byun1995}, which can be applied simultaneously to a large sample of galaxies. Many authors have applied 2D decomposition methods to separate discs and bulges (e.g. using \textsc{Gim2D} - \citealt{Allen2006}, \textsc{Gasp2D} - \citealt{Mendez2008}, \textsc{BUDDA} - \citealt{Gadotti2009}). The largest two-band image bulge+disc decomposition, of over a million galaxies in SDSS, was carried out by \citet{Simard2011}. However, simple bulge+disc decompositions can give inaccurate fits when applied to strongly barred galaxies, with the bar flux being erroneously assigned primarily to the bulge, as shown by \citet{Laurikainen2005}. Using detailed decompositions of 15 barred galaxies \citet{Laurikainen2006} showed that the bulge-to-total luminosity ratios can be significantly overestimated when only the discs and bulges are accounted for. 

\citet{Laurikainen2007} decomposed 216 nearby disc galaxies in detail, including bars, and found strong evidence for pseudobulges across all Hubble types. \citet{Reese2007} also attempted to decompose the light of 68 disc galaxies into discs, bulges and bar components. \citet{Weinzirl2009} decomposed 143 bright $H$-band galaxies, $\sim$80 including a bar component, and studied the correlations between bulges of barred and unbarred galaxies concluding that bulges are likely to have been built by a combination of secular processes and minor mergers in the recent Universe. Using the \textsc{BUDDA} software \citep{Souza2004}, \citet{Gadotti2009} performed disc+bulge+bar decomposition in three bands (\textit{g}, \textit{r} and \textit{i}) on a sample of 291 barred galaxies from SDSS and studied their properties in \citet{Gadotti2011}. More recently, \citet{Salo2015} decomposed 2,352 nearby (<40 Mpc) galaxies from the S$^{4}$G survey \citep{Sheth2010} (out of which $\sim$800 included a bar component), while \citet{Kim2015} fitted 144 face-on barred galaxies from S$^{4}$G with a bar component.

The aim of this paper is to get meaningful physical parameters for the bulges, discs and bars of the largest sample of barred galaxies to date ($\sim$3,500) and compare them with unbarred galaxies using the most complete multi-wavelength data for nearby galaxies from the SDSS. The bulge-to-total ($B/T$), bar-to-total ($Bar/T$) luminosity ratios, component colours and S\'ersic indices are analysed with the aim of understanding the effect of bars on the evolution of barred galaxies. The fits from this paper have already been used to identify a sample of 271 galaxies with an off-centre bar and study their properties \citep{Kruk2017}.

The paper is structured as follows. In Section \ref{data} we discuss the sample selection and identification of barred galaxies, 
and also describe the method used in the multi-wavelength 2D photometric decomposition. In Section \ref{barred} we present the main results on the properties of barred galaxies, while in Section \ref{barred_vs_unbarred} we compare the properties of mass-matched volume-limited samples of barred and unbarred galaxies. In Section \ref{lenses} we consider the properties of non-barred galaxies with inner lenses. Finally, in Section \ref{discussion} we discuss our findings in the context of secular evolution of barred galaxies. Throughout the paper we adopt the WMAP Seven-Year Cosmological parameters \citep{Jarosik2011} with ($\Omega_{M},\Omega_{\Lambda},h) = (0.27,0.73,0.71)$.

\section{Data and Methods}
\label{data}

\subsection{Galaxy Zoo and SDSS}

All the galaxies used in the study are drawn from the Sloan Digital Sky Survey (SDSS) \citep{Gunn1998,York2000,Eisenstein2011} DR7  \citep{SDSSDR7}. Morphological classification of galaxies are taken from the Galaxy Zoo 2\footnote{\url{http://zoo2.galaxyzoo.org}} project (GZ2) \citep{Lintott2008,Willett2013} which asked citizen scientists to provide detailed information about the visual appearance of galaxies. Each galaxy was inspected by at least 17 volunteers and the mean number of classification per galaxy is $\sim$42. 

From the subset of 240,419 galaxies classified in GZ2\footnote{Data available from \url{http://data.galaxyzoo.org}} which have stellar masses available from the MPA-JHU catalogue \citep{Kauffmann2003a}, available inclinations and measured spectroscopic redshifts, we have selected all the galaxies with redshifts 0.005<$z$<0.06. This redshift range provides reliable GZ2 morphological classifications and suitable SDSS image resolution. Identifying bars in highly inclined galaxies is challenging, thus we selected only galaxies with an axis ratio of $b/a>0.5$ given by the exponential model fits in SDSS \citep{Stoughton2002}, corresponding to inclinations $i\lesssim60^{\circ}$.

In order to reach the bar question a user must first classify a galaxy as a non edge-on galaxy with a disc or features. Following \citet{Masters2011} and the recommendation of \citet{Willett2013}, we only selected galaxies for which there were at least 10 answers to the question `Is there a sign of a bar feature through the centre of the galaxy?'. To quantify the likelihood that a galaxy is barred, GZ2 calculates the ratio of the number of volunteers who identified a galaxy being barred and the total number of votes to the bar question. These raw likelihoods are then adjusted to account for the inconsistency of users, as well as for the deterioration of image quality with redshift, as detailed in \citet{Willett2013}. Finally, we are left with a debiased bar likelhood, denoted as $p_{\mathrm{bar}}$, which will be used throughout this paper. A galaxy was classified as being barred if the number of volunteers identifying it as having a bar is larger than, or equal to the number identifying it as not having a bar, i.e. $p_{\mathrm{bar}}\geq0.5$. Furthermore, to avoid problems with the deblending of galaxy images we exclude merging or overlapping galaxies, which according to \citet{Darg2010}, can be achieved with a cut of the GZ1 \citep{Lintott2011} merging parameter $p_{\mathrm{merger}} < 0.4$. All galaxies in GZ2, considered in this study, are included in GZ1 and, although using a different classification tree, $p_{\mathrm{merger}}$ parameter has a strong correlation with the projected galaxy separation \citep{Casteels2013}. Our final, large sample of barred galaxies contains 5,282 galaxies, with a mean number of users who answered the bar question of 22.

The bars detected by GZ volunteers agree well with expert classifications made by \citet{Nair2010}. Using a sample size of ~14,000 galaxies and with an overlap of 90\% with GZ2, \citet{Nair2010} detected a bar fraction of $\sim$30\% \citep{Nair2010b} and classified the bars according to their strength as strong, intermediate or weak, depending on their sizes relative to the sizes of the discs and on the bars' prominence. Nevertheless, their classification corresponds to subclasses of the strong bar classification in RC3 \citep{deVauc1993}, as \citet{Nair2010} point out. Comparing the sample of barred galaxies in GZ2 and the one in \citet{Nair2010}, GZ tends to identify strong and intermediate bars with the threshold $p_{\mathrm{bar}}\geq0.5$ and weak bars with $0.2<p_{\mathrm{bar}}<0.5$, as discussed in \citet{Skibba2012,Masters2012} and shown in Figure 10 of \citet{Willett2013}. \citet{Masters2012} also show in their Appendix A that Galaxy Zoo detects 90\% of the strong and intermediate bars with $p_{\mathrm{bar}}\geq0.5$, while 92\% of their unbarred galaxies have $p_{\mathrm{bar}}<0.5$, suggesting that $p_{\mathrm{bar}}\geq0.5$ is adequate for selecting a clean sample of strong and intermediate bars. This cut has been adopted by several other Galaxy Zoo studies of barred galaxies \citep{Masters2011,Masters2012,Melvin2014,Cheung2015,Cheung2015b}. A further discussion regarding the implications of including the weak bars in this study can be found in Appendix \ref{appendixA}.

The main GZ2 spectroscopic sample contains only SDSS galaxies brighter than $r<17$ \citep{Willett2013}. Therefore, to study the statistical distribution and properties of these systems, we selected a volume-limited sample of barred galaxies brighter than $M_{\mathrm{r}}<-20.15$, which is the \textit{r}-band Petrosian absolute magnitude corresponding to the GZ2 completeness magnitude of 17, at a redshift of $z=0.06$. To construct a comparison sample of galaxies without bars, we have used a similar criteria (same cut for inclination and at least 10 answers to the question `Is there a sign of a bar feature through the centre of the galaxy?') in order to select disc galaxies, but with $p_{\mathrm{bar}}<0.2$, in this case, to select a volume-limited sample of unbarred galaxies between 0.005<$z$<0.06 and brighter than $M_{\mathrm{r}}<-20.15$. There are 3,547 and 8,689 galaxies in the \textsc{volume-limited barred} and \textsc{volume-limited unbarred} samples, respectively. Examples of barred and unbarred galaxies in the volume-limited samples can be seen in Figure \ref{mosaic}. The larger number of unbarred galaxies allows us to select a subsample of unbarred galaxies with a mass distribution matching the one of barred galaxies, which is described further in Section \ref{barred_vs_unbarred}. Eliminating the mass-dependence \citep{Kauffmann2003b} enables us to study secondary effects due to the presence of bars. The selection criteria and sample sizes are summarized in Table \ref{selection}. 

\begin{table}
 \caption{Selection criteria and sample size.}
 \label{selection}
 \begin{tabular}{|p{3.5cm}|p{2.3cm}|p{1.2cm}|}
  \hline
  Desciption & Criterion & No. \\
  \hline
  GZ2 & all GZ2$^{\textrm{a}}$ & 243,500 \\
  MPA-JHU Catalogue & match & 240,419 \\
  Nearby & 0.005<z<0.06 & 81,736\\
  Face-on & $i < 60 ^{\circ}$ & 52,851 \\
  Discs &  $N_{\mathrm{bar}}\geq 10$ & 24,478 \\
  Non-interacting &  $p_{\mathrm{merger}}<0.4$ & 23,388 \\
  Barred Discs &  $p_{\mathrm{bar}}\geq 0.5$ & 5,282 \\
  Volume-limited Barred &  $M_{\mathrm{r}}<-20.15$ & 3,547 \\
   & & \\
  Unbarred Discs$^{\textrm{b}}$ &  $p_{\mathrm{bar}}\leq0.2$ & 12,573 \\ 
  
  Volume-limited Unbarred &  $M_{\mathrm{r}}<-20.15$ & 8,689 \\
  \hline
  \end{tabular}
\small $^{\textrm{a}}$With spectroscopic redshifts.\\
\small $^{\textrm{b}}$The unbarred disc selection also follows the first 6 criteria.
\end{table}

\begin{figure*}
 \includegraphics[width=\textwidth]{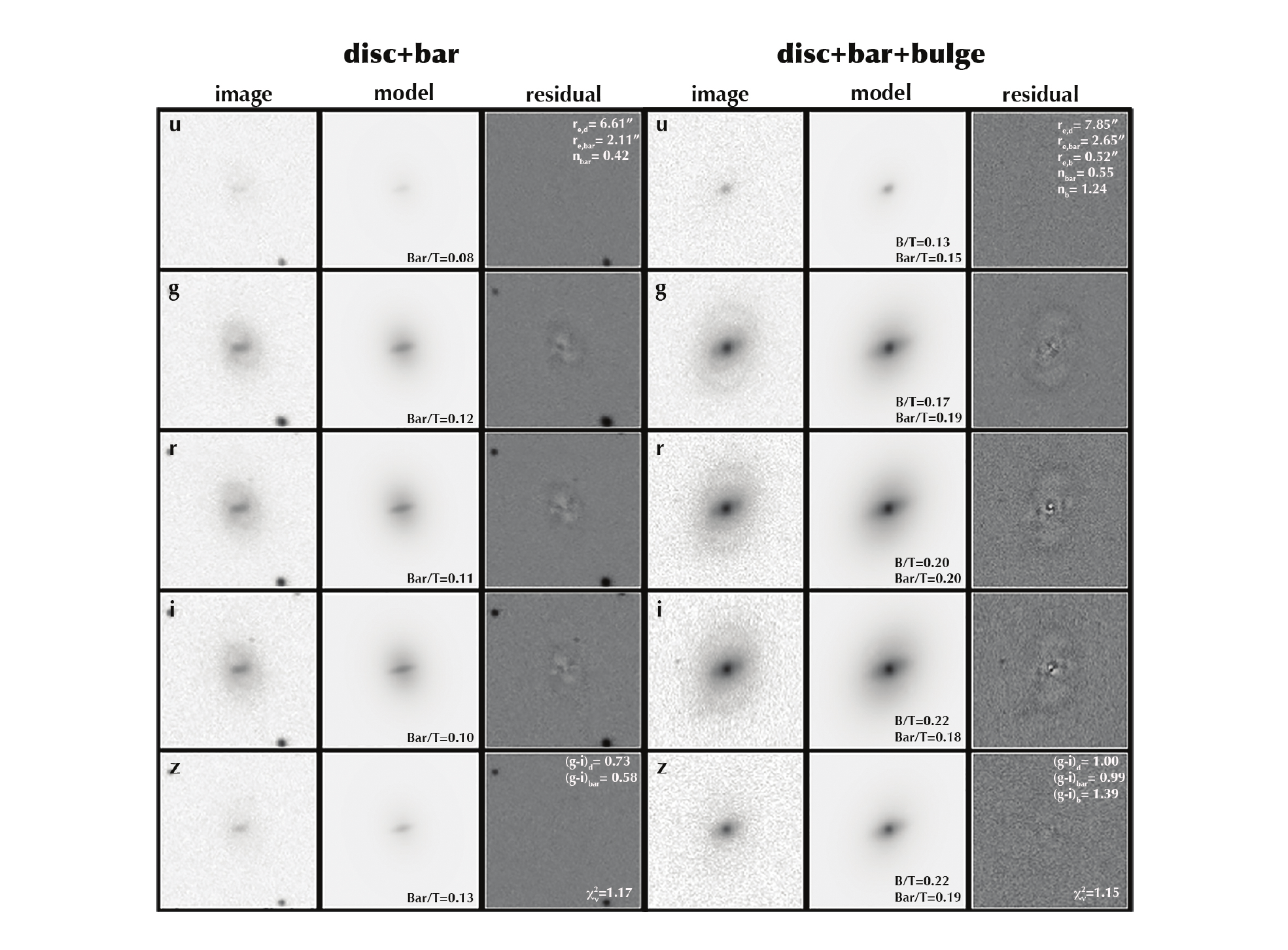}
 \caption{Example outputs from \textsc{GalfitM} of a two component (disc+bar) fit  (left) and a three component (disc+bar+bulge) fit (right), with an \textit{arcsinh} stretch. The first column shows the image in 5 bands, \textit{ugriz}, with the \textit{u} and \textit{z} bands having lower S/N compared to the rest. The second column shows the model fit from \textsc{GalfitM} with two and three components, respectively. The third column shows the residual (the model subtracted from the image). All panels have the same scale and their sizes are $40\arcsec \times 40\arcsec$, a zoom-in of the actual fitting regions to show greater detail. The disc, bar and bulge $r_{e}$ (in arcsec) and $n$ are shown at the top of the image, the $Bar/T$ and $B/T$ luminosity ratios are shown at the bottom right of each model. The $(g-i)$ colours for each component, corrected for Galactic extinction are shown at the top of the $z$-band residual. The reduced $\chi^{2}_{\nu}$ is also shown. }
 \label{example_fitting}
\end{figure*}

\subsection{Galaxy image decomposition}

A key observable is the spatial distribution of light in a galaxy, which can be measured using parametric functions such as the S\'ersic profile. The generalized S\'ersic profile can be expressed as an intensity profile, such that \citep{Sersic1968}:
\begin{equation}
 I(r) = I_{e}\exp \bigg \{ {-b_{n} \bigg [ \bigg ( \frac{r}{r_{e}} \bigg )^{\frac{1}{n}}-1} \bigg ]\bigg \}
\end{equation}
where $I_{e}$ is the intensity at the effective radius $r_{e}$ that encloses half of the total light from the model. $b_{n}$ is a constant depending on the model chosen and the S\'ersic index $n$ describes the shape of the light profile. For a de Vaucouleurs profile $n=4$, while for an exponential profile $n=1$.

In this paper we use a modified version of \textsc{Galfit}3.0 \citep{Peng2010} called \textsc{GalfitM}\footnote{\textsc{GalfitM} is publicly available at \url{http://www.nottingham.ac.uk/astronomy/megamorph/}} developed by the MegaMorph project \citep{Bamford2011,Heussler2013,Vika2013}, to perform automatic 2D disc+bar+bulge, disc+bar and disc+bulge decompositions. In contrast to \textsc{Galfit} which can fit only one band at a time, \textsc{GalfitM} makes use of the full wavelength coverage of surveys \citep{Heussler2013}. It enables fitting across multiple wavelengths to increase the accuracy of the measured parameters, as well as improving magnitudes and effective radii estimation in low S/N bands, by constraining the parameters to Chebyshev polynomials as a function of wavelength. Since the aim of this study is to extract as much physical information for each galaxy component as possible across the optical spectrum, it is the ideal software to use.  

\subsection{Images}

In this study, we use publicly available FITS images from SDSS DR10 \citep{Ahn2014}, in 5 bands: \textit{u, g, r, i} and $z$. For the galaxy images we use the corrected and background-subtracted SDSS fields\footnote{From \url{http://data.sdss3.org/fields}} in which the galaxy appears. To deal with galaxies that are at the edges of fields, we combine the frames into a single mosaic using \textsc{MONTAGE} \citep{Jacob2010}.  \textsc{MONTAGE} combines different fields into a single image by performing the required rebinning, reprojections and background transformations. We created cutouts of the galaxies with a square with a side length of 8 times the $r$-band Petrosian radius of the galaxy, as given by SDSS.

In SDSS-III all the fluxes are expressed in terms of nanomaggies, which is a linear unit of flux. In order for \textsc{GalfitM} to create a good $\sigma$ image we converted the images to electron counts by using an average of the \textit{nanomaggiespercount} factors in the \textsc{fits} Headers of all the frames, assuming an average gain for each band across the whole survey and an exposure time of 53.91 sec to calculate the zero-point magnitudes. 

\textsc{GalfitM} requires a PSF to correct the images for seeing effects, especially in the central regions of the galaxies. We constructed a PSF for each galaxy, in each band, at the position of the galaxy using the corresponding SDSS \texttt{psFields}\footnote{As explained in \url{http://www.sdss3.org/dr10/imaging/images.php}} frames. The estimation of the background level is also important for a successful fit \citep{Haeussler2007}. The SDSS pipeline sky subtraction is inevitably imperfect, therefore we used concentric elliptical annuli around the galaxy to extract the background value at the point where the surface brightness gradient is flat, as further detailed in \citet{Barden2012}, and we kept the sky value fixed throughout the fitting process. Finally, using \textsc{Sextractor} \citep{Bertin1996} segmentation maps we created a mask for each galaxy field in the \textit{r}-band by masking out all the bright sources (stars and galaxies), except the target. The same mask was used for all the 5 bands in the fitting process. We remind the reader that interacting or overlapping galaxies were excluded in this study, thus the galaxies should not have many bright close neighbours. 

\subsection{Model}

\textsc{GalfitM} can fit a wavelength-dependent model with multiple components to images in different bands. It uses the Levenberg-Marquardt algorithm to minimise the $\chi^2$ residual between an image and the PSF-convolved model, by changing the free parameters. The $\chi^2$ is calculated using a weighted sigma ($\sigma$) map created internally by \textsc{GalfitM}.

\textsc{GalfitM} fits all the five bands simultaneously and the user has the choice of varying all the parameters between the bands or fixing some of them. The reasons for fitting the bands simultaneously are: 1) to increase the overall signal-to-noise (S/N), 2) to use the colour differences between the components to help the decomposition, 3) measure consistent colours for each component. In the fitting procedure we constrain some of the parameters such as the centre $(x_{c},y_{c})$, the effective radius ($r_{e}$), the S\'ersic index ($n$), the axis ratio ($b/a$) and the position angle ($\theta$) of each component to be the same in all 5-bands. The only parameter that was allowed to vary freely, independent of wavelength, was the magnitude. This approximation ignores colour and, hence stellar population gradients within the independent models of each component, which is a simplified picture of galaxy structure. Nevertheless, \citet{McDonald2011} have shown that there is no significant variation of the S\'ersic index of the bulge and the effective radii of the disc and bulge with wavelength. We also test for the variations of the fitted parameters in Section \ref{test1}.

To fit the barred galaxies we used an iterative process, in which we added one component at a time. The process we used for fitting is as follows:

I. \textit{One component}. Firstly, we fitted a single S\'ersic profile for each galaxy, with the purpose of providing initial values for the parameters for the subsequent fits, as well as to measure the luminosity of the galaxy. As initial estimates for this fit, we used $n=1$ and magnitudes, $r$-band Petrosian radii, ellipticities and position angles from SDSS \citep{Stoughton2002}. 

II. \textit{Two components}. We then used the values from the single S\'ersic fit as input into a two component model: an exponential disc and a bar. For the bar component we used a slightly dimmer initial magnitude, an initial effective radius of 60\% that of the disc in the one component fit, an initial S\'ersic index of $n=0.7$ and axis ratio $b/a=0.2$, since the bar is an elongated feature, which according to \citet{Kormendy2004} has an ellipticity between 0.2-0.4. We modeled the bar using a free S\'ersic model rather than a Ferrers function \citep{Binney1987} as an approximation of the true bar intensity profile, in order to allow more flexibility in the bar profiles due to the mix of late and early-type galaxies in our sample, as well as to avoid the Ferrers function converging to a different component. We set the position angle of the bar to be at $90^{\circ}$ to the disc. We also tried fitting a boxy bar instead of an ellipse, which should be closer to the shape of a real bar; some bars are either boxy or peanut shapped \citep{Bureau1999, Athanassoula2002}. We found it almost impossible to automatically fit a boxy S\'ersic profile to a highly elliptical component, as the boxiness parameter, was rapidly diverging. Therefore we decided to use a pure ellipse model for the bar ($C_{0}=0$ in \textsc{GalfitM}). For the galaxies with a significant bulge present, the second component that was fitted did not appear to be a real bar. The light from the bulge and from the bar were modeled together in one component by \textsc{GalfitM}, yielding a component with a high S\'ersic index (68\% with $n\geq1$) and an $r_{e}$ larger than the typical $r_{e}$ of the bulge, but smaller than that of the bar.

\begin{figure}
 \includegraphics[width=\columnwidth]{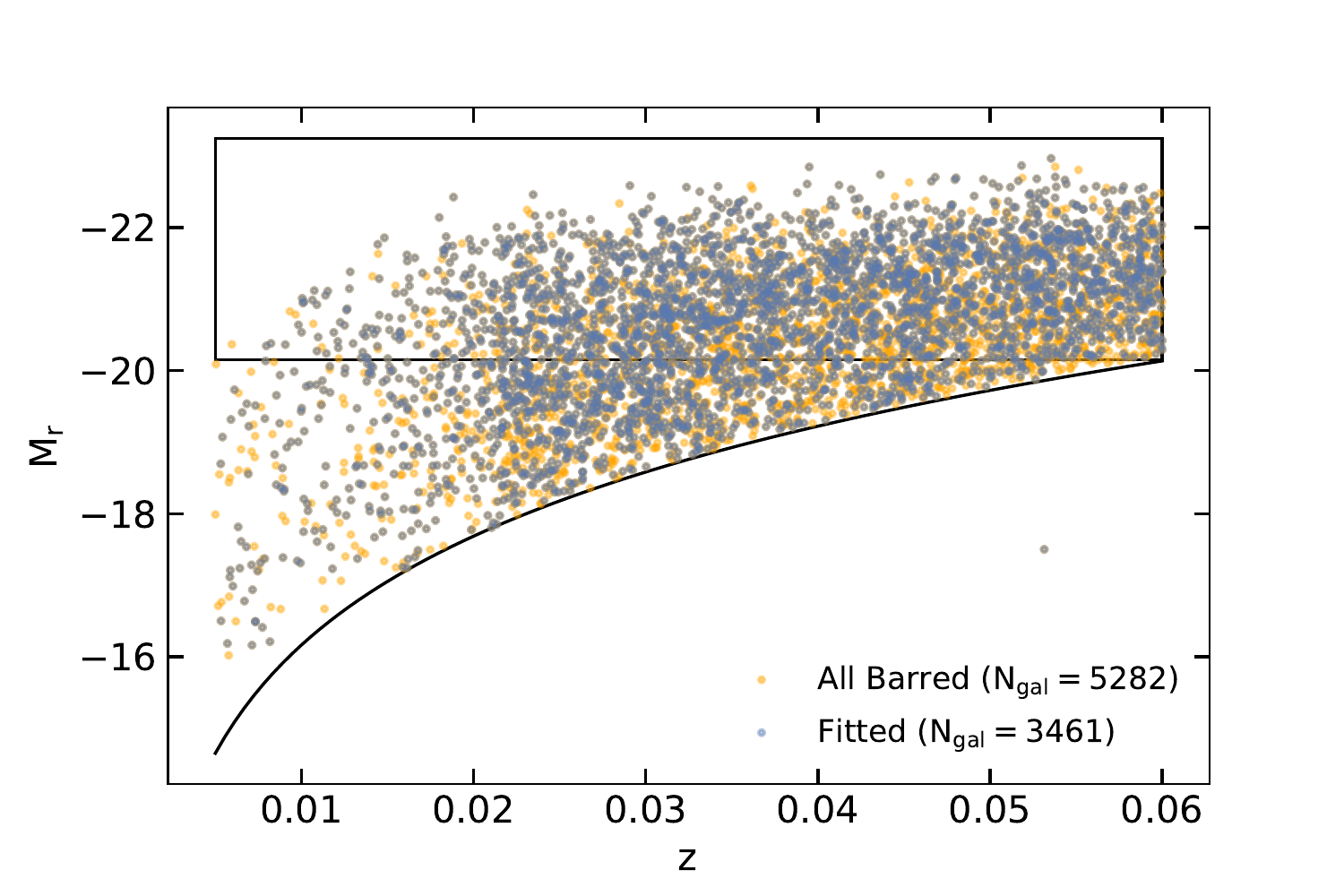}
 \caption{The \textit{r}-band Petrosian absolute magnitudes versus redshift of all barred galaxies in our sample and the barred galaxies that were successfully fitted. The curved line corresponds to the GZ2 completeness limit of $r=17$ magnitudes, at a particular redshift. The rectangle indicates the limit of our volume-limited barred sample, containing a total of 2,435 successfully fitted barred galaxies.}
 \label{magz}
\end{figure}

\begin{figure}
 \includegraphics[width=1.1\columnwidth]{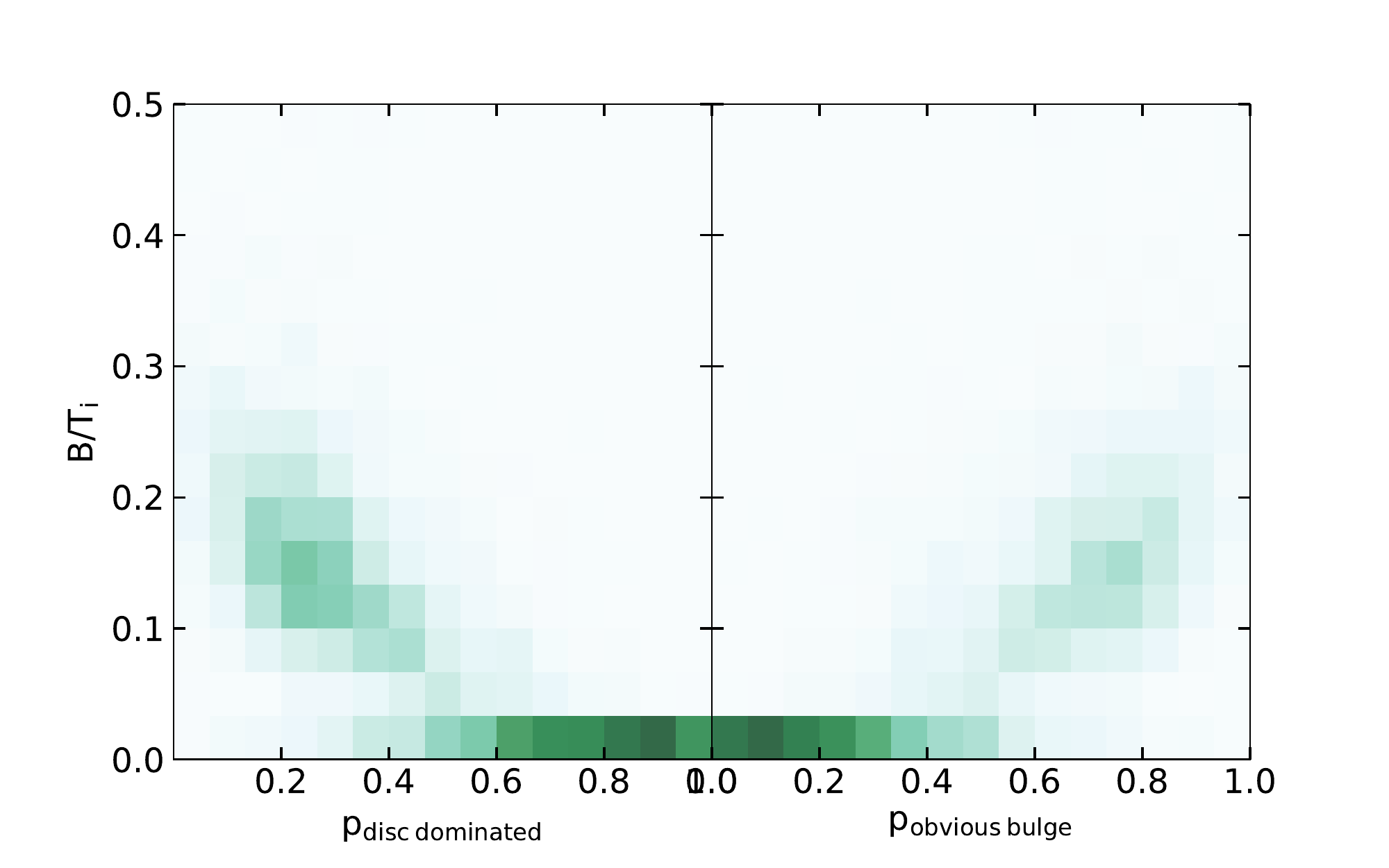}
 \caption{We have chosen to fit disc dominated barred galaxies with 2 components (disc+bar) and barred galaxies with obvious bulges with 3 components (disc+bar+bulge) using visual inspection of the fits and residuals. This correlates well with the GZ volunteers classification of the bulges into: \textsc{No Bulge}, \textsc{Just noticeable}, \textsc{Obvious}, \textsc{Dominant}. In this plot we compare our $B/T$ with the volunteers classification which was split into: $p_{\textrm{disc\:dominated}}=p_{\textrm{No\:Bulge}}+p_{\textrm{Just\:noticeable}}$ and $p_{\textrm{obvious\:bulge}}=p_{\textrm{Obvious}}+p_{\textrm{Dominant}}$.}
 \label{BT_gz}
\end{figure}

III. \textit{Three components}. Only the galaxies for which the two components successfully converged were fitted with three components. Based on the parameters from the two component fit as initial guesses, we added a third component, a bulge, also modeled with a free S\'ersic profile. We started with an initial disc having slightly larger (125\%) $r_{e,\textrm{disc}}$ than the $r_{e}$ of the disc in the two component fit and a bar with an $r_{e,\textrm{bar}}$ of 50\% the $r_{e}$ of the disc in the two component fit. As an initial estimate for the bulge effective radius we used 25\% of the $r_{e}$ of the second component in step II, while for the initial axis ratio of the bulge, we used a value of $b/a=0.8$, since the bulge should be a nearly round feature. The initial position angle was initially set to that of the disc. The initial S\'ersic index of the bulge was set to $n=2$, so that it is sufficiently different from the other components. This is also the boundary noted by \citet{Fisher2008} to distinguish pseudobulges from classical bulges.

We also tested adding the components in the order disc-bulge-bar, but since a large fraction of the barred galaxies in our sample lack a significant bulge (as discussed later), we found that the second component often converged to a bar. Therefore, we chose to add the components in the order disc-bar-bulge. For 523 galaxies the second component in step II converged to a model closer to that of a bulge, as discussed in the following subsection, hence we added the bar at the third iteration. To reduce the chances of \textsc{GalfitM} converging to an unphysical fit, we provided several constraints: the magnitude was required to be within 6 magnitudes of the input value, the effective radii between 0.5 and 500 pixels and the S\'ersic indices smaller than 8. We also required the bar and bulge components to have the same centre, in order to avoid one of the components converging to a clump, or overlapping star. However, the discs and the bar or bar+bulge components were not constrained to have the same centre, they were allowed to vary within $12\arcsec$, which is the median $r$-band radius containing 90\% of the Petrosian flux of the galaxies ($r_{\mathrm{Petro90}}$) in our sample. In \citet{Kruk2017} we discuss the case of the galaxies which have the disc-bar offsets larger than the FWHM of the PSF, where we categorise them as `offset systems' and discuss their properties in more detail.

The constraints mentioned above are reasonable and useful to guide the fitting process, but occasionally one or more of the fitted parameters converges to a limit imposed by a constraint. In such cases, the resulting fit is probably wrong and it is reasonable to discard it from further analysis. Finally, as in \textsc{GalfitM} the three components can interchange, we identified the disc has being the component with the largest effective radius, the bar being the elongated component and the bulge the component with the smallest effective radius at the end of step III.

In the case of unbarred galaxies, we used a similar method of fitting two components, a disc and a bulge. For the bulge, we used $10\%$ of the disc component's effective radius (from step I) as an initial guess, an initial S\'ersic index of $n=2$ and an initial axis ratio of $b/a=0.8$. The absolute values of the initial bulge $r_{e}$ were, on average, similar to the initial guesses in the case of barred galaxies, so the models for barred and unbarred galaxies are consistent.

We stress that the galaxies modeled in this paper are simple representations of galaxy structures, in which the galaxies can be represented by a bulge, bar and an exponential disc. In reality, galaxies are more complex, showing complex structures such as spiral arms, rings etc. Although fitting all these different features in \textsc{GalfitM} is possible, it would require much more detailed attention for each galaxy, which is beyond the scope of this paper. Our aim was to keep the models relatively simple and uniform over a large range of angular sizes and surface brightness, while also quantifying bar structural parameters for a large sample of barred galaxies. We further discuss the addition of another component for unbarred galaxies, lenses, in Section \ref{lenses}.

\subsection{Inspecting the models}
\label{model}

\begin{figure}
 \includegraphics[width=1.05\columnwidth]{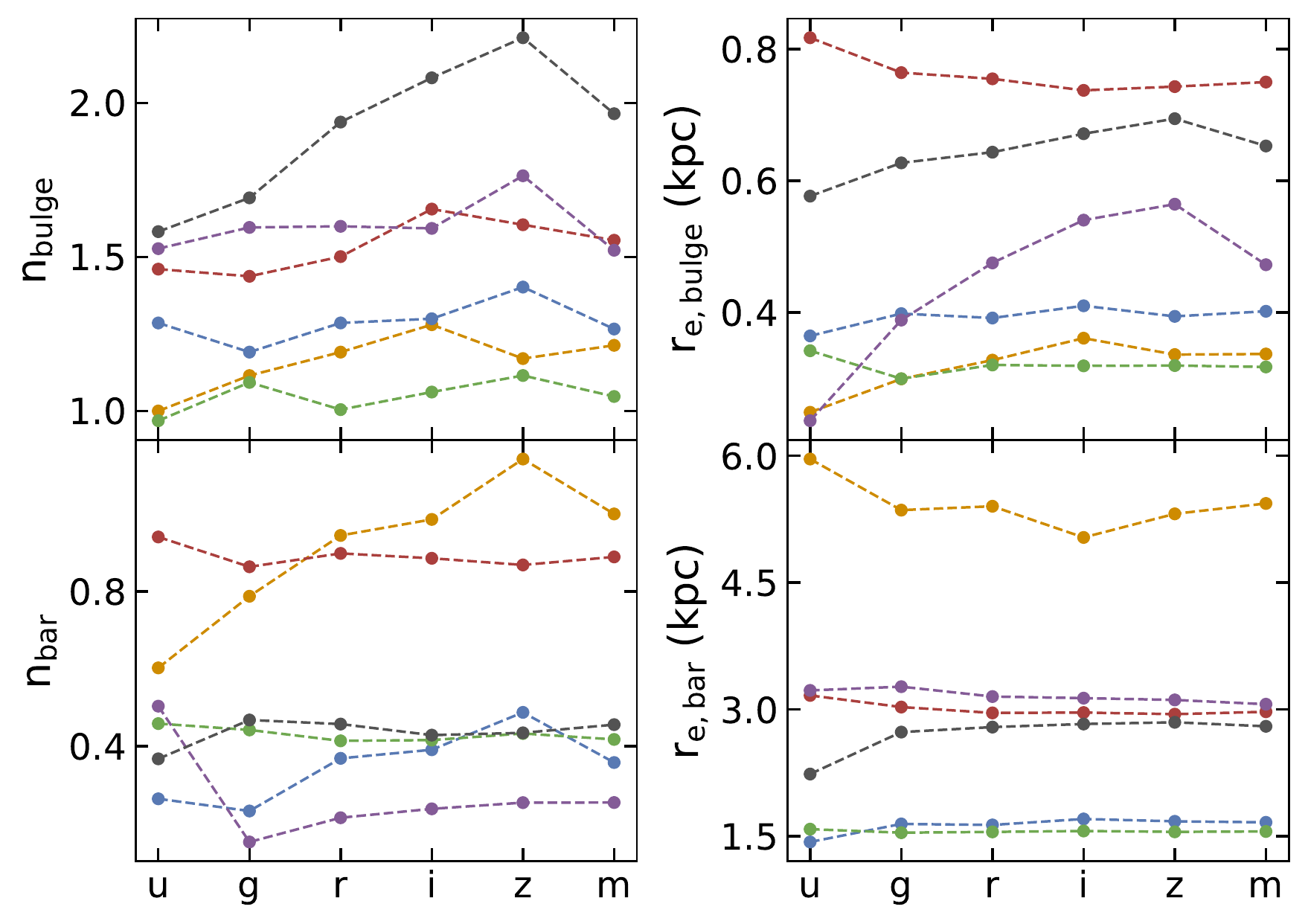}
 \caption{The dependence of the bulge and bar parameters with wave band for 6 randomly selected galaxies, fitted independently in the \textit{ugriz} bands. For comparison, we also plotted the parameters in the multi-band fitting, denoted by \textit{m}, where the $r_{e}$'s and $n$'s were kept constant with wavelength. Thus, there is a single value for $r_{e}$ and $n$ for all the bands.}
 \label{multi}
\end{figure}

\begin{figure}
 \includegraphics[width=\columnwidth]{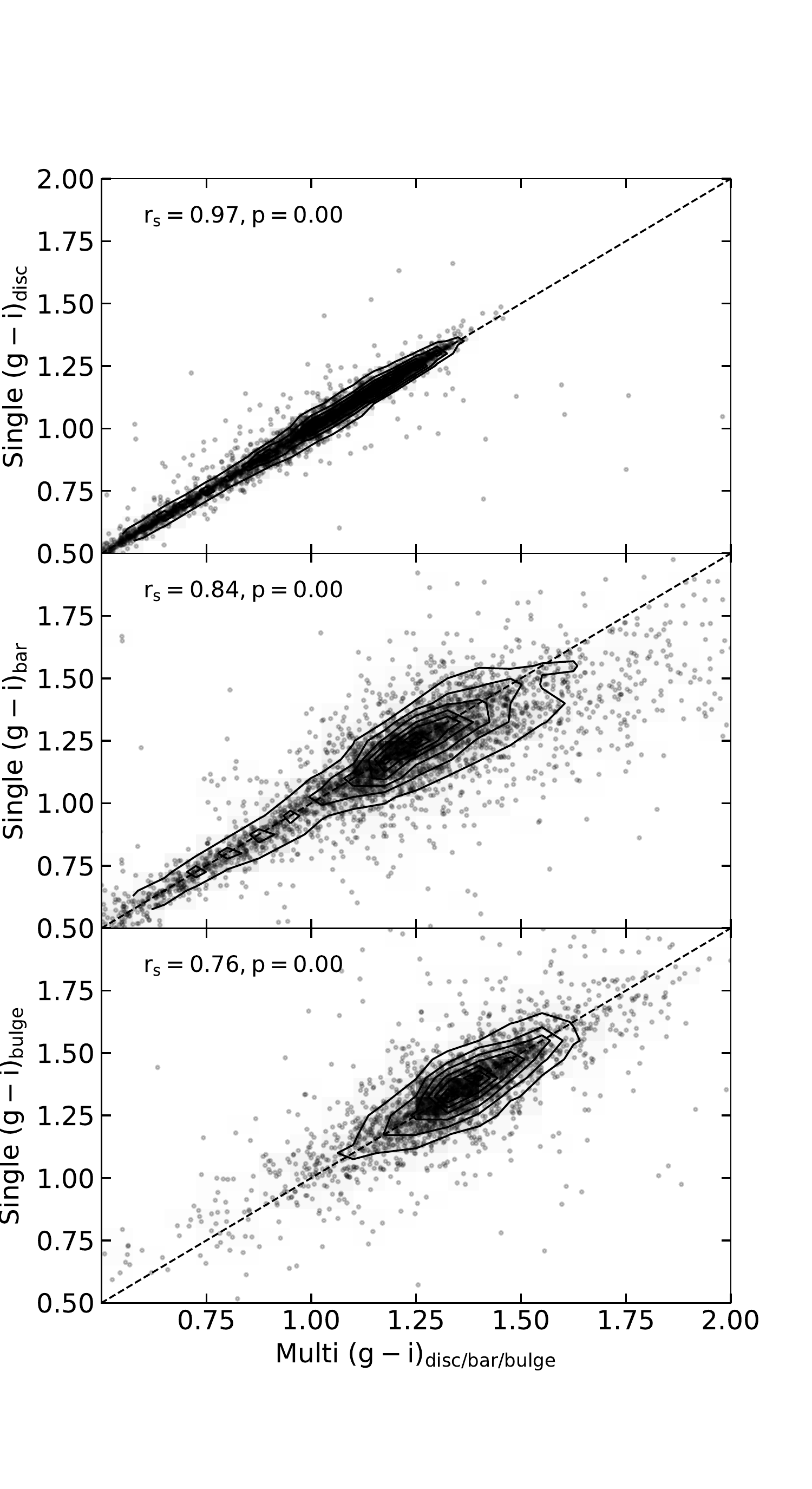}
 \caption{The correlation of (g-i) colours for the disc, bar and bulge components with the bands fitted independently (y-axis) against the same colours in multi-band fitting (x-axis). The Spearman $r_{s}$-coefficient is shown at the top and the 1-1 line is drawn. The two values are clearly correlated and lie on the 1-1 line, with the bar component showing the largest spread.}
 \label{gi_multi}
\end{figure}

The output of \textsc{GalfitM} is a \textsc{fits} file with 15 layers ($3\times5$ bands): the image, the model and the residual as seen in the example in Figure \ref{example_fitting}. \textsc{GalfitM} converged for 4,492 of 5,282 barred galaxies, or 85\% of the initial galaxies. In most of the cases where it failed, \textsc{GalfitM} either failed to converge for one of the parameters\footnote{one of the parameters was problematic, marked by *...* in the \textsc{GalfitM} output, and hence considered not to be reliable.}- for example the bar or bulge axis ratio being too small - or the low S/N made it impossible to extract a magnitude in one of the 5 bands.

To have a reliable sample of fitted galaxies, we selected only the fits with the following physical constraints: discs, bars and bulges having $r_{e}<200$ pixels, as all the components of disc galaxies should have effective radii smaller than $1.20\arcmin$(corresponding to $\sim$$10$ kpc at the lowest redshift of the sample). We also selected only bulges with $n_{\mathrm{bulge}}<7.8$ and axis ratios >0.3 and bars with $n_{\mathrm{bar}}<7.8$ to avoid components converging to a constraint, as discussed in the previous subsection. Finally, one of the authors (SK) visually inspected all the fits and compared the two component (disc+bar) to the three component fit (disc+bar+bulge), by looking at the image, model and residuals. Even though \textsc{GalfitM} returns a goodness-of-fit reduced $\chi^2$ value, $\chi^2_{\nu}$,  this is an indicator if one model is favoured compared to another, but not if the model has a physical meaning. In general, because of the complex morphology of galaxies, adding a further component always decreases the $\chi^2_{\nu}$ of the model, as the number of degrees of freedom is increased. In our fits, 98\% of the $\chi^2_{\nu}$ values varied between 1 and 2, with a median $\chi^2_{\nu}\sim1.2$. The models with two or three components need to be inspected to check if they are physically relevant for the galaxy in the images.

For 1,246 galaxies, the two component (disc+bar) fit proved to be a better fit (when judged by eye), given the lack of a significant third component (a bulge) in the galaxy images and in the residuals. There were 1,692 galaxies with good three component (disc+bar+bulge) fits. For 523 galaxies, the second stage of the fitting process (disc+bar model) converged to a disc+bulge model instead (the axis ratio of the second component was $b/a>0.6$, which is larger than the typical axis ratio of a bar). Since a bar was present in the galaxy images, we refitted these galaxies with three components, adding a bar, with the same initial parameters as in the second step of the fitting procedure. Furthermore, there were 1,031 galaxies for which \textsc{GalfitM} converged, but were discarded because the models were unphysical and did not represent a suitable disc+bar+bulge nor disc+bar model: in some cases a spiral arm, brighter star formation knot (clump) or overlapping star was fitted instead of one of the components. In other cases, nearby stars or galaxies had not been masked out and one of the components converged to their position, rather than to the galaxy which we tried to fit. Finally, 3,461 barred galaxies have meaningful fits out of the initial 5,282 (66\%), which is a significantly large sample to study the properties of barred galaxies. 

The magnitude-redshift distribution of the initial and successfully fitted sample of barred galaxies can be see in Figure \ref{magz}. We will refer to the 1,246 galaxies with disc and bar components as the \textsc{Disc dominated} sample, and to the 2,215 galaxies with discs, bars and bulges as the \textsc{Obvious bulge} sample. We only take the most suitable model for each galaxy (disc+bar or disc+bar+bulge), therefore the two samples of \textsc{Disc dominated} and \textsc{Obvious bulges} do not overlap. From this large sample of successfully fitted barred galaxies we select a \textsc{volume-limited} subsample of 2,435 \textsc{barred} galaxies, brighter than $M_{\mathrm{r}}<-20.15$.

It is important to note that only 315 out of the 1,401 (22\%) low mass barred galaxies ($M_{\star}<10^{10.25} M_{\odot}$) required a bulge component to achieve a good fit. The growth of inner stellar concentrations (bulges) is thought to occur around the mass of $\sim10^{10.5} M_{\odot}$, noticed by \citet{Kauffmann2003b}, who showed that the properties of galaxies in the low redshift universe change significantly at this mass. 

In Galaxy Zoo, citizen scientists were asked to visually classify the prominence of bulges of galaxies into four categories: \textsc{No-bulge}, \textsc{Just-noticeable}, \textsc{Obvious}, \textsc{Dominant} (further description in \citet{Willett2013} and \citealt{Simmons2013}). We can compare our structural classification with the GZ classification debiased vote fractions. Based on the bulge question, we divide the sample into:  `disc dominated' (having debiased likelihoods \textit{no-bulge}+\textit{just-noticeable} > \textit{obvious}+\textit{dominant}) and `obvious bulges' (having \textit{no-bulge}+\textit{just-noticeable} < \textit{obvious}+\textit{dominant}). The fraction of galaxies differing between our visual classification of bulges, based on the structural decomposition of the fits and residuals, and the Galaxy Zoo volunteers' classification of the bulge prominence is only 15\%. This is mostly due to galaxies fitted with disc+bar+bulge components being classified as being disc dominated (10\%) and 5\% of galaxies fitted with two components being identified as having obvious bulges. In Figure \ref{BT_gz} we plot the $B/T$ luminosity ratio versus the GZ vote fractions for bulge prominence. There is a significant correlation between the volunteers' classification, our inspection and the decomposition, therefore we proceed using our split into disc+bar and disc+bar+bulge fits in the following section.

\subsection{Tests}

\subsubsection{Fitting the \textit{ugriz} bands independently}
\label{test1}

We tested the reliability of the multi-band fitting compared to single band fitting, by decomposing all the fitted barred galaxies with two (disc+bar) and three components (disc+bar+bulge) independently in the 5 (\textit{ugriz}) SDSS bands. For this, we used the parameters from the multi-band fitting as initial guesses for all bands and refitted the five bands by allowing the $r_{e}$, $n$ and the centre to vary freely with band. We kept the axis ratio and position angle to be constant with band in all fits to prevent the components from interchanging. In this case of fitting 5 single bands independently, the fits to only 3,102 galaxies converged to meaningful values, showing that constraining parameters in multi-band fitting increases the number of reliable fits.

As shown in Figure \ref{multi} for 6 randomly selected galaxies out of the 3,102 fitted galaxies, the structural parameters for the bars and for the bulges of the fitted galaxies vary slightly with wavelength, but do not change significantly (typically much less than a factor of two). For the 6 galaxies, we also compare the parameters in the single-band fitting to the multi-band fitting, denoted with $m$ in Figure \ref{multi}, the multi-band parameters agreeing well with the parameters fitted in individual bands. The $\chi^2$ minimization in \textsc{GalftiM} uses the measured pixel-by-pixel noise as a weight, so although individual bands are not given different weights, those which are noisier ($u$, $z$) will have lower weights. This is seen clearly in Figure \ref{multi}, where the multi-band parameters trace the $g$, $r$, $i$ features closer than $u$ and $z$. In particular, the converged values are more similar to the values in the \textit{i}-band, which is the deepest image in SDSS data and, hence, the band in which the decompositions are most reliable.

A similar multi-band fitting procedure was applied to bulge-disc decompositions of 163 artificially redshifted nearby galaxies and shown to improve the measurements of structural parameters \citep{Vika2014}. Figure 1 in \citet{Vika2014} shows a similar trend for the measured parameters of a two component fit with wavelength. 

To check whether the estimated magnitudes are similar between the single and multi-band fitting, we plot the ($g-i$) colours in Figure \ref{gi_multi} for all 3,102 galaxies. There is a clear 1-1 correlation for all the three components, with the discs showing the smallest spread and the bars showing the largest spread in colours. Even though the magnitudes for the components of individual galaxies do not match exactly, the advantage of using multi-band fits is that they effectively use the same aperture in each band, while the colours of the single-band fits vary due to inconsistent decompositions in different bands. Furthermore, the sample size using single band fits will be considerably smaller, due to the larger proportion of fits that failed. Therefore, in what follows, we will use only the parameters from multi-band fitting.

\subsubsection{\textsc{MONTAGE} versus single frames}
\label{Montage}

We next check the effect of using \textsc{MONTAGE} to coadd the images. Using \textsc{MONTAGE} and multiple fields has some obvious advantages: being able to create images of galaxies close to the edges of the fields, with sufficient background around them, while also increasing the S/N ratio. It also has some disadvantages, such as combining PSFs from different observations when coadding the frames. To test the effect of using \textsc{MONTAGE} to create the images, we fit three components, using the same method as before, to $\sim$1,500 barred galaxies with obvious bulges which were not situated at field edges. There is a higher failure rate for the galaxies where \textsc{MONTAGE} is not used compared to the stacked images, because of the lower S/N ratio of the images in the overlap region. We compared all the fitted parameters between the single frames and multi frames and find a clear correlation and no systematics in most parameters. The only parameters for which we notice a systematic change between the single band and MONTAGE are for the bulges, which might be due to the modified PSFs: 1) the bulge S\'ersic indices, $n_{\mathrm{bulge}}$, in the single frames are 1.3 times higher than in the \textsc{MONTAGE} frames; 2) the bulge effective radii, $r_{\mathrm{e,bulge}}$ are $\sim$10\% smaller in the single frames. However, this is the same effect observed for the two parameters. The $n$ and $r_{e}$ are related for a component with fixed flux, therefore we expect that a change in one parameter to result in a change for the other. The median colours of the three components change insignificantly: $\Delta(u-r)=0.04$, $\Delta(g-i)=0.02$, $\Delta(r-z)=0.01$. We expect these small effects to occur in both barred and unbarred galaxies.

Therefore, it is advantageous to use \textsc{MONTAGE} to recover the parameters of a higher fraction of galaxies, with the expense of smoothing the data to a small extent, having the main effect of possibly estimating 1.3 times smaller bulge S\'ersic indices.

\subsubsection{Uncertainties}

\textsc{GalfitM} computes statistical errors (typically of $\sim$few \%) internally based on the covariance matrix produced during the least-squares minimisation by the Levenberg-Marquardt algorithm. They are known to underestimate the true error because it assumes that the only source of error is Poisson noise \citep{Haeussler2007}. In reality, uncertainties are underestimated because they do not take into account the errors due to sky measurements, improper masking, correctness of the PSF, the assumed models for the galaxy and parameter degeneracy. Uncertainties in the background level are one of the main sources of errors, especially for components with high S\'ersic indices, as these have extended wings \citep{Peng2010}.

\citet{Vika2013} showed that the uncertainties in a single S\'ersic fit with \textsc{GalfitM} of images similarly created with \textsc{MONTAGE} in the $ugriz$ bands are typically: for $magnitude$
($\pm0.13, \pm0.09, \pm0.10, \pm0.11, \pm0.12$ mags), $r_{e}$ ($\pm12\%,\pm±11\%, \pm12\%, \pm14\%, \pm15\%$) and $n$ ($\pm9\%, \pm11\%, \pm14\%, \pm15\%, \pm17\%$). These were based on the uncertainties in estimating the sky flux, which dominates the error budget. The uncertainties on fitting multiple components are more complex, \cite{Vika2014} shows that the bulge $n$ and $r_{e}$ can vary by up to 25\%, while the uncertainties in the disc components in the disc+bulge decompositions are similar to the uncertainties in the single S\'ersic fits. Since we used the same software and images of the same quality, we believe our uncertainties in the disc, bar and bulge parameters are similar to those found by \cite{Vika2014} in disc+bulge decompositions. Even though the individual fits can have substantial scatter, the median values for the entire population are robust.

\section{Disc, Bar and Bulge properties}
\label{barred}
\begin{table*}
\caption{Structural parameters of discs, bars and bulges for 10 randomly selected barred galaxies out of the 3,461 galaxies fitted with disc+bar or disc+bar+bulge components. Columns (3), (7), (11) show the integrated $i$-band magnitudes (from fits, not corrected for Galactic extinction), columns (4), (8), (12) show the S\'ersic indices, columns (5), (9), (13) the effective radii in pixels and columns (6), (7), (14) the axis ratios from the multi-band fits of the three components. We remind the reader that the measured bulge S\'ersic indices in the coadded frames are $\sim$1.3 times smaller than those in single frames, as discussed in Section \ref{Montage}. The magnitudes in the $u$, $g$, $r$, $z$ bands are also available. Full table is available in the electronic version of the paper.}
\begin{tabular}{cc|r|r|r|r|r|r|r|r|r|r|r|r|r|r|}
\hline
\hline
SDSS DR8 id & Fit components & \multicolumn{4}{c}{Disc} & \multicolumn{4}{c}{Bar} &  \multicolumn{4}{c}{Bulge}\\
 & & $mag$ &\textit{n}& \textit{$r_{e}$} & \textit{$b/a$} & $mag$ &\textit{n}& \textit{$r_{e}$} & \textit{$b/a$} & $mag$&\textit{n}& \textit{$r_{e}$} & \textit{$b/a$}\\

\hline
  1237668312168202669 & disc+bar & 15.57 & 1.0 & 26.96 & 0.84 & 18.68 & 0.33 & 3.84 & 0.38 & - & - & - & - \\
  1237668335787901378 & disc+bar & 16.78 & 1.0 & 11.28 & 0.86 & 16.76 & 1.01 & 2.86 & 0.60 & - & - & - & - \\
  1237668335787704768 & disc+bar+bulge & 14.75 & 1.0 & 19.48 & 0.98 & 17.17 & 0.39 & 5.29 & 0.33 & 17.15 & 0.63 & 1.48 & 0.83 \\
  1237667783395508535 & disc+bar & 15.81 & 1.0 & 24.59 & 0.87 & 17.24 & 1.05 & 9.47 & 0.33 & - & - & - & - \\
  1237668272988487820 & disc+bar+bulge & 14.69 & 1.0 & 28.37 & 0.79 & 15.78 & 0.43 & 13.58 & 0.43 & 16.05 & 1.13 & 3.35 & 0.68 \\
  1237665230522351799 & disc+bar & 15.59 & 1.0 & 21.92 & 0.72 & 17.95 & 1.90 & 4.81 & 0.19 & - & - & - & - \\
  1237665231059091845 & disc+bar & 16.58 & 1.0 & 13.09 & 0.75 & 18.44 & 0.38 & 7.28 & 0.31 & - & - & - & - \\
  1237665565007151489 & disc+bar+bulge & 15.44 & 1.0 & 20.92 & 0.86 & 16.03 & 0.38 & 11.76 & 0.29 & 17.02 & 1.25 & 3.35 & 0.49 \\
  1237667782857195666 & disc+bar+bulge & 15.65 & 1.0 & 21.04 & 0.68 & 18.27 & 0.10 & 14.42 & 0.17 & 19.63 & 0.30 & 1.22 & 0.34 \\
  1237648721790697923 & disc+bar+bulge & 16.29 & 1.0 & 16.25 & 0.94 & 16.57 & 0.74 & 7.53 & 0.47 & 16.96 & 0.66 & 1.43 & 0.69 \\
\hline\end{tabular}
\label{table2}
\end{table*}

\begin{table*}
\caption{Properties of the same 10 galaxies as in Table \ref{table2}, fitted with disc+bar or disc+bar+bulge components. Redshifts and $r$-band Petrosian absolute magnitudes ($M_\mathrm{r}$) are drawn from SDSS DR7 and the stellar masses are drawn from average values in the MPA-JHU catalogue \citep{Kauffmann2003a}. Column (5) shows the debiased bar likelihood of the galaxies from the GZ2 catalogue \citep{Willett2013}, based on the volunteers' visual inspection. Disc-to-total, bar-to-total and bulge-to-total luminosity ratios in the $i$-band are given in columns (6), (8), (10). Columns (7), (9), (11) show the $(g-i)$ colours of the three components, corrected for Galactic reddening and extinction using the maps from \citet{Schlegel1998} and \textit{k}-corrected \citep{Blanton2007}. Finally, column (12) shows the reduced-$\chi^2$ value of the fits. Full table is available in the electronic version of the paper. Luminosity ratios in the $u$, $g$, $r$, $z$ bands, as well as other colours ($u-r$ and $r-z$) are available in the online table. }
\begin{tabular}{|c|c|c|r|r|r|r|r|r|r|r|r|}
\hline
\hline
SDSS DR8 id & Redshift & $M_\mathrm{r}$ & $\log(M_{\star})$ & $p_{\mathrm{bar}}$ & \multicolumn{2}{c}{Disc} & \multicolumn{2}{c}{Bar} &  \multicolumn{2}{c}{Bulge}\\
 & & &$[M_{\odot}]$& & $D/T$ &$(g-i)_{\mathrm{d}}$ & $Bar/T$ & $(g-i)_{\mathrm{bar}}$ & $B/T$ & $(g-i)_{\mathrm{b}}$ & $\chi_{\nu}^{2}$\\
\hline
  1237668312168202669 & 0.036 & -20.11& 9.92 &0.56 & 0.95 & 0.67 & 0.05 & 0.88 & - & - & 1.19\\
  1237668335787901378 & 0.035 & -19.64 & 9.80 &0.61 & 0.49 & 0.45 & 0.51 &  0.88 & - & - & 1.19\\
  1237668335787704768 & 0.047 &-21.72 & 10.90 & 0.72 & 0.82 & 0.98 & 0.09 & 1.27 & 0.09 & 1.50 & 1.20\\
  1237667783395508535 & 0.035 & -19.85 & 9.24 & 0.69 & 0.79  & 0.48 & 0.21 & 0.33 & -  & - & 1.22\\
  1237668272988487820 & 0.035 & -21.38 & 10.85 & 0.81 & 0.60 & 1.19 & 0.22 & 1.18 & 0.17 & 1.18 & 1.15\\
  1237665230522351799 & 0.058 & -21.29 & 10.31 & 0.59 & 0.90 & 0.52 & 0.10 & 0.92 & - & - & 1.25\\
  1237665231059091845 & 0.031 & -19.11 & 9.42 & 0.82 & 0.85 & 0.79 & 0.15 & 0.54 & -  & - & 1.14\\
  1237665565007151489 & 0.037 &-20.98 & 10.53 & 0.96 & 0.55 & 0.88 & 0.32 & 0.90 & 0.13   & 1.19 & 1.22\\
  1237667782857195666 & 0.035  &-20.14 & 10.00 & 0.65 & 0.90 & 0.82 & 0.08 & 0.96 & 0.02  & 1.60 & 1.20\\
   1237648721790697923 & 0.039 &-20.53 & 10.37 & 0.56 & 0.43 & 1.00 & 0.33 & 1.07 & 0.23 & 1.15 & 1.13\\
\hline\end{tabular}
\label{table3}
\end{table*}

First, in Section \ref{component_colors} we study the colour distribution of discs, bars and bulges of barred galaxies, the differences in component colours within individual galaxies, and trends with stellar mass. Then, in Section \ref{bars}, we look at the properties of bars and how they vary with different galaxy properties. Furthermore, in each subsection we compare our findings with other published studies on barred galaxies. The structural parameters, luminosity ratios and colours of the discs, bars and bulges for the successfully fitted 3,461 galaxies are given in Table \ref{table2} and \ref{table3}.

\begin{figure*}
 \includegraphics[width=\textwidth]{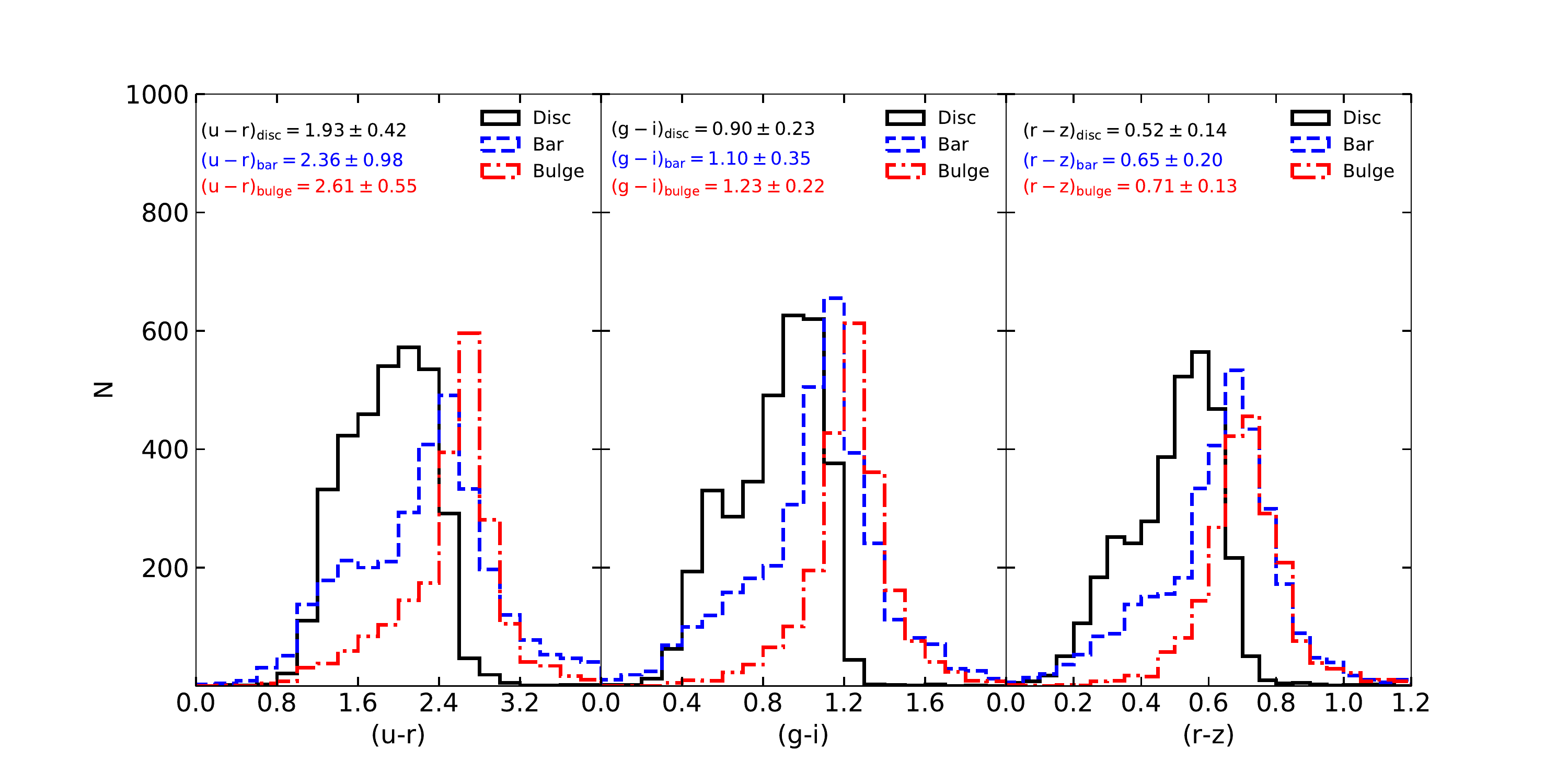}
 \caption{The $(u-r)$, $(g-i)$ and $(r-z)$ colours of the different galaxy components for all the fitted barred galaxies (3,461 galaxies). The discs are bluer than the bars, which in turn are slightly bluer than the bulge. The median colours and their corresponding $1\sigma$ spreads are represented for each component, since the median is less sensitive than the mean to outliers.}
 \label{colours}
\end{figure*}

 \begin{figure}
 \includegraphics[width=1.05\columnwidth]{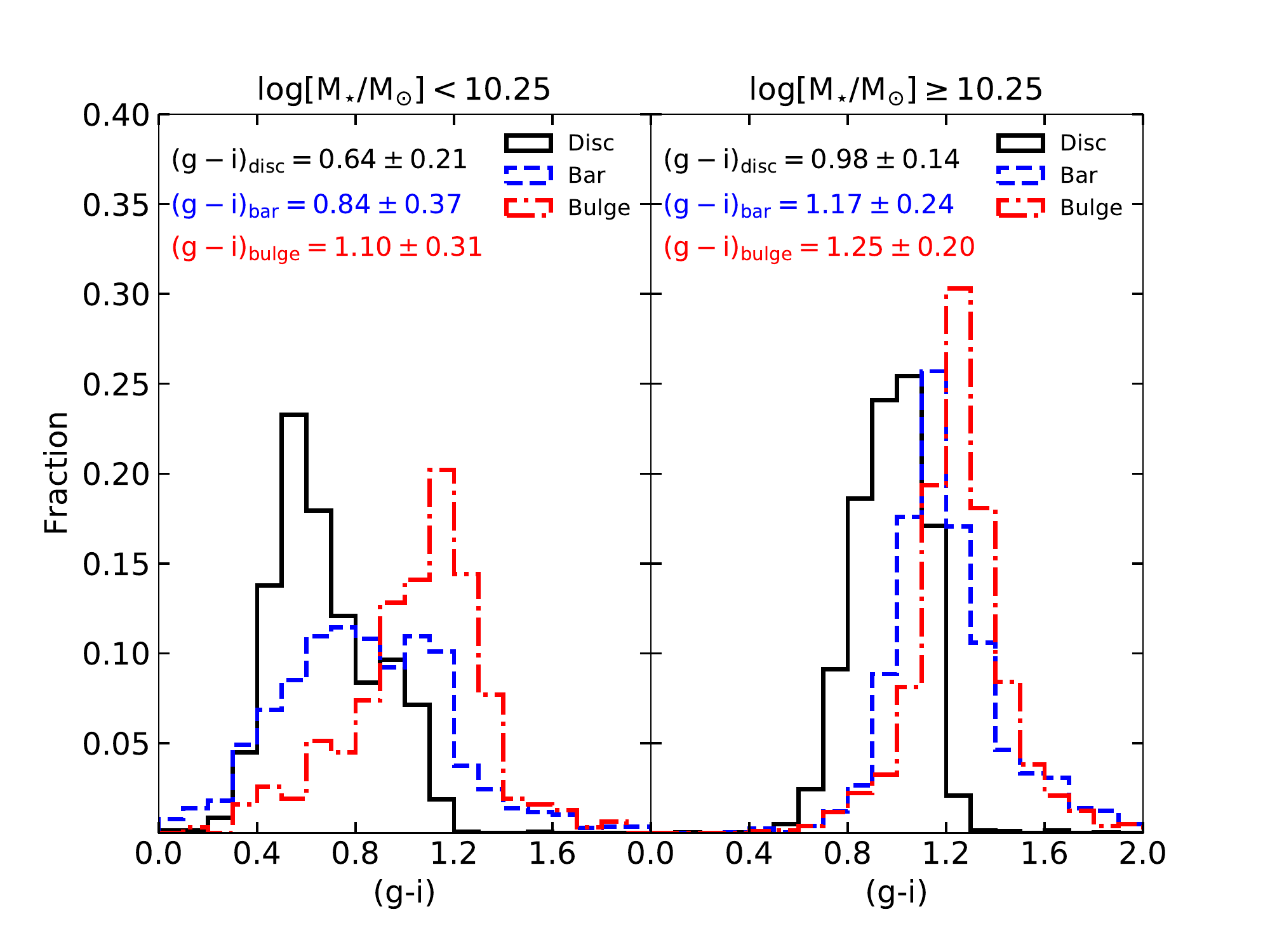}
 \caption{Normalised histograms of the $(g-i)$ colours of the different galaxy components, split by galaxy mass. There were 1,086 low mass galaxies fitted with disc+bar, 315 with disc+bar+bulge. Similarly, there were 1,900 high mass galaxies fitted with disc+bar+bulge and 160 with disc+bar. The discs and bars of lower mass galaxies are significantly bluer than those of high mass galaxies, while the bulges are only moderately bluer compared to their high mass counterparts. The median colours and their corresponding $1\sigma$ spreads are represented for each component.}
 \label{barred_colours_mass}
\end{figure}

\subsection{Component colours}
\label{component_colors}

\begin{figure}
 \includegraphics[width=1\columnwidth]{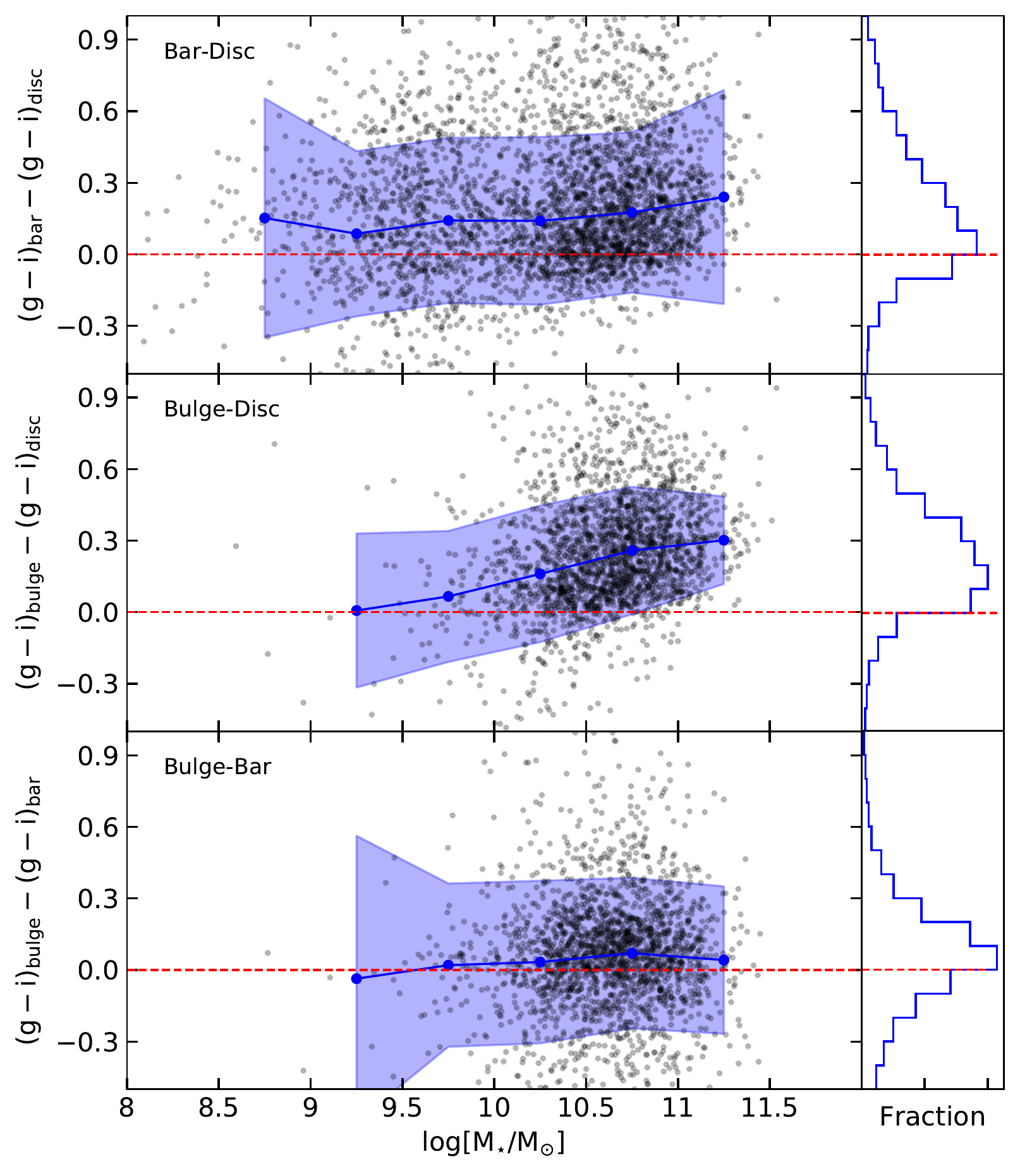}
 \caption{The differences in $(g-i)$ colours of the three galaxy components, showing the change in colour for each individual galaxy. This plot contains all the successfully fitted barred galaxies (3,461) with disc+bar (1,246 galaxies) and disc+bar+bulge (2,215 galaxies) components. The median $(g-i)$ colour is plotted with blue in stellar mass bins of $\log{(\frac{M_{\star}}{M_{\odot}}})=0.5$ (excluding $>10\sigma$ outliers) and the shaded band represents the $1\sigma$ scatter. }
 \label{col_gradient}
\end{figure}

One key result of our work is the distribution of colours of the three components, this being possible with multi-band fitting. The colours of the individual components are important because they reflect the distribution of stellar populations within galaxies. In Figure \ref{colours}, three different colour distributions, $(u-r)$, $(g-i)$, $(r-z)$, are plotted for the discs, bars and bulges. The colours were corrected for Galactic dust reddening and extinction, using the maps from \citet{Schlegel1998}\footnote{Using \url{https://github.com/rjsmethurst/ebvpy}}. The magnitudes were \textit{k}-corrected \citep{Blanton2007}; these corrections are small given the proximity of our sample.

As seen in Figure \ref{colours}, there is a clear difference between the colours of the three components of barred galaxies. The discs are clearly bluer than the bars, which in turn are slightly bluer than the bulges, in $(u-r)$, $(g-i)$ and $(r-z)$ colours. In what follows, we will focus on the $(g-i)$ colours because they are less prone to dust extinction, while the bands are sufficiently separated in wavelength to probe both star forming and quiescent stellar populations.

In $(g-i)$ colours, the median difference between bulges and discs is $\Delta(g-i)_{\mathrm{b,d}}=0.33$ and between the bars and discs $\Delta(g-i)_{\mathrm{bar,d}}=0.20$. Our sample of barred galaxies contains galaxies of stellar masses between $10^{8} M_{\odot}$ and $10^{11.5} M_{\odot}$. Since the colours and properties of galaxies are dependent on mass, we split the sample by galaxy stellar mass into low-mass, $M_{\star}<10^{10.25} M_{\odot}$ (1,401 galaxies), and high-mass $M_{\star}\geq10^{10.25} M_{\odot}$ (2,060 galaxies) and we plot the distribution of component $(g-i)$ colours, in Figure \ref{barred_colours_mass}. As expected, the colours of the components of lower mass galaxies, especially the discs and bars, are bluer compared to high-mass galaxies. The shift in colours is less significant for the bulges, which still appear red in colour, but their $(g-i)$ colour spread increases, although there are only 315 low mass galaxies fitted with a bulge. At high masses, the discs and bar components appear to be much redder compared to the lower mass counterparts. For high mass galaxies, the colours of bars and bulges are more similar, suggesting that they host similar, old stellar populations.

Another study using disc+bar+bulge decomposition with available $(g-i)$ colours from fits is the work by \citet{Gadotti2009, Gadotti2010a} who fitted 291 face-on (with axial ratio $b/a\geq0.9$) barred galaxies with masses $M_{\star}>10^{10} M_{\odot}$ from SDSS. They find median values of $(g-i)_{\mathrm{disc}}=1.04\pm0.20$, $(g-i)_{\mathrm{bar}}=1.27\pm0.42$, $(g-i)_{\mathrm{bulge}}=1.26\pm0.39$ for the individual components. The main differences between our study and \citet{Gadotti2009} are the higher stellar masses and that their colours from the fits were not corrected for Galactic extinction. Furthermore, they fitted each band individually without constraining the parameters in different bands, therefore not measuring colours within the same $r_{e}$ as done in this study. Applying a similar selection for galaxy masses and not correcting for Galactic extinction, we find similar values for the discs and bars (in this study, $(g-i)_{\mathrm{disc}}=1.07\pm0.16$, $(g-i)_{\mathrm{bar}}=1.27\pm0.27$), and only slightly redder bulges in our study ($(g-i)_{\mathrm{bulge}}=1.35\pm0.22$). The size of our sample, which is roughly an order of magnitude larger, the slight difference in the sample selection (the galaxies in \citet{Gadotti2009} have lower inclinations, thus being less affected by internal dust extinction), different PSFs or the different fitting softwares used (as discussed further in Section \ref{bulges}) might account for the differences in the bulge colours and other bulge parameters.

Other authors have reported similar differences in colour between bulges and discs in disc+bulge decompositions. For example, in a multi-band bulge+disc decomposition of 163 galaxies, \citet{Vika2014} found a difference in the colours of discs and bulges of $\Delta(g-i)_{\mathrm{b,d}}\sim0.3$ for all late-type Sa-Sm galaxies, well in agreement with our study. Furthermore, \citet{Kennedy2016} using bulge+disc decompositions on galaxies from the GAMA survey also found that regardless of morphology, bulges are consistently redder than their corresponding discs. Finally, \citet{Head2014} found a bulge-disc colour difference of $(g-i)=0.09$ for S0 galaxies. Nevertheless, our observations that bulges are, in the vast majority of cases (91\% for both barred and unbarred galaxies), redder than their discs seems to be in contradiction with the spectroscopic observations of \citet{Johnston2014}, who found that bulges of S0 galaxies are consistently younger and more metal rich than their corresponding discs. Although differences might arise because our sample contains a mix of Hubble types. 

Next, instead of looking at the distributions of component colours for the entire population of barred galaxies, we can look at the component colours for individual galaxies. This should show in more detail how the colours of components are related.  As shown above, galaxy colours depend strongly on total stellar mass. Therefore, we plot the colour difference between each two of the three fitted components against the stellar mass \citep[drawn from average values in the MPA-JHU catalogue;][]{Kauffmann2003a}, in Figure \ref{col_gradient}. First, we notice that the bars are consistently redder than their accompanying discs (top panel), by $\Delta(g-i)\sim0.2$. There is a slight trend with stellar mass, higher mass galaxies having the reddest bars compared to their corresponding discs. Secondly, bulges are almost always redder than their associated discs, as suggested by Figure \ref{col_gradient} (middle panel), by $\Delta(g-i)\sim0.25$ on average, but appear to become more similar in colour to discs in lower mass galaxies, where they are much less common. Thirdly, we have already seen in Figure \ref{colours} that bars are bluer than bulges when comparing the fitted sample of barred galaxies, however Figure \ref{col_gradient} (bottom panel) shows that within the same galaxy they have similar colours. Disc dominated galaxies have bluer bars compared to the bars in galaxies fitted with a bulge component, therefore shifting the histogram corresponding to the bar component in Figure \ref{colours} to bluer colours. The trend in Figure \ref{col_gradient} (bottom panel) is relatively flat with stellar mass, suggesting a common evolution for the stellar populations of bars and bulges. 

Converting from colours of individual components to stellar ages is not trivial. Galaxy colours become redder as the stars in the galaxy age and at the same time the stellar metallicity increases as the surface temperature decreases and stars becomes less opaque. Using a simple model for a single stellar population with solar metallicity, an initial burst of star formation and optical colours predicted by the \citet{Bruzual2003} stellar population synthesis code we find that most of the stellar populations in bulges are consistent with being formed at $z\sim2$ (10 Gyr ago) with no significant rejuvenation. Only a small proportion of bulges extend to bluer colours, and hence having stellar population ages < 5 Gyr. Discs have stellar population ages of a few Gyrs, while bars host, in general, older stellar populations, having similar ages as the bulges. The ages discussed in this paragraph are the average ages of the stellar populations that dominate the light of the components, not the dynamical ages of the disc, bars or bulges. To study the stellar populations of barred galaxies in greater detail and to break the observed colour degeneracy, one has to use spatially resolved spectroscopy. Luckily, large scale IFU surveys such as MaNGA \citep{Manga2016} are in progress, which will allow us to better model the stellar populations in these galaxies. Disentangling stellar population ages and metallicities directly with MaNGA data and using 2D image decomposition will be the subject of future work. One such step in separating the spectra of bulges and discs using \textsc{GalfitM} and MaNGA data was achieved by \citet{Johnston2017}.

Furthermore, the effect of internal dust reddening should be considered when comparing the colours of different galaxy components. \citet{Masters2010} showed that the dust effects are systematic with the inclination of spiral galaxies, finding a total extinction from face-on ($i=0^\circ$) to edge-on ($i=90^\circ$) galaxies of 0.7, 0.6, 0.5, 0.4 and 0.3 mag for the $ugriz$ passbands. The extinction is much smaller from completely face-on ($i=0^\circ$) to moderately face-on ($i=60^\circ$) (0.17, 0.12, 0.07, 0.04 for the $ugri$ bands, using Equation (3) in \citet{Masters2010} and assuming no extinction in the $z$ band). The galaxies in our sample were selected to be moderately face-on ($i\lesssim60^{\circ}$), thus we do not expect the effect of dust to be significant. 

We also checked for systematic trends with inclination in our sample, by assuming that the fitted $b/a_{\textrm{disc}}$ can be easily translated to an inclination ($\cos^{2}{i}= \frac{{b/a}^2_{\textrm{disc}}-q^2}{1-q^2}$, where $q=0.2$, the intrinsic thickness of an edge-on disc, \citealt{Unterborn2008}). We find only a small trend of colours with inclination, such that at $i\sim60^{\circ}$, the $(g-i)$ colours of the bulges, bars and discs given by the lines of best fit are 1.29, 1.10 and 0.83, while for completely face-on galaxies ($i\sim0^{\circ}$) they are 1.19, 1.04 and 0.87, respectively. Hence the colour excesses between $60^{\circ}$ and $0^{\circ}$ are: $\Delta(g-i)_{\textrm{bulge}}\sim0.1$, $\Delta(g-i)_{\textrm{bar}}\sim0.06$ and $\Delta(g-i)_{\textrm{disc}}\sim-0.04$. We find that bulges suffer from more attenuation with inclination than discs, as also shown by \citet{Pierini2004} and \citet{Tuffs2004}. Perhaps counter-intuitive we find a negative dust attenuation for the discs, such that the face-on discs are redder compared to the slightly inclined ones. This can be an optical depth effect - for the more inclined galaxies we can better observe the outer stellar populations which are likely bluer, while for the face-on galaxies we better observe the inner disc which is intrinsically redder. \citet{Gadotti2010b} also found that the dust attenuation in the discs at low inclinations can be negative, suggesting that this is probably due to scattering of photons propagating parallel to the plane of the galaxy into the line of sight. 

Our sample contains both late and early-type galaxies, which contain different amounts of dust. Ideally, one should correct for the internal dust extinction, however even the different components of galaxies (discs, bars and bulges) contain different amounts of dust and hence suffer different dust extinctions \citep{Driver2008}. Considering the diversity of our sample, as well as the large range of masses in our study $10^{8}-10^{11.5}\:M_{\odot}$, it is impossible to correct for internal dust extinction using a simple relation. Therefore, the colours discussed in this paper were corrected only for Galactic extinction.

Dust might also affect the measured parameters of the components, as discussed in more detail by \citet{Pastrav2013a,Pastrav2013b}, especially at lower wavelengths. However, considering the face-on sample chosen for the decomposition and the multi-wavelengths used in this study ($ugriz$), its effects should be minimised.

\subsection{Properties of bars}
\label{bars}

Having identified the bars, we consider their properties in this subsection.

\begin{figure}
 \includegraphics[width=\columnwidth]{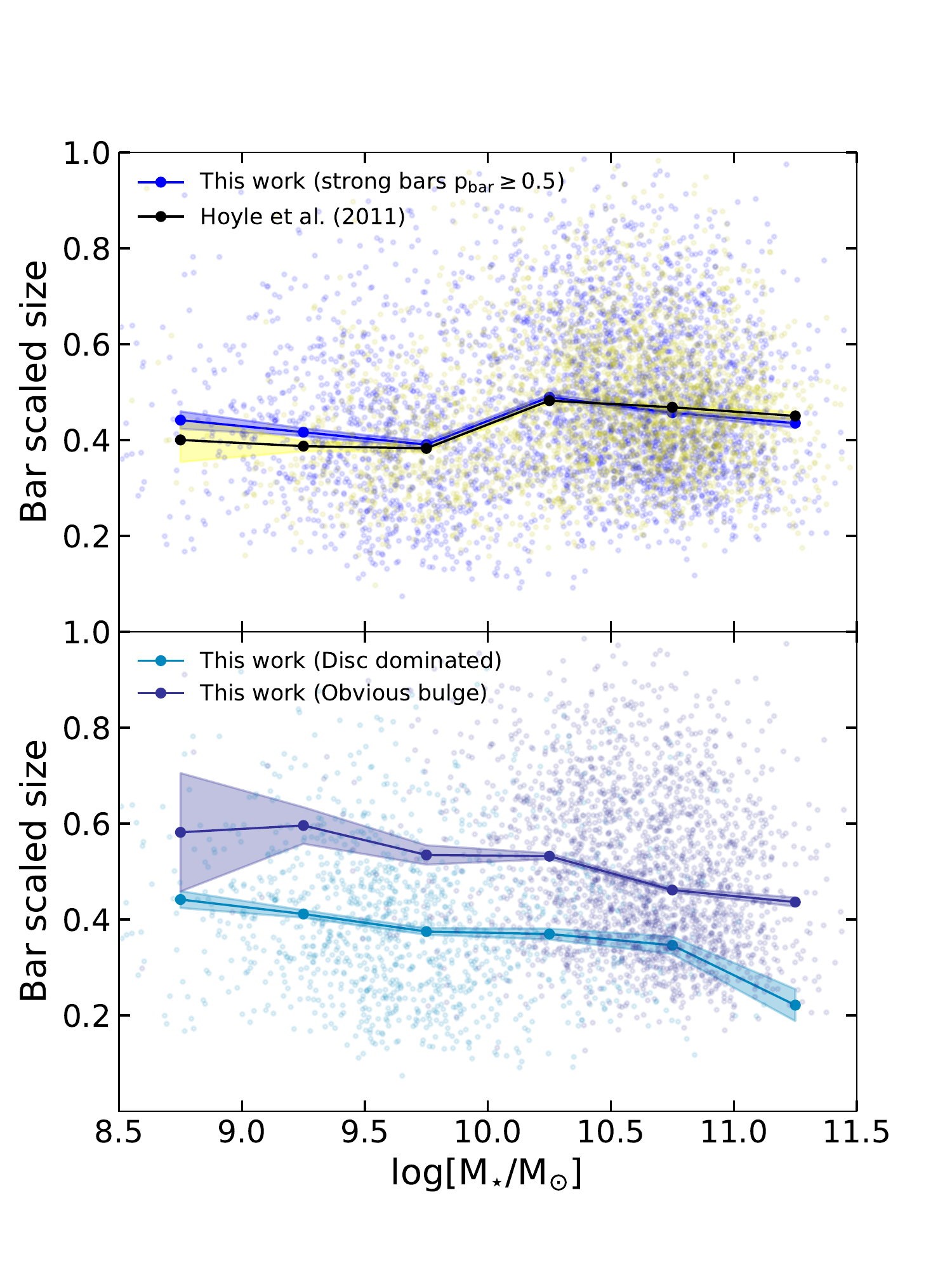}
 \caption{The scaled bar length, $r_{\mathrm{e,bar}}/r_{\mathrm{e,disc}}$ in this work, and $L/2R_{\textrm{Petro90}}$ in \citet{Hoyle2011}, as a function of stellar mass (top). The median bar size compared to galaxy size is constant at low masses and reaches a maximum of 0.5 at $M_{\star}\sim10^{10.25} M_{\odot}$, then the scaled size declines slightly with mass. In the bottom plot, the scaled bar length $r_{\mathrm{e,bar}}/r_{\mathrm{e,disc}}$ in this work split into `disc dominated' (fitted with disc+bar) and `obvious bulges' (fitted with disc+bar+bulge) is shown. Galaxies with significant bulges have consistently larger bar scaled lengths. Median values in stellar mass bins of $\log{(\frac{M_{\star}}{M_{\odot}}})=0.5$ are plotted and the shaded areas represent the $1\sigma/\sqrt{N}$ error on the mean per bin.}
 \label{scaled_barlength}
\end{figure}

In a previous Galaxy Zoo project, \citet{Hoyle2011}, volunteers were asked to measure bar lengths and widths of 3,150 local galaxies with strong bars using a Google Maps interface. Our sample of galaxies that was successfully fitted contains 1,700 barred galaxies that are also found in \citet{Hoyle2011}. Even though we do not measure the length of the bar in our study, the effective radii that we measure for the bar are correlated with the visually measured average bar lengths in \citet{Hoyle2011}. We find that the $r_{\mathrm{e,bar}}$ increases with stellar mass, but so does $r_{\mathrm{e,disc}}$. To investigate how the size of the bar changes compared to the size of the galaxy, we plot the ratio of the bar and disc effective radii (defined as the bar scaled size) as a function of stellar mass in Figure \ref{scaled_barlength}. As a comparison, we also plot the scaled bar length from \citet{Hoyle2011}, who, although used a different measure (the length of the bar divided by two times the radius containing 90\% of the Petrosian flux, $L/2R_{\mathrm{Petro90}}$) found a similar trend with stellar mass.  Across all stellar masses, strong bars identified in Galaxy Zoo are 20-80\% of the size of the discs and the bar scaled sizes are constant with stellar mass, at a first approximation. The median scaled bar size is $\sim40-50\%$ in both our measurements and those of \citet{Hoyle2011}. Both papers observe a peak in the relative bar size of $\sim50\%$ at $10^{10.25}\:M_{\odot}$ for strong bars, which is similar to the transition mass between disc dominated and galaxies with obvious bulges. In the bottom plot of Figure \ref{scaled_barlength}, the scaled bar length is plotted for the sample split into disc dominated galaxies and galaxies with obvious bulges showing that the peak in the relative bar size is due to the increasing prominence of bulges in our sample. Galaxies with obvious bulges have $\sim$25\% longer bar scaled sizes when compared to disc dominated galaxies. The fact that \citet{Hoyle2011} observe a similar trend using a different measure for the bar length suggests that it is not an artifact of the additional component (+bulge) added to our disc+bar fits around the same galaxy mass. Finally, at masses higher than $10^{10.25}\:M_{\odot}$, the bar scaled size drops to $\sim0.45$.

\begin{figure}
 \includegraphics[width=1.05\columnwidth]{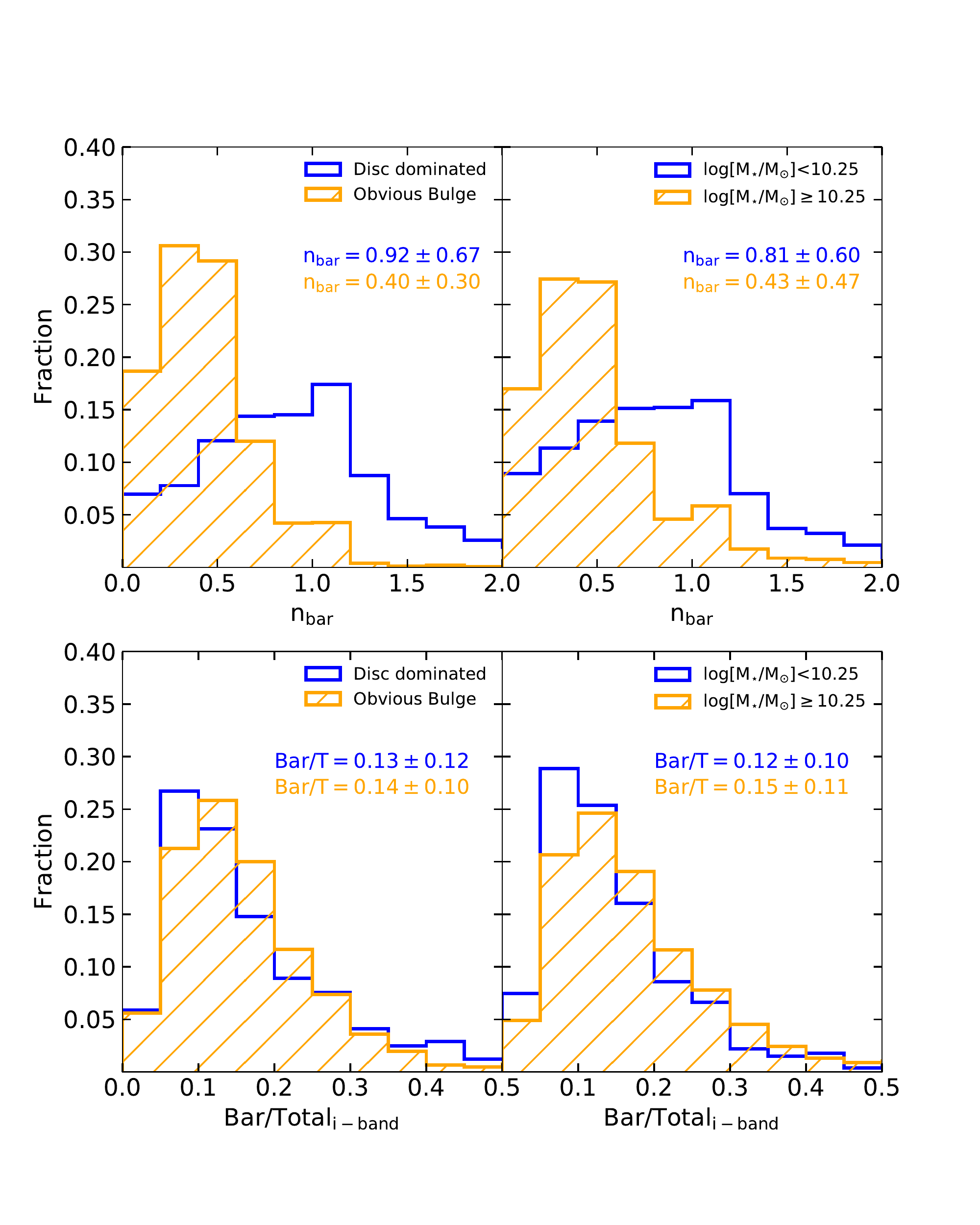}
 \caption{The first two figures (top) show the bar S\'ersic indices split into \textsc{Disc dominated} (modeled with disc+bar) and \textsc{Obvious bulge} (modeled with disc+bar+bulge) (left) and stellar mass bins (right). Low mass, disc dominated barred galaxies have bars with a broad distribution of profiles, with a large fraction of bars having exponential profiles, while high mass galaxies with prominent bulges have flatter profiles. The median S\'ersic indices of the bars are represented in the plot. The bottom plots show the bar-to-total luminosity in the \textit{i}-band. The bar-to-total luminosity ratio is consistent for \textsc{Disc dominated} and \textsc{Obvious bulge} galaxies (left), as well as for low and high mass galaxies.}
 \label{sersic_index}
\end{figure}

The measured axis ratio of the bar varies between 0.1 and 0.6, with a median and $1\sigma$ scatter of $b/a=0.31\pm0.12$, in good agreement with the expected values of 0.2-0.4 \citep{Kormendy2004}. 

Our measured axis ratios, in general, correspond well with other studies, but note that alternative measurement methods may lead to minor differences.  Our median axis ratio is $\sim$30\% higher than the axis ratio found by \citet{Hoyle2011}, $\left< b/a \right>=0.24\pm0.07$, but in this case the axis ratio was calculated as the ratio of the measured bar width to bar length. \citet{Gadotti2011} found a higher axis ratio of $\left< b/a \right>=0.37\pm0.10$, using a boxy fit, which is closer to the real shape of bars \citep{Athanassoula1990}. 

We now turn to the radial light profile of bars, as measured by their S\'ersic index. We notice a a significant difference (Kolmogorov-Smirnov test $k=0.52,\;p_{\textrm{KS}}<10^{-15}$) when the sample is split into `disc dominated' (disc+bar fit) and `obvious bulge' (disc+bar+bulge fit), as shown in the top-left panel of Figure \ref{sersic_index}. A similar difference, but less pronounced (K-S test $k=0.36,\;p_{\textrm{KS}}<10^{-15}$) is seen when the sample is split into low mass and high mass (top-right panel of Figure \ref{sersic_index}).  We remind the reader that there is a significant overlap between the `disc dominated' and low mass samples (and `obvious bulge' and high mass, respectively) as more disc dominated galaxies tend to have lower masses. We find that disc dominated, low mass galaxies have stellar bars with a S\'ersic index of $n_{\mathrm{bar}}=0.92\pm0.67$. On the other hand, high mass galaxies, many with obvious bulges, have bars with shallower, Gaussian, light profiles with $n_{\mathrm{bar}}=0.40\pm0.30$. We notice that 80\% of the galaxies with $n_{\mathrm{bar}}>0.8$ and almost all with $n_{\mathrm{bar}}$ between 1 and 2 are disc dominated, suggesting that the presence of a significant bulge is the most important factor in the bar light profile. Alternatively it might be possible that a faint bulge is not separable from the bar, but its presence acts to steepen the apparent bar profile. However, we find only a very weak correlation between $n_{bar}$ and $B/T$ (Spearman $r_{s}$-correlation test $r_{s}=0.09,\:p=0.0001$).

One of the first authors to observe a difference in the bar light profiles, in a sample of 11 barred galaxies, was \citet{Elmegreen1996}, who noticed that bars in early-type galaxies have a flatter profile compared to late-type ones, which have exponential profiles. They suggested that flat profiles arise from the overcrowding of old and young stars at the bar ends. Furthermore, \citet{Elmegreen1993} found in simulations that these differences arise because of a difference in their resonance locations. \citet{Kim2015} found a similar difference in the light profiles of bars has been observed, in a sample of 144 nearby galaxies, suggesting that the flatness of the bar profile can be used as a bar age indicator. In their study, galaxies with obvious bulges have a median S\'ersic index of $\left<n_{\mathrm{bar}} \right>\sim0.3$, while disc dominated systems have $\left< n_{\mathrm{bar}} \right>\sim0.85$. We agree with these previous findings, albeit using a much larger sample, therefore strengthening the result that bars in late and early-type galaxies have different radial light profiles.

Using the fits, another quantity that can be measured is the bar-to-total luminosity ($Bar/T$). In Figure \ref{sersic_index} (bottom panels) the $Bar/T$ luminosity ratio can be seen for the \textit{i}-band. The distribution of $Bar/T$ luminosity is consistent within all the five SDSS bands, hence only one band is shown. The $Bar/T$ ratio appears to be similar (K-S test $k=0.07,\;p_{\textrm{KS}}=0.002$) for `disc dominated'  galaxies and galaxies with `obvious bulges', as well as for low mass and high mass barred galaxies ($Bar/T\sim0.14$). There is only a $\sim$$10\%$ difference in the median $B/T$ value for galaxies with $M_{\star}<10^{10.25} M_{\odot}$, compared to higher mass galaxies, implying a mostly mass-independent bar growth. Hence, the $Bar/T$ luminosity ratio does not correlate with the bulge prominence or the mass of the galaxy. 

For galaxies with $M_{\star}>10^{10} M_{\odot}$, \citet{Gadotti2011} found a median $Bar/T\sim0.10$, 40\% smaller than in this study. We find a better agreement with \citet{Weinzirl2009}, who also used a similar decomposition method, although their sample comprised of only 80 barred galaxies and the images were in the \textit{H}-band.
 
\section{Comparison of barred and unbarred galaxies}
\label{barred_vs_unbarred}

\begin{figure}
 \includegraphics[width=0.95\columnwidth]{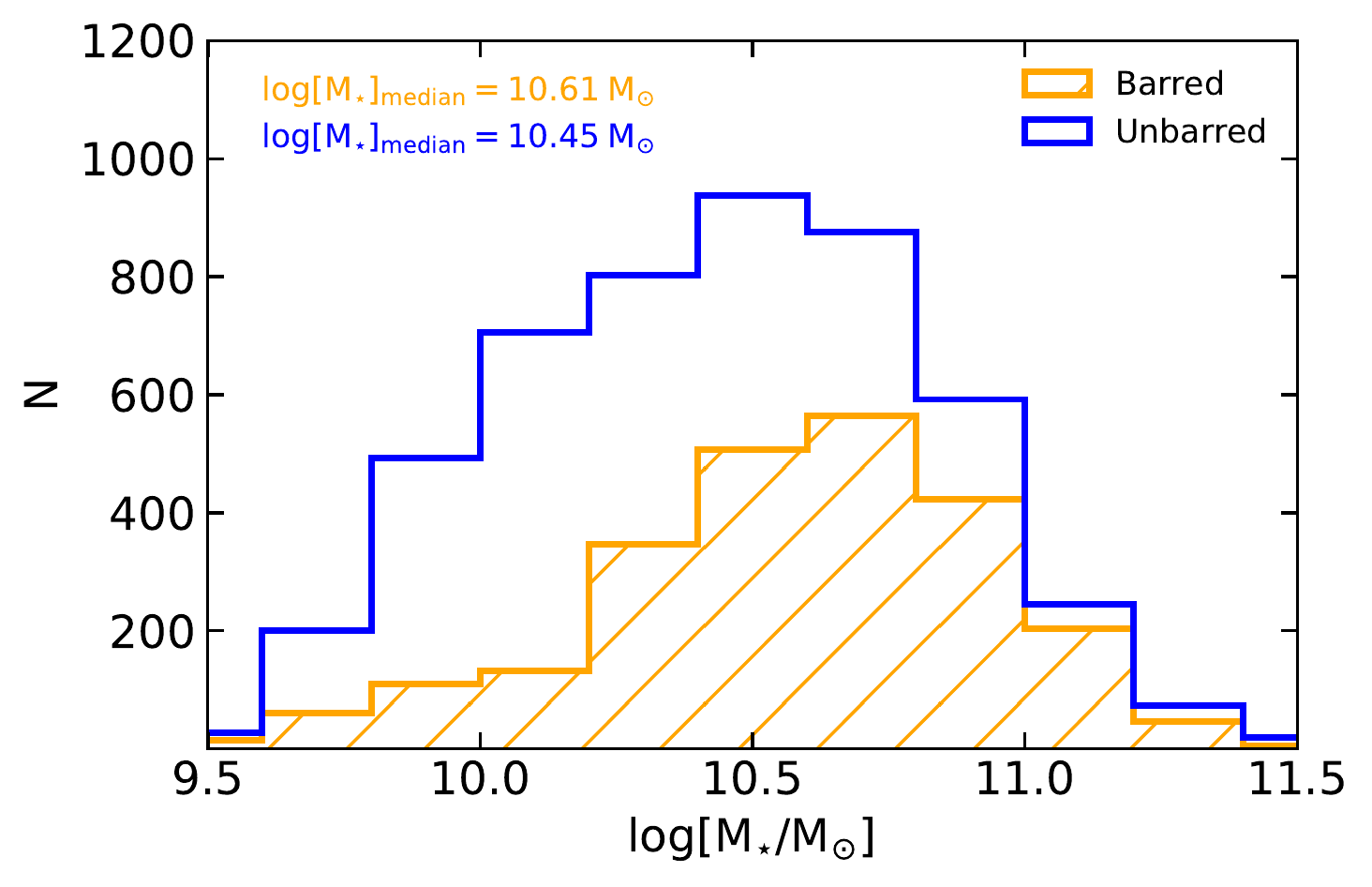}
 \caption{The distribution of masses of the successfully fitted \textsc{volume-limited samples} of \textsc{barred} vs \textsc{unbarred} galaxies. Barred galaxies, although lower in number, have, on average, higher masses. From this distribution we selected a mass-matched sample of unbarred galaxies.}
 \label{mass_distribution}
\end{figure}

\begin{figure}
 \includegraphics[width=1.05\columnwidth]{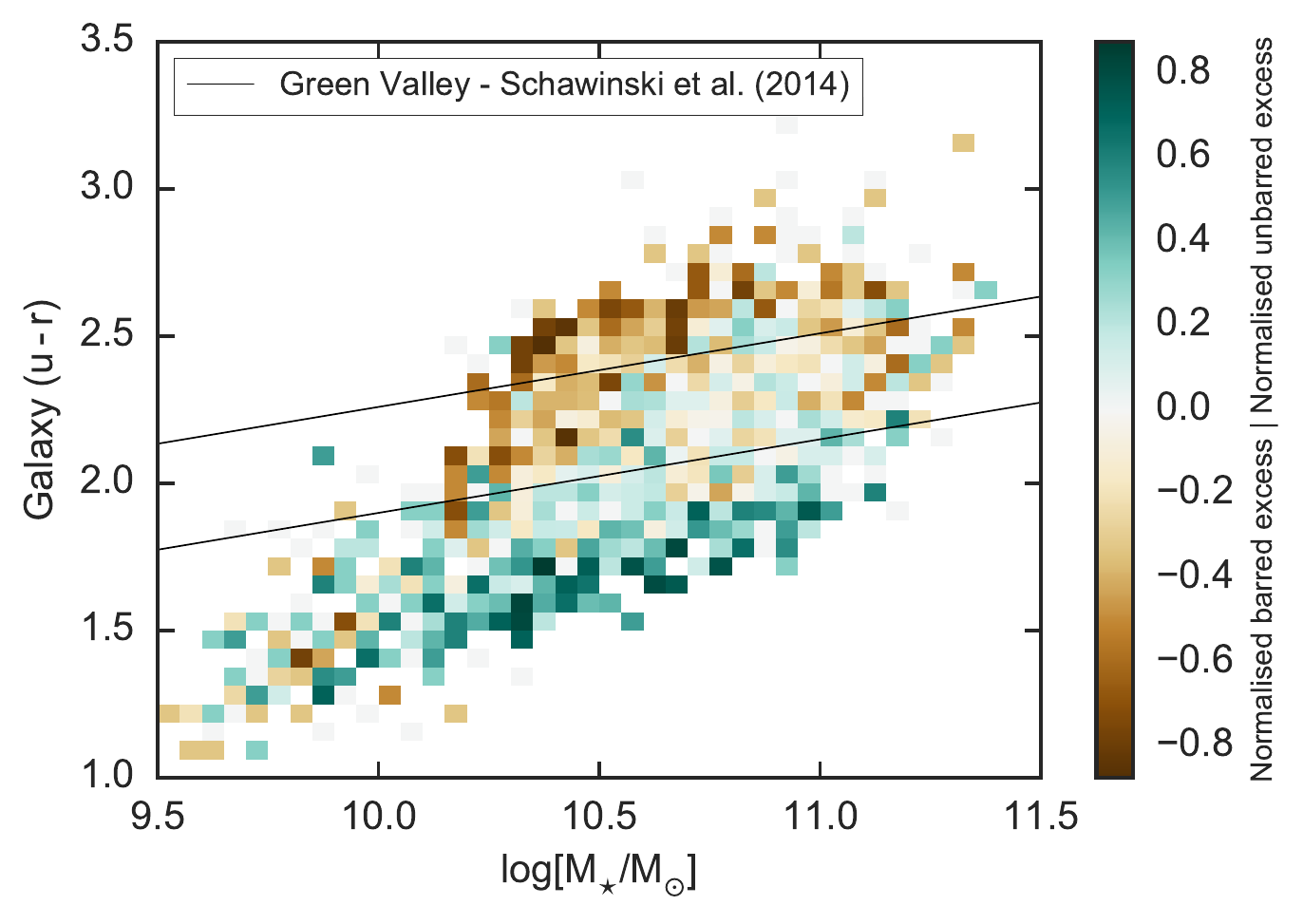}
 \caption{Colour-mass diagram of the mass-matched \textsc{volume limited samples} of \textsc{barred} and \textsc{unbarred} galaxies. Instead of overlaying the two distribution, they were subtracted and normalised by the total number of galaxies in each bin. A darker red colour suggests an excess of barred galaxies, while dark blue colour an excess of unbarred ones. It is clear that the barred galaxies tend to be redder, while unbarred ones tend to be bluer, at the same stellar masses. Although mass is thought to drive most of the evolution of a galaxy, the main physical difference between the two populations in this plot is the presence of a strong bar. The two lines show the definition of the `green valley' from \citet{Schawinski2014}.}
 \label{cmd}
\end{figure}

Another aim of this paper is to compare the properties of barred and unbarred galaxies to infer the effect the bar has on its host galaxy. In order to have a statistically meaningful comparison, we selected a \textsc{volume-limited} subsample of \textsc{barred} galaxies, and a similar \textsc{volume-limited unbarred} sample, based on the Galaxy Zoo users' classifications, as described in Section \ref{data}. 

There are 8,689 galaxies in a \textsc{volume-limited unbarred} sample, selected with $p_{bar}\leq0.2$. There are 4,692 (57\%) unbarred galaxies with negligible bulges (disc dominated galaxies) and 3,587 (43\%) unbarred galaxies with obvious bulges, according to the Galaxy Zoo volunteers' classification described in Section \ref{model}.

We have fitted all the galaxies in the \textsc{volume-limited unbarred} sample with two (disc+bulge) components, which converged for 6,314 galaxies. Furthermore, as for the barred galaxies, we excluded bulges with low axis ratios $b/a_{\textrm{bulge}}<0.3$, yielding a total of 5,080 successful fits (a 58\% success rate). This sample contains both disc dominated and unbarred galaxies with obvious bulges, in proportions of 44\% and 56\%, respectively. Therefore, a higher fraction of disc dominated galaxies failed the two component fits, which is expected. We have one component fits available for these galaxies, but we used the two component fits (disc+bulge) in our analysis of unbarred galaxies so that we do not bias the comparison with single versus multi component fits.  

The mass distribution of the two successfully fitted, \textsc{volume-limited} samples can be seen in Figure \ref{mass_distribution}. The distribution of masses of the two samples are clearly different (K-S test $k=0.19,\;p_{\textrm{KS}}<10^{-15}$); galaxies with strong bars have significantly higher masses compared to unbarred galaxies. 

Most of the differences between the barred and unbarred galaxies are driven by stellar mass. Thus, to study mass independent effects, we selected a mass-matched subsample of 2,435 unbarred galaxies (matched in bins of $\log{(\frac{M_{\star}}{M_{\odot}}})=0.1$). The mass-matched sample of unbarred galaxies contains 1,570 (64\%) galaxies with obvious bulges and 868 (36\%) disc dominated galaxies. This is different from the distribution of bulges in the \textsc{volume-limited} subsample of \textsc{barred} galaxies, according to the Galaxy Zoo volunteers: 74\% strongly barred galaxies with obvious bulges and 26\% galaxies with negligible bulges.

Figure \ref{cmd} shows the colour-mass diagram for both the mass-matched unbarred and barred galaxies (for 2,435 galaxies of each type). At the same mass, barred galaxies (denoted by the darker red colours) are more common than unbarred disc galaxies in the `red sequence' and `green valley', while unbarred galaxies are more common in the `blue cloud'. We note, however, that due to the luminosity limit of Galaxy Zoo ($r<17$), our \textsc{volume-limited samples} are incomplete for red (and hence passive) galaxies at $M_{\star}\lesssim10^{10} M_{\odot}$. Therefore, our sample is complete only for $M_{\star}\gtrsim10^{10} M_{\odot}$.

\subsection{Bulges}
\label{bulges}

\begin{figure}
  \includegraphics[width=1\columnwidth]{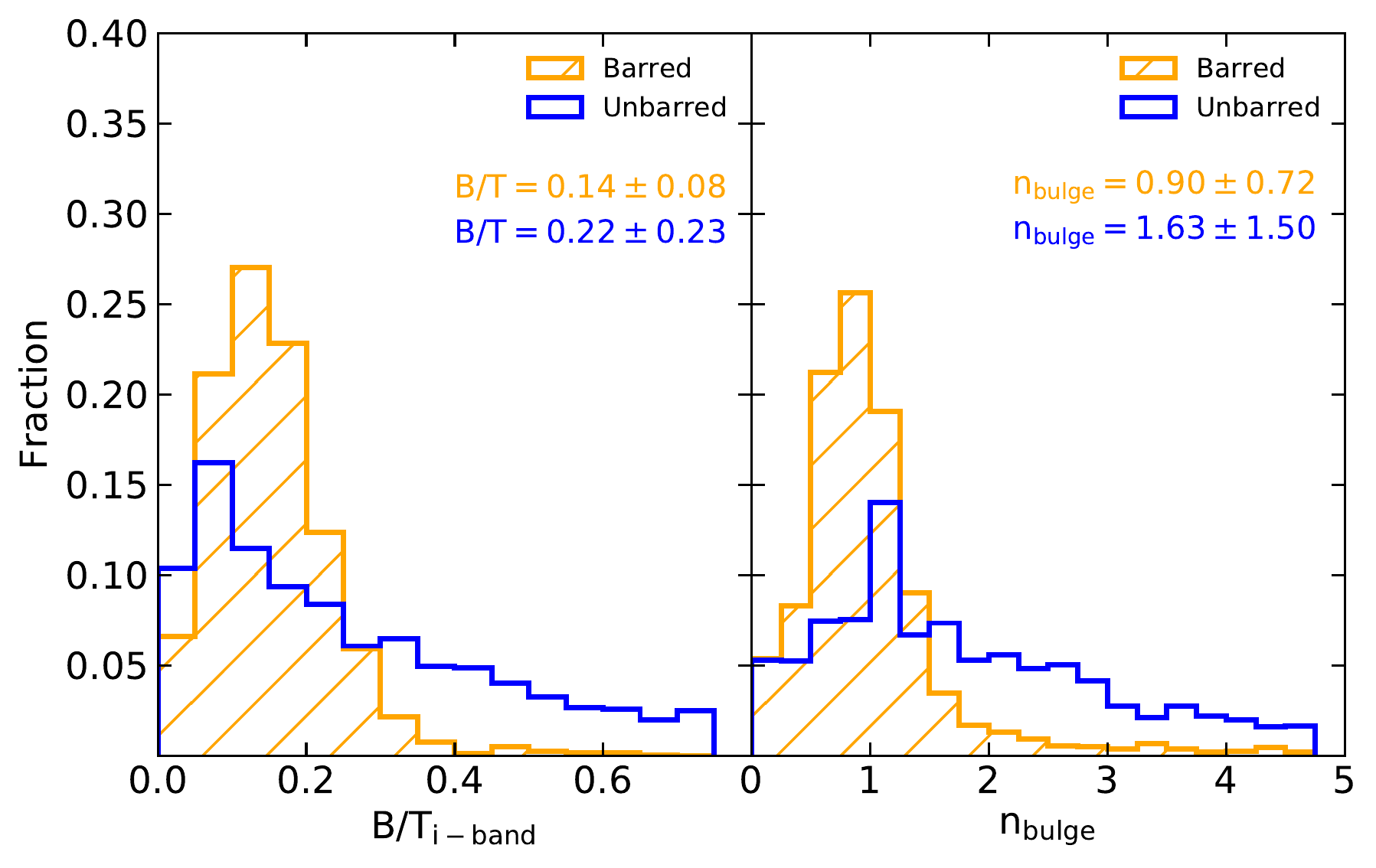}
  \caption{Left panel - The  $i$-band $B/T$ ratio for barred and unbarred galaxies. Right panel - The bulge S\'ersic indices of galaxies with and without bars. The two distributions for barred and unbarred galaxies clearly different. The bulges of barred galaxies have low S\'ersic indices (exponential on average, typical of pseudobulges), while unbarred galaxies have a large spread of bulge S\'ersic indices, with a higher fraction of classical bulges. The median and $1\sigma$ scatter for each distribution is given in the plot.}
 \label{n_distribution}
\end{figure}

As described in Section \ref{data}, we fitted bulges to 2,040 galaxies in the \textsc{volume-limited barred} sample and to 2,435 in the \textsc{volume-limited unbarred} sample. In \citet{Gadotti2008}, the authors argue that bulges can be well fit if their effective radius is at least $80\%$ of the half width at half maximum (HWHM). For our sample, 92\% of the barred galaxies and 99\% of the unbarred galaxies have $r_{\mathrm{e,bulge}}>0.8\times$HWHM, therefore it is reasonable to assume that the bulges are well resolved. 

The bulge-to-total luminosity ratio, $B/T$, for the \textsc{volume-limited} and \textsc{mass-matched} samples of barred and unbarred galaxies, in the \textit{i}-band, can be seen in Figure \ref{n_distribution} (left panel). For the barred galaxies, the median $B/T$ in \textit{i}-band is 0.14, a vast majority of 83\% of the galaxies having $B/T\leq 0.2$, in good agreement with other studies of barred galaxies, with smaller samples (e.g. \citealt{Laurikainen2007,Weinzirl2009}). The bulge-to-total luminosity increases with wavelength from the $u$-band to the $z$-band, which is expected if bulges host an older population of stars.

The median \textit{i}-band $B/T=0.22$ for the bulges of unbarred galaxies is significantly higher than that of barred galaxies. We have investigated the images of unbarred galaxies with high $B/T$. In the majority of cases this is due to another component present in the proximity of the bulge: a `lens' or `oval', which was also fit by the bulge model component. We discuss this further in Section \ref{lenses}. 

\begin{figure}
 \includegraphics[width=1\columnwidth]{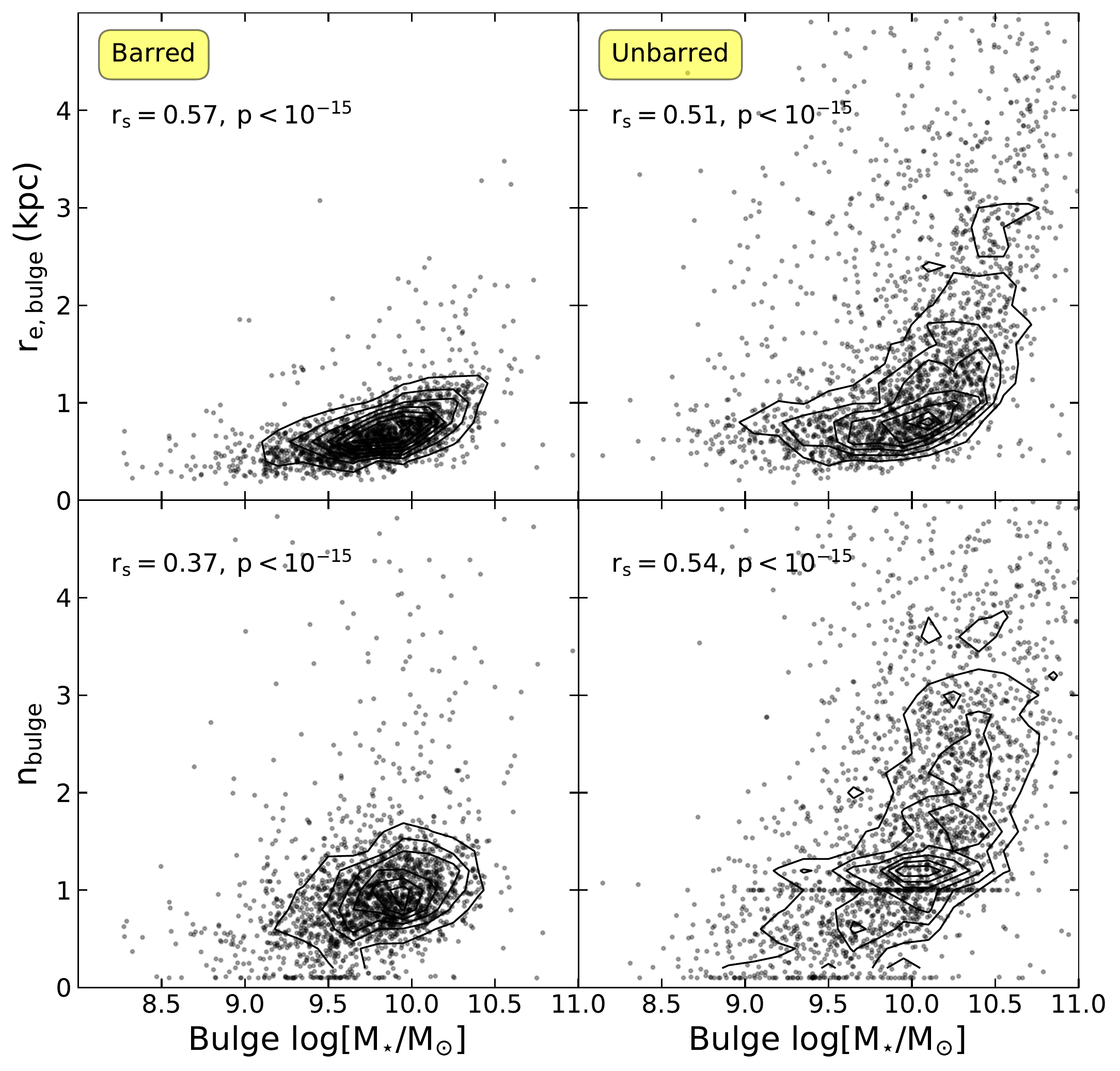}
 \caption{The mass-size and mass-\textit{n} scaling relations for the bulges of both barred and unbarred galaxies. Although the bulge parameters for the two samples are different, they clearly increase with the bulge mass for both samples. The higher concentration of $n_{\mathrm{bulge}}=1$ for the unbarred galaxies is due to the discs (fitted with a fixed $n=1$ profile) and the bulges (fitted with a free S\'ersic index) interchanging in the fitting procedure. The components were identified as discs and bulges, respectively, by comparing the $r_{e}$ of the components. The Spearman $r_{s}$ correlation coefficient is shown.}
 \label{mass_size_relation}
\end{figure}

We use Equation (8) in \citet{Taylor2011} to convert from $(g-i)$ colours and $M_{i}$, \textit{i}-band absolute magnitudes, to stellar masses for each components. As can be seen in Figure \ref{mass_size_relation}, the bulge $r_{e}$ and S\'ersic index, $n_{\textrm{bulge}}$, are correlated with the bulge stellar mass, for both barred and unbarred galaxies. For both samples these measured bulge parameters increase with the bulge mass, which is expected, more massive bulges being physically bigger (see e.g. \citealt{Fisher2010}). Recovering this scaling relation also indicates that our decompositions are reliable. However, the bulge sizes and S\'ersic indices for the two samples are clearly different. 

For the bulges, the median axis ratios are 0.77 for barred and 0.68 for unbarred galaxies. The lower axis ratio for the bulges of unbarred galaxies suggests the presence of a more elongated component in the proximity of the bulge. As also seen in the right panel of Figure \ref{n_distribution}, the S\'ersic index of the bulge varies between 0.1 and 4, with a median of $n_{\mathrm{bulge}}=0.90$ for the barred galaxies and $n_{\mathrm{bulge}}=1.63$ for unbarred galaxies. As discussed in Section \ref{Montage}, due to image stacking, the bulge S\'ersic indices are underestimated by $\sim$30\%. Correcting for this, the median values are $n_{\mathrm{bulge}}\sim1.2$ for barred galaxies and $n_{\mathrm{bulge}}\sim2.1$ for unbarred galaxies. 

The low bulge S\'ersic indices in the barred sample suggest that the bulges in these galaxies are overwhelmingly `disc-like' pseudobulges, in contrast to `elliptical-like' classical bulges, which are rare. The distinction is not clear, but authors generally agree that, in a statistical sense, on large populations, bulges with $n\leq2$ are pseudobulges and with $n>2$ are classical bulges \citep{Fisher2008}. \citet{Graham2008} and \citet{Graham2011} argue that there is no bimodality in the bulge S\'ersic index, and thus we cannot reliably separate between classical bulges and pseudobulges using the S\'ersic index alone. For the purpose of comparing the bulge properties of barred and unbarred galaxies, as well as to compare our results of the bulge parameters with different studies, we will make use of this division. In our sample of barred galaxies, only 10\% have classical bulges whereas a large majority of 90\% have pseudobulges. In contrast, 40\% of unbarred galaxies have classical bulges and 60\% pseudobulges. 

Some previous studies using disc+bar+bulge decompositions disagree on the properties of bulges in barred galaxies. Using the \textsc{BUDDA} software, \citet{Gadotti2011} found a median $n_{\mathrm{bulge}}=2.5$ (39$\%$ pseudobulges, 61$\%$ classical bulges, according to the threshold by \citealt{Fisher2008}\footnote{\citet{Gadotti2009} uses the Kormendy relationship to separate pseudobulges from classical bulges. For this work, we chose to use the simple cut of $n_{\mathrm{bulge}}\sim2$ by \citet{Fisher2008} to be consistent in our comparison with other studies. \citet{Graham2011} suggests against using the Kormendy relation to differentiate the two types bulges.}) in their disc+bar+bulge decomposition of 291 barred SDSS galaxies. The main differences between our fitting procedure and the one in \citet{Gadotti2009} are the coadded versus single frames and the different PSFs used: star PSFs versus circular Gaussian PSF. We have tested the effects of using a circular Gaussian PSF (with the FWHM given by SDSS), on single $i$-band frames (not coadded with \textsc{MONTAGE}, to be consistent with \citealt{Gadotti2009}), for 50 barred galaxies fitted with disc+bar+bulge components, in common between \citet{Gadotti2009} and our study, and found that the shape of the PSF has a small effect on the bulge S\'ersic index. The median bulge S\'ersic index and the correlation with $n_{\mathrm{bulge}}$ measured by \citet{Gadotti2009} increases slightly when using a circular Gaussian PSF (from median $n_{\mathrm{bulge}}=1.3$ to $n_{\mathrm{bulge}}=1.6$ and from $r_{s}=0.3$ to $r_{s}=0.5$, where $r_{s}$ is the Spearman rank correlation coefficient), however, the majority of the values are still $n\sim1$ lower than the ones measured by \citet{Gadotti2009} (median $n_{\mathrm{bulge}}=2.8$ for the 50 barred galaxies). Nevertheless, \citet{Gadotti2009} discusses in their Appendix A (Figure A1) that the bulge S\'ersic index is the least robust parameter and hardest to constraint when varying the input parameter.

Our findings are, however, consistent with other studies. For example, \citet{Laurikainen2004b} find a typical barred galaxy bulge S\'ersic index of $n_{\mathrm{bulge}}=1.4$ (74\% pseudobulges, 26\% classical bulges). Similarly, \citet{Weinzirl2009} found that 76\% of bright spirals have $n_{\mathrm{bulge}}\leq2$ in the $H$-band; the bar fraction of galaxies with $n_{\mathrm{bulge}}\leq2$ is 65\% and the mean bulge S\'ersic index of barred galaxies is $n_{\mathrm{bulge}}\sim1.3$ across all Hubble types. The median $n_{\mathrm{bulge}}$ for the nearby and well resolved barred galaxies in the CALIFA survey is 1.6 \citep{Califa2017} (66\% pseudobulges, 34\% classical bulges). Furthermore, the median $n_{\mathrm{bulge}}$ for barred galaxies in the decomposition of S$^{4}$G galaxies \citep{Salo2015} is also 1.6 (63\% pseudobulges, 37\% classical bulges). In contrast, \citet{Kim2015} find a median $n_{\mathrm{bulge}}=2.1$ (37$\%$ pseudobulges, 63$\%$ classical bulges) in a similar decomposition of 144 barred galaxies from the S$^{4}$G survey. Even though the two S$^{4}$G studies concern the same data set, there is a discrepancy in their measured bulge properties. The main differences between the two studies (\citealt{Kim2015} and \citealt{Salo2015}) are the softwares used: \textsc{BUDDA} versus \textsc{Galfit} (which is the same software this work is based on) and the fitting procedures: S\'ersic versus Ferrers bar profiles, boxy versus ellipse bar shapes, disc breaks versus single exponential disc profile, different input parameters etc. The presence of a possible systematic difference between the two common codes or fitting procedures deserves further study. 

\begin{figure}
  \includegraphics[width=\columnwidth]{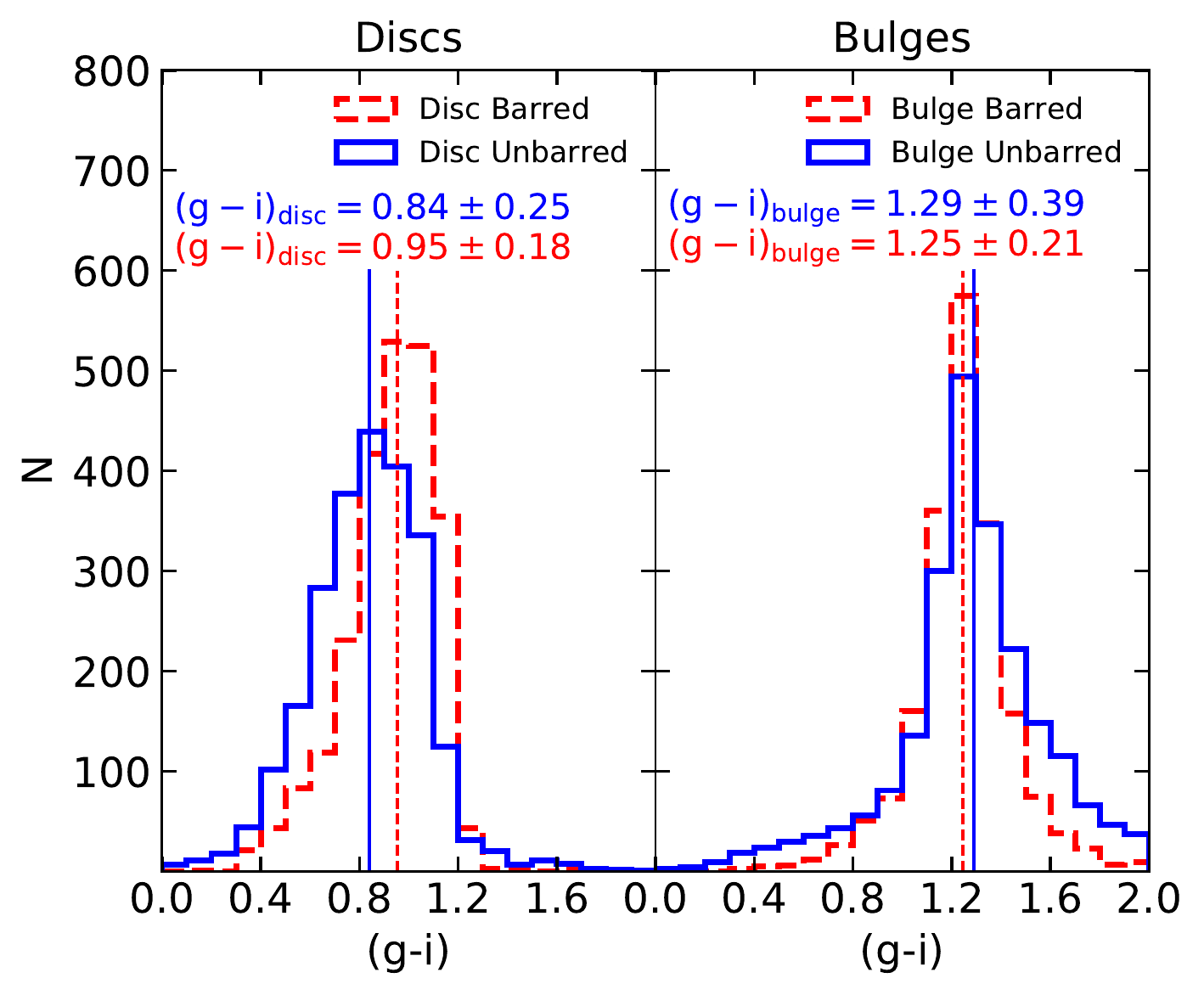}
  \caption{The $(g-i)$ colours of discs (left) and bulges (right) of barred (with red) and unbarred (with blue) mass-matched galaxies. The discs of unbarred galaxies are clearly redder than the ones of unbarred galaxies, while their bulges have more similar colours to those of unbarred galaxies. Median values for the colours and the $1\sigma$ spread are shown. }
 \label{colour_difference}
\end{figure}

Finally, the colours of the discs and bars of barred and unbarred galaxies can be seen in Figure \ref{colour_difference}. The discs of barred galaxies are clearly redder compared to the unbarred galaxies by $\Delta(g-i)\sim0.11\pm0.01$ (the error represents the standard error on the mean, in quadrature). This is consistent with studies which find barred galaxies to be redder overall \citep{Masters2011}, since we see that the disc dominates the total luminosity of these galaxies. On the other hand, the colours of bulges of barred galaxies are more similar to their unbarred equivalents ($\Delta(g-i)\sim0.04\pm0.01$). The scatter in bulge colour of unbarred galaxies to very red colours possibly reflects a greater presence of dust in unbarred galaxies, consistent with higher gas content and specific SFR. Similar colour differences for discs and bulges are found when comparing galaxies with obvious bulges only.

This result on the colour of discs is in contrast with the work of \citet{Sanchez2013} who found similar colours for discs in barred and unbarred galaxies in the sample of \citet{Gadotti2009}. The modes of their colour distributions actually suggest that barred discs are bluer than their unbarred counterparts. However, they also find that discs with the bluest colours $(g-i)_{\textrm{disc}} < 0.8$ are mostly unbarred. We compared the two samples and the main difference arises due to a large number of unbarred galaxies in \citet{Sanchez2013} having $(g-i)_{\textrm{disc}} \sim1.25$, which does not exist in our sample. Of the unbarred galaxies common in \citet{Sanchez2013} and GZ2 \citep{Willett2013} (325 out of 390), 53\% are classified as `smooth' as opposed to `discs' by Galaxy Zoo (having debiased likelihoods $p_{\textrm{smooth}}\geq0.5$), therefore being categorized as elliptical galaxies rather than unbarred discs. 

\section{Unbarred galaxies with lenses}
\label{lenses}

\begin{figure}
  \includegraphics[width=\columnwidth]{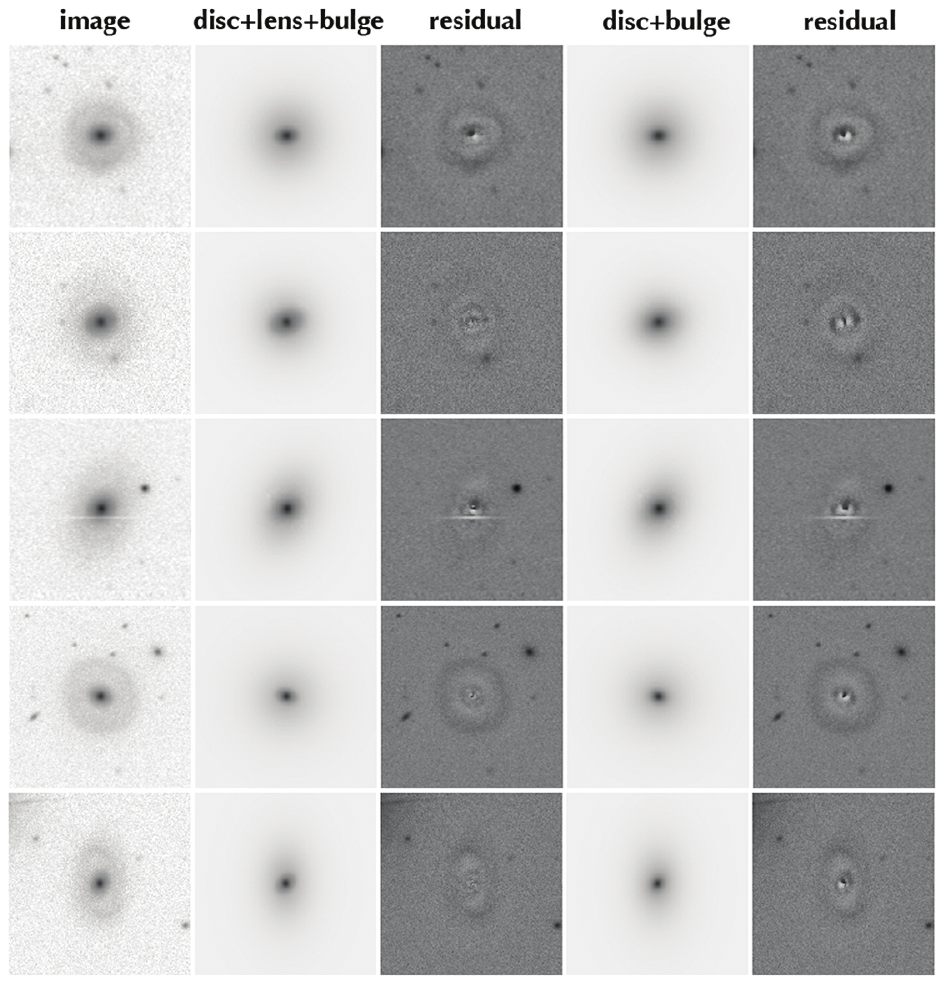}
  \caption{Examples of galaxies with inner lenses which were initially part of the unbarred sample. The image on the left is the $i$-band SDSS image, the second and third columns are the disc+lens+bulge model and residual, while the fourth and fifth columns are the disc+bulge model and the corresponding residuals. The disc+lens+bulge models are a better representation for the light distribution of these galaxies than the simple disc+bulge models. The properties of galaxies with inner lenses are more similar to those of barred galaxies. Galaxies with inner lenses were identified using the Galaxy Zoo answers to the `ring' question, therefore all the fitted galaxies with inner lenses show an outer ring feature in the residuals. The size of the images is $40\arcsec \times 40\arcsec$. }
 \label{lens}
\end{figure}

While inspecting the images and the fits, we notice a significant number of galaxies with inner lenses (morphological components of the galaxies themselves rather than gravitational lenses of background sources) or ovals in the unbarred sample, which might account for the higher observed $B/T$ and increased bulge S\'ersic index of the unbarred sample in the two component fits. An inner lens is a region around the bulge with little variation of brightness with radius \citep{Buta2007}. Lenses are frequently observed in S0 galaxies and in early-type spirals \citet{Laurikainen2005,Laurikainen2007,Laurikainen2009}. In the Near-Infrared S0 Survey (NIRS0S) \citep{Laurikainen2011} found that 61\% of the barred and 38\% of the unbarred S0 galaxies host lenses. Ovals are observed in late-type galaxies and they look similar to lenses in early-type galaxies. However there is no clear evidence whether or not they are physically similar \citep{Kormendy2004}.

It has been shown that the presence of a component which is not accounted for in a fit (specifically, in this case the inner lens/oval) can increase the S\'ersic index of another component (the bulge, in this case) while also contributing to the increase of the $B/T$ ratio \citep{Peng2010, Laurikainen2013}. Therefore, not accounting for these components in the disc+bulge fits can lead to measuring erroneous properties for the bulges.

\citet{Kormendy1979,Kormendy2013} suggested that as strong bars weaken, the stars escape from the bar and migrate into lenses, therefore pointing to an evolutionary scenario leading to the formation of lenses/ovals. His conclusion is based on observations of barred galaxies without lenses, galaxies with bars embedded in lenses and galaxies with lenses and no bars.

While the Galaxy Zoo project did not ask the volunteers a question about the presence of `inner lenses', it did enquire about the presence of `rings' in a galaxy. \citet{Willett2013} discusses the ring classification in comparison with the expert classification of rings in \citet{Nair2010}. As can be seen in Figure 12 of  \citet{Willett2013}, by requiring a threshold of $p_{\mathrm{ring}}\geq0.5$ the volunteers reliably identify rings when compared to expert classifications. \citet{Nair2010} also notes that the inner lenses are most easily and often identified when they have are accompanied by ring, noticing a correlation between rings and lenses. Recently, \citet{Buta2017} also noticed that there are many inner lenses in a sample identified with outer rings in GZ2, with the question: `Is the odd feature a ring?'. 41.2\% of the 3962 ringed galaxies identified in Galaxy Zoo 2 by \citet{Buta2017} have inner lenses. 

Because we are identifying lenses using the `ring' question in Galaxy Zoo, we cannot distinguish between inner rings and lenses. They tend to occupy similar locations in a galaxy and are believed to be related; often a ring is a subtle enhancement at the edge of a lens \citep{Buta1996}. \citet{Buta2017} also argues that it is difficult to distinguish between lenses and inner rings if the resolution of SDSS image is poor. Furthermore, we cannot fully exclude the presence of a weak bar inside the inner lenses.

To identify unbarred galaxies with inner lenses and to quantify the bias introduced by having an additional third component in the models of barred galaxies (the bar) compared to only two components in the case of unbarred galaxies, we added a third component to the fits of unbarred galaxies, modeled in a similar fashion to a bar (with the same initial parameters). We found that adding a third component to the unbarred galaxies can decrease the bulge S\'ersic index, effective radius and $B/T$ by factors of $\sim$1.4, 1.5 and 1.8, respectively. Therefore, not accounting for the additional components (bars, lenses/ovals), if they are present, can lead to significantly overestimating the parameters of the bulge in traditional disc+bulge decompositions. 

Of the 6,314 unbarred galaxies fitted with a third component, only 4,534 converged to a value for all the parameters. Following the selection criteria for meaningful fits as in the case of barred galaxies (as described in Section \ref{model}), there are 3,957 good fits, out of the initial 8,689 unbarred galaxies. Furthermore, we noticed that for a large fraction of the fitted galaxies (1,129 galaxies), the S\'ersic index of one of the components converged to the lower limit of \textsc{GalfitM}, of $n=0.10$. Therefore, we excluded these fits as they are probably unphysical. Finally, only 2,828 fits out of the initial 8,689 converged to a meaningful three component model, a success rate of only 33\%, which, as expected, is smaller than that for barred galaxies. Thus, a simple disc+bulge model is, in general, more appropriate for the unbarred galaxies. Nevertheless, we noticed that in many cases, the third component of the unbarred galaxies had a physical meaning, representing the lens/oval.
 
From the unbarred galaxies with meaningful 3 component fits we selected a clean sample of galaxies with inner lenses (therefore, fitted with a disc+lens+bulge model) by selecting galaxies with $p_{\mathrm{ring}}\geq0.5$ and also requiring that at least 5 volunteers classified the galaxy as having a `ring', $N_{\mathrm{ring}}\geq5$. This resulted in 674 unbarred galaxies with inner lenses, 609 of them having `obvious' bulges according to the volunteers' classification. One of the authors (SK) inspected the fits and residuals and selected 394 with realistic disc+lens+bulge fits. Five examples of galaxies with inner lenses, images, disc+lens+bulge fits and residuals are shown in Figure \ref{lens}. For comparison, the disc+bulge fits and corresponding residuals for the same galaxies are also shown. 

We would also like to select a clean volume-limited sample of unbarred galaxies which does not contain galaxies with inner lenses. This is more difficult to achieve since just excluding galaxies with $p_{\mathrm{ring}}<0.5$ does not guarantee a sample with high purity. The $p_{\mathrm{ring}}$ is largely bimodal, with most galaxies having either $p_{\mathrm{ring}}=0$ or $p_{\mathrm{ring}}=1$. Therefore, we choose only galaxies with $p_{\mathrm{ring}}=0$ to select 1,837 unbarred galaxies with no lenses. 619 , or only 34\% of these have `obvious' bulges. Similarly, SK inspected the fits and residuals and selected 447 unbarred galaxies with `obvious' bulges and with good disc+bulge fits, without lenses.

Finally, we compare the properties of a the following volume-limited samples: barred galaxies (fitted with disc+bar+bulge), non-barred galaxies (fitted with disc+bulge) and non-barred galaxies with lenses (fitted with disc+lens+bulge). All three samples were selected to have `obvious' bulges (so that a bulge is significantly bright and included in the fit in all cases), as classified by Galaxy Zoo users. Due to the small sample sizes, the three volume-limited samples were not mass-matched. A mass-match is also not possible while preserving a statistically useful sample size because the three samples have different mass distributions: the masses of galaxies with lenses are similar to the barred sample, and different from the purely unbarred sample. The median values and 1$\sigma$ standard deviations of the colours, the S\'ersic indices, the axis ratios, luminosity ratios and the scaled effective radii of the three components can be seen in Table \ref{params}.

One important result is the similarity of the colours of the discs and bulges of barred galaxies and galaxies with lenses, and the clear difference from purely unbarred galaxies. Galaxies with inner lenses show properties such as masses, S\'ersic indices and luminosity ratios that are, in general, similar to barred galaxies with obvious bulges. The only small differences between the unbarred galaxies with lenses and barred galaxies are the slightly different colours of the bars and lenses, the lenses being bluer than the bars. This is possible due to the presence of rings at the end of lenses, which are usually defined by recent star formation \citet{Buta2007}. This result suggests that galaxies with inner lenses should not be considered in the same category as unbarred galaxies. 

The lenses are also $\sim40\%$ shorter than the bars, in terms of their sizes normalized to the effective radius of the discs, and rounder, with an axis ratio of $\sim0.67$ compared to the median axis ratio of $\sim0.35$ of the bars. \citet{Laurikainen2013} found that lenses in unbarred galaxies have similar sizes to lenses in barred galaxies suggesting that they may be lenses in former barred galaxies.

The samples of unbarred galaxies with and without inner lenses discussed in this section are clean samples, but not complete. The properties of inner lenses and other galaxy substructures should be examined further in a future work. 

\begin{table}
 \caption{Median parameters and 1$\sigma$ standard deviation for the fitted unbarred galaxies, unbarred galaxies with inner lenses and barred galaxies. All galaxies were selected from a volume-limited sample, based on the volunteers classifications for the presence of bars, rings and having an `obvious' bulge. The total stellar masses are \citep[drawn from average values in the MPA-JHU catalogue;][]{Kauffmann2003a}, while the stellar masses of the components were calculated from the optical colours, based on Equation (8) in \citet{Taylor2011}. }
 \label{params}
 \begin{tabular}{|p{2.1cm}||p{1.45cm}|p{1.45cm}|p{1.45cm}|}
  \hline
  Parameter& bulge+disc & +lens & +bar \\
  \hline
$N_{gal}$ & 447  & 394 & 1699 \\ \hline
$\log(M_{\star}/M_{\odot})$ &  $10.42\pm1.46$ & $10.70\pm1.34$ & $10.67\pm1.19$ \\ \hline
$(u-r)$ disc & $1.65\pm0.39$ &  $2.14\pm0.29$ & $2.14\pm0.32$ \\
$(u-r)$ bar/lens & - &  $2.38\pm0.52$ & $2.55\pm0.65$  \\
 $(u-r)$ bulge & $2.69\pm1.32$ &  $2.70\pm0.36$ & $2.64\pm0.48$ \\
$(g-i)$ disc & $0.74\pm0.23$ &  $1.01\pm0.14$ & $1.00\pm0.14$  \\
$(g-i)$ bar/lens &  - &  $1.11\pm0.24$ & $1.16\pm0.22$  \\
 $(g-i)$ bulge &  $1.33\pm0.44$ &  $1.26\pm0.17$ & $1.24\pm0.19$  \\
$(r-z)$ disc &   $0.45\pm0.15$ &  $0.56\pm0.09$ & $0.57\pm0.09$ \\
$(r-z)$ bar/lens &  - &  $0.67\pm0.15$ & $0.69\pm0.12$ \\
$(r-z)$ bulge & $0.74\pm0.19$ &  $0.72\pm0.13$ & $0.71\pm0.12$ \\ \hline
Discs & & & \\ \hline
$\log(M_{\star}/M_{\odot})$ &  $10.07\pm0.39$  & $10.22\pm0.31$ & $10.20\pm0.65$ \\
n & 1 &  1 & 1 \\
$b/a$ & $0.73\pm0.16$ &  $0.71\pm0.18$ & $0.80\pm0.13$ \\
$D/T_{\mathrm{i-band}}$ & $0.82\pm0.18$ &  $0.58\pm0.16$ & $0.67\pm0.14$ \\
$r_{e}$ (kpc) & $6.06\pm2.27$  &  $7.93\pm3.67$ & $6.80\pm2.87$ \\ \hline
Bar/Lens & & & \\ \hline
$\log(M_{\star}/M_{\odot})$ &  - & $9.88\pm1.13$ & $9.78\pm0.55$ \\
n & - &  $0.37\pm0.32$ & $0.42\pm0.26$ \\
$b/a$ &- &  $0.67\pm0.15$ & $0.35\pm0.11$ \\
$Bar/T_{\mathrm{i-band}}$ & - &  $0.19\pm0.15$ & $0.15\pm0.10$ \\
$r_{e}/r_{\mathrm{e,disc}} $ & - &  $0.31\pm0.12$ & $0.46\pm0.16$ \\ \hline
Bulge & & & \\ \hline
$\log(M_{\star}/M_{\odot})$ &  $9.92\pm0.62$ & $9.98\pm0.47$ & $9.83\pm0.38$ \\
n & $1.28\pm1.23$ &  $1.00\pm1.07$ & $0.92\pm0.64$ \\
$b/a$ & $0.69\pm0.15$ &  $0.77\pm0.14$ & $0.78\pm0.13$ \\
$B/T_{\mathrm{i-band}}$ & $0.18\pm0.18$ &  $0.20\pm0.09$ & $0.15\pm0.08$ \\
$r_{e}/r_{\mathrm{e,disc}} $ & $0.17\pm0.11$ &  $0.08\pm-4$ & $0.08\pm0.04$ \\
 
  \hline
 \end{tabular}
\end{table}

\section{Discussion}
\label{discussion}

Detailed studies involving large samples of nearby galaxies, such as this work, are necessary since they allow us to investigate - in a statistically reliable fashion - both the qualitative morphology via visual classifications and a more quantitative morphology by the means of photometric decompositions. In this study we find that the bulges of barred galaxies are predominantly pseudobulges, with a typical S\'ersic index of $n_{\mathrm{bulge}}\sim1$. We find two types of bar S\'ersic profiles: bars in low mass disc dominated galaxies have approximately exponential profiles ($n_{\mathrm{bar}}\sim0.9$), while bars in higher mass galaxies with obvious bulges have flatter profiles ($n_{\mathrm{bar}}\sim0.4$). With the multi-band fitting we measure the colours of the individual components and find that the bars and bulges of barred galaxies are redder compared to the galaxy discs by $\Delta(g-i)\sim0.2$ and $\Delta(g-i)\sim0.3$, respectively. Furthermore, when comparing to a mass-matched sample of galaxies without bars, the discs of barred galaxies are redder by $\Delta(g-i)\sim0.1$ than the corresponding discs, while their bulges are bluer by $\Delta(g-i)\sim0.04$ than the corresponding bulges of unbarred galaxies. Finally, we find a subsample of galaxies with inner lenses/ovals within the unbarred sample of galaxies that have similar structural properties to barred galaxies. In this Section we discuss these findings in the context of secular evolution of disc galaxies.\\

$\bullet${\textit{Are bars responsible for building central mass concentrations?}}\\
Bars are thought to be efficient in transporting gas to the central regions, and possibly leading to the growth of bulges \citep{Kormendy2004}. We find that a large fraction (>90\%) of the `obvious' bulges of galaxies with strong bars are pseudobulges, or `discy' bulges. Classical bulges are believed to be formed by major and minor merger events, while pseudobulges form by slow secular evolution. Our results support the scenario of bulges built from the disc material. However, it is surprising that such a high fraction of galaxies with strong bars have `discy' bulges, given the high fraction of mergers suggested by hierarchical galaxy formation, as noted before by \citet{Kormendy2010}. Nevertheless, this result is in agreement with simulations of barred galaxies \citep{Scannapieco2010}, who also found that almost all barred galaxies host bulges with $n_{\mathrm{bulge}}\leq1$, even though the galaxies have undergone minor mergers. Furthermore, the presence of low S\'ersic index bulges in unbarred galaxies is not evidence against them being formed by a bar, since the galaxies may have hosted a bar at an earlier time. 

Recent spectroscopic studies have shown that the current star formation is enhanced in the centres of barred galaxies \citep{Ellison2011} and that the bulges of barred galaxies contain a younger population of stars compared to the bulges of unbarred galaxies \citep{Coelho2011}. Other studies on quiescent galaxies have shown that there is no statistically significant difference in the stellar populations of the bulges of barred versus unbarred galaxies \citep{Cheung2015b}. Here we find modest differences in the colours of the bulges. The bulges of barred galaxies are only slightly bluer than the bulges of unbarred ones. While the colours cannot be translated to stellar populations directly without considering the effects of dust and metallicity, almost all the bulges have red colours. It is possible that the gas has been transported into the central regions during the formation of the bar, where it has all been consumed in a burst of star formation or accreted onto the supermassive black hole, leaving behind a gas depleted region as suggested by recent simulations \citep{Carles2016, Spinoso2017, Robichaud2017}. \citet{Perez2011} found that the bulges of barred galaxies are more metal rich and $\alpha$-enhanced implying that the bulges in barred galaxies formed in a starburst. The quick formation mechanism is a possible explanation why we observe quenched bulges.

Our observation of lower masses of the bulges in barred galaxies seems to be in contrast with the idea of the bar adding mass to the bulge. The fact that the $B/T$ ratio is smaller in the strongly barred galaxies compared to unbarred galaxies, which was also observed by \citet{Laurikainen2013}, might suggest a disagreement with bar induced bulge growth, unless the bulges in barred and unbarred galaxies have different formation scenarios. 

We find that barred galaxies have predominantly pseudobulges, while unbarred galaxies have a higher fraction of classical bulges (a median S\'ersic index of $n\sim1$ compared to $n\sim1.6$). Classical bulges are thought to form early in galaxy mergers \citep{Aguerri2001}, therefore it is reasonable that mergers form higher mass bulges than the bar-induced bulges. Simulations of minor mergers should address the issue of bulge formation and explore the bulge masses that arise in mergers with different mass ratios and their frequencies.

Another possibility is that massive bulges destroy bars in galaxies. Some simulations suggest that bars can be destroyed due to the buckling from angular momentum transport or from building large central concentrations \citep{Friedli1993,Bournaud2005}. However, other $N$-body simulations \citep{Shen2004,Athanassoula2013} show that bars are long-lived and the central mass concentration has to be significantly massive to be able to destroy a bar in a Hubble time. Recent observations \citep{Simmons2014,Perez2017} also suggest that some bars are long lived and have been in place for a long time ($\sim10$ Gyr). This is further supported by simulations \citep{Kraljic2012} showing that the epoch of bar formation is $z\sim0.8-1$.\\

$\bullet${\textit{How do bars relate to the quenching of star-formation in discs?}}\\
One of our key findings is that the discs of barred galaxies are significantly redder compared to their unbarred counterparts. We find this result even if we select a volume-limited mass-matched sample of barred and unbarred galaxies. Therefore, bars either have an effect on quenching the galaxies, or the processes that can lead to the formation of a bar also leads to galaxy quenching. Another possibility is that bar formation is suppressed in star forming discs. \citet{Masters2012} found that strong bars reside mainly in gas-poor discs consistent with the gas making the disc resilient to forming instabilities. Simulations by \citet{Athanassoula2013} suggest that large-scale bars form much later in gas-rich discs than in gas-poor ones, confirming the expectations that strong bars tend to reside in more massive red discs compared to blue spirals. Furthermore, \citet{Carles2016} found in gasdynamical simulations of disc galaxies that bars can drive a substantial amount of gas to the centre, quickly converting it to stars, which lowers the gas content of barred galaxies when compared to unbarred galaxies of the same stellar mass.

\citet{Cheung2013} also found evidence for `bar quenching' using similar data from Galaxy Zoo. However, in quantifying the dominance of bulges they split their sample into disc pseudobulges and classical bulges based on the global S\'ersic index of the galaxies (with $n\sim2.5$ used as a discriminator). They have found an anti-correlation between $p_{\mathrm{bar}}$ and the specific SFR and a correlation between $p_{\mathrm{bar}}$, the length of the bar and the global S\'ersic index. We have shown that a high S\'ersic index does not necessarily suggest that the galaxy hosts a classical bulge, as the light from the bulge and from the bar are added together in single component or disc and bulge decompositions. Here we find that most barred galaxies, including quiescent disc galaxies host discy pseudobulges, so perhaps this is the strong evidence for `bar quenching' having acted in these galaxies, suggested by \citet{Cheung2013}.

\citet{Skibba2012} noticed an environmental dependence of barred and bulge dominated galaxies, such that they tend to be found in denser environments than their unbarred counterparts. Even though some of this dependence can be explained by a colour and mass-environment dependences, up to half of the bar-environment correlation must be explained by another environmental influence. \citet{Smethurst2017} also found an increasing bar fraction towards the central regions in galaxy groups which coincides with an increasing of the time since the galaxies were quenched. This suggests that bars may be at least partly responsible for the relation between quenched galaxies and denser environments. We also need to consider that bars may be triggered in interactions in denser environments \citep{Noguchi1988,Moore1996}. One possibility is that the process of `strangulation' in dense environments - in which gas from the discs is stripped, removing fuel for future star formation - also contributes to galaxies growing strong bars \citep{Berentzen2007}. It is therefore difficult to disentangle if the quenching is driven by morphology or environment and probably these two processes are not independent of each other, as suggested by \citet{Smethurst2017}.

Another possibility why the galactic discs of barred can have redder colours compared to unbarred discs is if the bar is efficient in mixing the stars in the galaxy \citep{Sellwood2002,Minchev2010}. This reduces the colour gradient across the components, a possible evidence for it being the lack of metallicity gradients in barred galaxies \citep{Friedli1994,DiMatteo2013}, however some of this evidence is conflicting (see e.g. \citealt{Blazquez2014}). Another potential effect is a higher dust obscured star formation in the discs of barred galaxies compared to the discs of unbarred galaxies. \citet{Hart2017} found that two-armed spirals have an additional $\sim10\%$ obscured star formation compared to many-armed spirals, while 50\% of the two-armed spirals host strong bars compared to only 20\% of the many-armed spirals. However, it is improbable that the small difference in dust obscured star formation can account for the large difference in colour that we observe for the galaxy discs. \\ 

Recent simulations of a Milky Way model by \citet{Aumer2017} show that the presence of a hot, thick disc delays the formation of a bar. More massive discs form bars early, when the disc mass dominates the gravitational field over the dark matter halo in the central parts of the galaxy \citep{Aumer2016}. Thus, it is also possible that what we observe is a timing effect, massive galaxies, that are now red, formed their bars first, and are now observed as strong bars, while gas-rich galaxies, which are blue, are currently in their process of forming a bar.\\

Our result that the central regions of barred galaxies (bulges and bars) are redder compared to the galactic discs, across almost all stellar masses is consistent with simulations of `bar quenching' and  observations of star formation ceasing from inside out \citep{Tacchella2015}. Cosmological `zoom-in' hydrodynamical simulations by \citet{Spinoso2017} find that strong bars are efficient in driving gas inflows, from within the bar corotation radius to the centre where it is consumed in star formation, while the central $\sim$2 kpc is gas depleted. They suggest that observations would identify the bar at a stage when the galactic central regions in already quenched. Therefore, it is plausible that the disc region within the bar corotation radius is gas depleted, the star formation is suppressed, yielding the redder colours of the discs that we observe. This is also supported by the work of \citet{Gavazzi2015} who found that strong bars contribute significantly to the red colors observed in the inner parts of massive galaxies. Evidence for inside out quenching has been supplied recently by spatially resolved data from the MaNGA survey \citep{Belfiore2017}.\\

$\bullet${\textit{How do the properties of bars change with galaxy mass?}}\\
We find that the bar profiles depend on the prominence of the bulge and the stellar mass of the galaxy. Strong bars in low mass disc dominated galaxies have a flatter profile compared to bars in massive galaxies, but they contain similar $Bar/Total$ flux-ratios and have similar sizes scaled to the size of the discs. This is consistent with the findings of \citet{Elmegreen1985}. Bars are believed to be born out of disc material, which has an exponential profile, and in their evolution, they trap stars in the bar orbits \citep{Sellwood1993, Sellwood2014, Athanassoula2013} which can flatten the light profile. This change in the light profile of bars also coincides with the mass at which galaxies change significantly. At $M_{\star}\sim10^{10.5} M_{\odot}$ galaxies start growing central concentrations, their surface mass density and colour changes \citep{Kauffmann2003b}. At a similar mass, the bars start to buckle and form boxy-peanut bulges \citep{Erwin2017}. We find that bars become increasingly redder with stellar mass, being more similar in colour to bulges and almost always redder than the discs, suggesting that there is little star formation occuring in the bars.\\

$\bullet${\textit{Do bars evolve into lenses?}}\\
\citet{Kormendy1979} proposed a scenario in which the lenses of unbarred galaxies are the end products of bar evolution and transformation. Stars are scattered out from the bar forming a more circular feature with a roughly flat brightness profile. Simulations by \citet{Bournaud2002} show that in the case of an isolated barred galaxy, it may consume all its gas in stars and when the disc is hot enough, the bar weakens, leaving behind the lens it was embedded in, while the galaxy evolves to be an early-type system. This lens can be observed for $\sim$10 Gyr. \citet{Laurikainen2013} suggested that inner lenses in unbarred S0 galaxies are barlenses (i.e. lens-like features embedded in bars, believed to be the vertically thick part of the bar - the boxy/peanut bulge - seen face-on, as proposed by \citealt{Athanassoula2015}) in formerly barred galaxies, where the ends of the bar evolve into ansae (i.e. bright enhancements at the bar ends) and slowly dissolve with time.

Other studies such as \citet{Athanassoula1983} suggest that lenses form similarly to bars, due to an instability in the galactic discs, but in hot discs instead of cool discs. 

We find the properties (stellar masses, red colours, bar/lens S\'ersic indices) of unbarred galaxies with lenses to be similar to the properties of barred galaxies, and different from purely unbarred galaxies, suggesting a connection between the first two. Unless the lenses and bars are formed through exactly the same mechanism and the lenses have the same impact on the evolution of galaxies as bars do (rearranging angular momentum, transporting stars and gas), our preferred scenario is the one described by \citet{Kormendy1979}, \citet{Bournaud2002} or \citet{Laurikainen2013}. The bulges and lenses of unbarred galaxies with lenses have slightly larger masses (by $\sim$0.1 dex) compared to the bulges and bars of barred galaxies, compatible with a later evolutionary stage. The lenses have, on average, shorter sizes compared to bars, but higher ellipticities, possibly due to the scattering of stars in the perpendicular direction to the bars. As the mass of the central component, the bulge, increases, the bars might weaken (over a long period of time) and dissolve into a lens feature. This process was also noticed in simulations by \citet{Heller2007} who found that bars are formed early (in the first few Gyrs of disc formation), strengthen and then weaken over time. We see many lenses already present along the bar major axes, which might be a snapshot of this process in action \citep{Kormendy2013}. 

We further stress the importance of accounting for components such as lenses/ovals and bars when fitting galaxies, as these features appear frequently in galaxies and can significantly influence the derived properties of bulges. Using simple bulge+disc decompositions can lead to misleading results, especially in overestimating the fraction of de Vaucouleurs (S\'ersic index $n_{\textrm{bulge}}=4$) bulges.

\section{Conclusion}

In this paper, we make use of morphological classifications from the Galaxy Zoo project and 2D photometric decomposition to study the properties of a local sample of $\sim$3,500 galaxies with strong bars. This is currently the largest sample of barred galaxies studied through image decomposition. Using a multi-wavelength galaxy fitting routine we decompose barred galaxies into bars, discs and bulges and recover the light from each component. Taking advantage of multi-band data, we determine the structural parameters of each component, such as their colours, S\'ersic indices, effective radii, axis ratios and the fraction of total light in each component.  

We find a clear difference in colour between the components in barred galaxies: discs are bluer than the bars, which in turn are bluer than the bulges, compatible with scenarios of inside-out quenching. This colour difference steepens with stellar mass, such that the most massive galaxies show the largest difference in colour between the components. We find that the properties of bars change with galaxy type. Low mass, disc dominated galaxies have bars with an almost exponential light profile, while high mass galaxies with obvious bulges have bars with a shallower, Gaussian-like light profile. These findings are compatible with scenarios in which the bars grow in time by trapping stars from the disc in bar orbits, flattening the bar profiles. 

By comparing the barred galaxies with a similar, volume-limited and mass-matched sample of unbarred galaxies, we find a clear difference between the colours of the discs of barred and unbarred galaxies, which does not depend on mass. Discs of unbarred galaxies are significantly bluer compared to discs of barred galaxies suggesting that bars are related to the quenching of star formation in galaxy discs. Barred galaxies also contain a large proportion of `disc-like' pseuodobulges, products of secular evolution via bars (through transfer of gas to the galaxy centers), in contrast to classical bulges believed to be built by mergers. 

In conclusion, this analysis on a large sample of barred galaxies shows that bars affect the evolution of their host galaxies by forming bulges at galaxy centres and by quenching the star formation across the galaxy. We found a good agreement between our observations and simulations of the formation and evolution of barred galaxies. Still, our findings need to be tested by studying the stellar populations of bars, discs and bulges using integral field spectroscopy and this will be the subject of future work. Furthermore, we have also found that galaxies with inner lenses around the galactic bulge have more similar properties to barred galaxies than to unbarred galaxies which points towards a connection between the two. Future theoretical and observational work should elucidate the formation and evolution of these galaxy components.

\section*{Acknowledgements}

The authors would like to acknowledge the useful discussions with Dimitri Gadotti about the details of the fitting procedure and about the possible discrepancies in the measured parameters between this work and \citet{Gadotti2009}. The authors would also like to thank the anonymous referee for their helpful and useful comments.

SJK acknowledges funding from the Science and Technology Facilities Council (STFC) Grant Code ST/MJ0371X/1. RJS acknowledges funding from the STFC Grant Code ST/K502236/1. Support for this work was provided by the National Aeronautics and Space Administration through Einstein Postdoctoral Fellowship Award Number PF5-160143 issued by the Chandra X-ray Observatory Center, which is operated by the Smithsonian Astrophysical Observatory for and on behalf of the National Aeronautics Space Administration under contract NAS8-03060. 

The development of Galaxy Zoo was supported in part by the Alfred P. Sloan Foundation and by The Leverhulme Trust. 

Funding for the SDSS and SDSS-II has been provided by the Alfred P. Sloan Foundation, the Participating Institutions, the National Science Foundation, the US Department of Energy, the National Aeronautics and Space Administration, the Japanese Monbukagakusho, the Max Planck Society, and the Higher Education Funding Council for England. The SDSS website is \url{http://www.sdss.org/}. The SDSS is managed by the Astrophysical Research Consortium for the Participating Institutions. The Participating Institutions are the American Museum of Natural History, Astrophysical Institute Potsdam, University of Basel, University of Cambridge, Case Western Reserve University, University of Chicago, Drexel University, Fermilab, the Institute for Advanced Study, the Japan Participation Group, Johns Hopkins University, the Joint Institute for Nuclear Astrophysics, the Kavli Institute for Particle Astrophysics and Cosmology, the Korean Scientist Group, the Chinese Academy of Sciences (LAMOST), Los Alamos National Laboratory, the Max-Planck Institute for Astronomy (MPIA), the Max-Planck-Institute for Astrophysics (MPA), New Mexico State University, Ohio State University, University of Pittsburgh, University of Portsmouth, Princeton University, the United States Naval Observatory and the University of Washington.

This research made use of NASA's Astrophysics Data System Bibliographic Services. This work made extensive use of \textit{Astropy}\footnote{\url{http://www.astropy.org/}}, a community-developed core Python package for Astronomy \citep{Astropy} and of the Tool for Operations on Catalogues And Tables, \citep[TOPCAT\footnote{\url{http://www.star.bris.ac.uk/~mbt/}};][]{Topcat}.




\bibliographystyle{mnras}
\bibliography{references} 

\begin{thebibliography}{}
\makeatletter
\relax
\def\mn@urlcharsother{\let\do\@makeother \do\$\do\&\do\#\do\^\do\_\do\%\do\~}
\def\mn@doi{\begingroup\mn@urlcharsother \@ifnextchar [ {\mn@doi@}
  {\mn@doi@[]}}
\def\mn@doi@[#1]#2{\def\@tempa{#1}\ifx\@tempa\@empty \href
  {http://dx.doi.org/#2} {doi:#2}\else \href {http://dx.doi.org/#2} {#1}\fi
  \endgroup}
\def\mn@eprint#1#2{\mn@eprint@#1:#2::\@nil}
\def\mn@eprint@arXiv#1{\href {http://arxiv.org/abs/#1} {{\tt arXiv:#1}}}
\def\mn@eprint@dblp#1{\href {http://dblp.uni-trier.de/rec/bibtex/#1.xml}
  {dblp:#1}}
\def\mn@eprint@#1:#2:#3:#4\@nil{\def\@tempa {#1}\def\@tempb {#2}\def\@tempc
  {#3}\ifx \@tempc \@empty \let \@tempc \@tempb \let \@tempb \@tempa \fi \ifx
  \@tempb \@empty \def\@tempb {arXiv}\fi \@ifundefined
  {mn@eprint@\@tempb}{\@tempb:\@tempc}{\expandafter \expandafter \csname
  mn@eprint@\@tempb\endcsname \expandafter{\@tempc}}}

\bibitem[\protect\citeauthoryear{{Abazajian} et~al.,}{{Abazajian}
  et~al.}{2009}]{SDSSDR7}
{Abazajian} K.~N.,  et~al., 2009, \mn@doi [\apjs]
  {10.1088/0067-0049/182/2/543}, \href
  {http://adsabs.harvard.edu/abs/2009ApJS..182..543A} {182, 543}

\bibitem[\protect\citeauthoryear{{Aguerri}, {Balcells}  \&
  {Peletier}}{{Aguerri} et~al.}{2001}]{Aguerri2001}
{Aguerri} J.~A.~L.,  {Balcells} M.,   {Peletier} R.~F.,  2001, \mn@doi [\aap]
  {10.1051/0004-6361:20000441}, \href
  {http://adsabs.harvard.edu/abs/2001A%26A...367..428A} {367, 428}

\bibitem[\protect\citeauthoryear{{Ahn} et~al.,}{{Ahn} et~al.}{2014}]{Ahn2014}
{Ahn} C.~P.,  et~al., 2014, \mn@doi [\apjs] {10.1088/0067-0049/211/2/17}, \href
  {http://adsabs.harvard.edu/abs/2014ApJS..211...17A} {211, 17}

\bibitem[\protect\citeauthoryear{{Allen}, {Driver}, {Graham}, {Cameron},
  {Liske}  \& {de Propris}}{{Allen} et~al.}{2006}]{Allen2006}
{Allen} P.~D.,  {Driver} S.~P.,  {Graham} A.~W.,  {Cameron} E.,  {Liske} J.,
  {de Propris} R.,  2006, \mn@doi [\mnras] {10.1111/j.1365-2966.2006.10586.x},
  \href {http://adsabs.harvard.edu/abs/2006MNRAS.371....2A} {371, 2}

\bibitem[\protect\citeauthoryear{{Astropy Collaboration} et~al.,}{{Astropy
  Collaboration} et~al.}{2013}]{Astropy}
{Astropy Collaboration} et~al., 2013, \mn@doi [\aap]
  {10.1051/0004-6361/201322068}, \href
  {http://adsabs.harvard.edu/abs/2013A%26A...558A..33A} {558, A33}

\bibitem[\protect\citeauthoryear{{Athanassoula}}{{Athanassoula}}{1983}]{Athanassoula1983}
{Athanassoula} E.,  1983, in {Athanassoula} E.,  ed.,  IAU Symposium Vol. 100,
  Internal Kinematics and Dynamics of Galaxies. pp 243--250

\bibitem[\protect\citeauthoryear{{Athanassoula}}{{Athanassoula}}{2000}]{Athanassoula2000}
{Athanassoula} E.,  2000, in {Alloin} D.,  {Olsen} K.,   {Galaz} G.,  eds,
  Astronomical Society of the Pacific Conference Series Vol. 221, Stars, Gas
  and Dust in Galaxies: Exploring the Links. p.~243 (\mn@eprint {}
  {astro-ph/0006403})

\bibitem[\protect\citeauthoryear{{Athanassoula} \& {Misiriotis}}{{Athanassoula}
  \& {Misiriotis}}{2002}]{Athanassoula2002}
{Athanassoula} E.,  {Misiriotis} A.,  2002, \mn@doi [\mnras]
  {10.1046/j.1365-8711.2002.05028.x}, \href
  {http://adsabs.harvard.edu/abs/2002MNRAS.330...35A} {330, 35}

\bibitem[\protect\citeauthoryear{{Athanassoula}, {Morin}, {Wozniak}, {Puy},
  {Pierce}, {Lombard}  \& {Bosma}}{{Athanassoula}
  et~al.}{1990}]{Athanassoula1990}
{Athanassoula} E.,  {Morin} S.,  {Wozniak} H.,  {Puy} D.,  {Pierce} M.~J.,
  {Lombard} J.,   {Bosma} A.,  1990, \mnras, \href
  {http://adsabs.harvard.edu/abs/1990MNRAS.245..130A} {245, 130}

\bibitem[\protect\citeauthoryear{{Athanassoula}, {Machado}  \&
  {Rodionov}}{{Athanassoula} et~al.}{2013}]{Athanassoula2013}
{Athanassoula} E.,  {Machado} R.~E.~G.,   {Rodionov} S.~A.,  2013, \mn@doi
  [\mnras] {10.1093/mnras/sts452}, \href
  {http://adsabs.harvard.edu/abs/2013MNRAS.429.1949A} {429, 1949}

\bibitem[\protect\citeauthoryear{{Athanassoula}, {Laurikainen}, {Salo}  \&
  {Bosma}}{{Athanassoula} et~al.}{2015}]{Athanassoula2015}
{Athanassoula} E.,  {Laurikainen} E.,  {Salo} H.,   {Bosma} A.,  2015, \mn@doi
  [\mnras] {10.1093/mnras/stv2231}, \href
  {http://adsabs.harvard.edu/abs/2015MNRAS.454.3843A} {454, 3843}

\bibitem[\protect\citeauthoryear{{Aumer} \& {Binney}}{{Aumer} \&
  {Binney}}{2017}]{Aumer2017}
{Aumer} M.,  {Binney} J.,  2017, \mn@doi [\mnras] {10.1093/mnras/stx1300},
  \href {http://adsabs.harvard.edu/abs/2017MNRAS.470.2113A} {470, 2113}

\bibitem[\protect\citeauthoryear{{Aumer}, {Binney}  \& {Sch{\"o}nrich}}{{Aumer}
  et~al.}{2016}]{Aumer2016}
{Aumer} M.,  {Binney} J.,   {Sch{\"o}nrich} R.,  2016, \mn@doi [\mnras]
  {10.1093/mnras/stw777}, \href
  {http://adsabs.harvard.edu/abs/2016MNRAS.459.3326A} {459, 3326}

\bibitem[\protect\citeauthoryear{{Bamford}, {H{\"a}u{\ss}ler}, {Rojas}  \&
  {Borch}}{{Bamford} et~al.}{2011}]{Bamford2011}
{Bamford} S.~P.,  {H{\"a}u{\ss}ler} B.,  {Rojas} A.,   {Borch} A.,  2011, in
  {Evans} I.~N.,  {Accomazzi} A.,  {Mink} D.~J.,   {Rots} A.~H.,  eds,
  Astronomical Society of the Pacific Conference Series Vol. 442, Astronomical
  Data Analysis Software and Systems XX. p.~479

\bibitem[\protect\citeauthoryear{{Barden}, {H{\"a}u{\ss}ler}, {Peng},
  {McIntosh}  \& {Guo}}{{Barden} et~al.}{2012}]{Barden2012}
{Barden} M.,  {H{\"a}u{\ss}ler} B.,  {Peng} C.~Y.,  {McIntosh} D.~H.,   {Guo}
  Y.,  2012, \mn@doi [\mnras] {10.1111/j.1365-2966.2012.20619.x}, \href
  {http://adsabs.harvard.edu/abs/2012MNRAS.422..449B} {422, 449}

\bibitem[\protect\citeauthoryear{{Belfiore} et~al.,}{{Belfiore}
  et~al.}{2017}]{Belfiore2017}
{Belfiore} F.,  et~al., 2017, \mn@doi [\mnras] {10.1093/mnras/stw3211}, \href
  {http://adsabs.harvard.edu/abs/2017MNRAS.466.2570B} {466, 2570}

\bibitem[\protect\citeauthoryear{{Berentzen}, {Shlosman}, {Martinez-Valpuesta}
  \& {Heller}}{{Berentzen} et~al.}{2007}]{Berentzen2007}
{Berentzen} I.,  {Shlosman} I.,  {Martinez-Valpuesta} I.,   {Heller} C.~H.,
  2007, \mn@doi [\apj] {10.1086/520531}, \href
  {http://adsabs.harvard.edu/abs/2007ApJ...666..189B} {666, 189}

\bibitem[\protect\citeauthoryear{{Bertin} \& {Arnouts}}{{Bertin} \&
  {Arnouts}}{1996}]{Bertin1996}
{Bertin} E.,  {Arnouts} S.,  1996, \mn@doi [\aaps] {10.1051/aas:1996164}, \href
  {http://adsabs.harvard.edu/abs/1996A%26AS..117..393B} {117, 393}

\bibitem[\protect\citeauthoryear{{Binney} \& {Tremaine}}{{Binney} \&
  {Tremaine}}{1987}]{Binney1987}
{Binney} J.,  {Tremaine} S.,  1987, {Galactic dynamics}

\bibitem[\protect\citeauthoryear{{Blanton} \& {Roweis}}{{Blanton} \&
  {Roweis}}{2007}]{Blanton2007}
{Blanton} M.~R.,  {Roweis} S.,  2007, \mn@doi [\aj] {10.1086/510127}, \href
  {http://adsabs.harvard.edu/abs/2007AJ....133..734B} {133, 734}

\bibitem[\protect\citeauthoryear{{Bournaud} \& {Combes}}{{Bournaud} \&
  {Combes}}{2002}]{Bournaud2002}
{Bournaud} F.,  {Combes} F.,  2002, \mn@doi [\aap]
  {10.1051/0004-6361:20020920}, \href
  {http://adsabs.harvard.edu/abs/2002A%26A...392...83B} {392, 83}

\bibitem[\protect\citeauthoryear{{Bournaud}, {Combes}  \& {Semelin}}{{Bournaud}
  et~al.}{2005}]{Bournaud2005}
{Bournaud} F.,  {Combes} F.,   {Semelin} B.,  2005, \mn@doi [\mnras]
  {10.1111/j.1745-3933.2005.00096.x}, \href
  {http://adsabs.harvard.edu/abs/2005MNRAS.364L..18B} {364, L18}

\bibitem[\protect\citeauthoryear{{Bruzual} \& {Charlot}}{{Bruzual} \&
  {Charlot}}{2003}]{Bruzual2003}
{Bruzual} G.,  {Charlot} S.,  2003, \mn@doi [\mnras]
  {10.1046/j.1365-8711.2003.06897.x}, \href
  {http://adsabs.harvard.edu/abs/2003MNRAS.344.1000B} {344, 1000}

\bibitem[\protect\citeauthoryear{{Bureau} \& {Freeman}}{{Bureau} \&
  {Freeman}}{1999}]{Bureau1999}
{Bureau} M.,  {Freeman} K.~C.,  1999, \mn@doi [\aj] {10.1086/300922}, \href
  {http://adsabs.harvard.edu/abs/1999AJ....118..126B} {118, 126}

\bibitem[\protect\citeauthoryear{{Buta}}{{Buta}}{2017}]{Buta2017}
{Buta} R.~J.,  2017, \mn@doi [\mnras] {10.1093/mnras/stx1829}, \href
  {http://adsabs.harvard.edu/abs/2017MNRAS.471.4027B} {471, 4027}

\bibitem[\protect\citeauthoryear{{Buta} \& {Combes}}{{Buta} \&
  {Combes}}{1996}]{Buta1996}
{Buta} R.,  {Combes} F.,  1996, \fcp, \href
  {http://adsabs.harvard.edu/abs/1996FCPh...17...95B} {17, 95}

\bibitem[\protect\citeauthoryear{{Buta}, {Corwin}  \& {Odewahn}}{{Buta}
  et~al.}{2007}]{Buta2007}
{Buta} R.~J.,  {Corwin} H.~G.,   {Odewahn} S.~C.,  2007, {The de Vaucouleurs
  Altlas of Galaxies}.
Cambridge University Press

\bibitem[\protect\citeauthoryear{{Byun} \& {Freeman}}{{Byun} \&
  {Freeman}}{1995}]{Byun1995}
{Byun} Y.~I.,  {Freeman} K.~C.,  1995, \mn@doi [\apj] {10.1086/175986}, \href
  {http://adsabs.harvard.edu/abs/1995ApJ...448..563B} {448, 563}

\bibitem[\protect\citeauthoryear{{Carles}, {Martel}, {Ellison}  \&
  {Kawata}}{{Carles} et~al.}{2016}]{Carles2016}
{Carles} C.,  {Martel} H.,  {Ellison} S.~L.,   {Kawata} D.,  2016, \mn@doi
  [\mnras] {10.1093/mnras/stw2056}, \href
  {http://adsabs.harvard.edu/abs/2016MNRAS.463.1074C} {463, 1074}

\bibitem[\protect\citeauthoryear{{Casteels} et~al.,}{{Casteels}
  et~al.}{2013}]{Casteels2013}
{Casteels} K.~R.~V.,  et~al., 2013, \mn@doi [\mnras] {10.1093/mnras/sts391},
  \href {http://adsabs.harvard.edu/abs/2013MNRAS.429.1051C} {429, 1051}

\bibitem[\protect\citeauthoryear{{Cheung} et~al.,}{{Cheung}
  et~al.}{2013}]{Cheung2013}
{Cheung} E.,  et~al., 2013, \mn@doi [\apj] {10.1088/0004-637X/779/2/162}, \href
  {http://adsabs.harvard.edu/abs/2013ApJ...779..162C} {779, 162}

\bibitem[\protect\citeauthoryear{{Cheung} et~al.,}{{Cheung}
  et~al.}{2015a}]{Cheung2015}
{Cheung} E.,  et~al., 2015a, \mn@doi [\mnras] {10.1093/mnras/stu2462}, \href
  {http://adsabs.harvard.edu/abs/2015MNRAS.447..506C} {447, 506}

\bibitem[\protect\citeauthoryear{{Cheung} et~al.,}{{Cheung}
  et~al.}{2015b}]{Cheung2015b}
{Cheung} E.,  et~al., 2015b, \mn@doi [\apj] {10.1088/0004-637X/807/1/36}, \href
  {http://adsabs.harvard.edu/abs/2015ApJ...807...36C} {807, 36}

\bibitem[\protect\citeauthoryear{{Cisternas}, {Sheth}, {Salvato}, {Knapen},
  {Civano}  \& {Santini}}{{Cisternas} et~al.}{2015}]{Cisternas2015}
{Cisternas} M.,  {Sheth} K.,  {Salvato} M.,  {Knapen} J.~H.,  {Civano} F.,
  {Santini} P.,  2015, \mn@doi [\apj] {10.1088/0004-637X/802/2/137}, \href
  {http://adsabs.harvard.edu/abs/2015ApJ...802..137C} {802, 137}

\bibitem[\protect\citeauthoryear{{Coelho} \& {Gadotti}}{{Coelho} \&
  {Gadotti}}{2011}]{Coelho2011}
{Coelho} P.,  {Gadotti} D.~A.,  2011, \mn@doi [\apjl]
  {10.1088/2041-8205/743/1/L13}, \href
  {http://adsabs.harvard.edu/abs/2011ApJ...743L..13C} {743, L13}

\bibitem[\protect\citeauthoryear{{Combes} \& {Elmegreen}}{{Combes} \&
  {Elmegreen}}{1993}]{Elmegreen1993}
{Combes} F.,  {Elmegreen} B.~G.,  1993, \aap, \href
  {http://adsabs.harvard.edu/abs/1993A%26A...271..391C} {271, 391}

\bibitem[\protect\citeauthoryear{{Combes} \& {Sanders}}{{Combes} \&
  {Sanders}}{1981}]{Combes1981}
{Combes} F.,  {Sanders} R.~H.,  1981, \aap, \href
  {http://adsabs.harvard.edu/abs/1981A%26A....96..164C} {96, 164}

\bibitem[\protect\citeauthoryear{{Croom} et~al.,}{{Croom}
  et~al.}{2012}]{Sami2012}
{Croom} S.~M.,  et~al., 2012, \mn@doi [\mnras]
  {10.1111/j.1365-2966.2011.20365.x}, \href
  {http://adsabs.harvard.edu/abs/2012MNRAS.421..872C} {421, 872}

\bibitem[\protect\citeauthoryear{{Darg} et~al.,}{{Darg}
  et~al.}{2010}]{Darg2010}
{Darg} D.~W.,  et~al., 2010, \mn@doi [\mnras]
  {10.1111/j.1365-2966.2009.15686.x}, \href
  {http://adsabs.harvard.edu/abs/2010MNRAS.401.1043D} {401, 1043}

\bibitem[\protect\citeauthoryear{{Debattista}, {Mayer}, {Carollo}, {Moore},
  {Wadsley}  \& {Quinn}}{{Debattista} et~al.}{2006}]{Debattista2006}
{Debattista} V.~P.,  {Mayer} L.,  {Carollo} C.~M.,  {Moore} B.,  {Wadsley} J.,
   {Quinn} T.,  2006, \mn@doi [\apj] {10.1086/504147}, \href
  {http://adsabs.harvard.edu/abs/2006ApJ...645..209D} {645, 209}

\bibitem[\protect\citeauthoryear{{Di Matteo}, {Haywood}, {Combes}, {Semelin}
  \& {Snaith}}{{Di Matteo} et~al.}{2013}]{DiMatteo2013}
{Di Matteo} P.,  {Haywood} M.,  {Combes} F.,  {Semelin} B.,   {Snaith} O.~N.,
  2013, \mn@doi [\aap] {10.1051/0004-6361/201220539}, \href
  {http://adsabs.harvard.edu/abs/2013A%26A...553A.102D} {553, A102}

\bibitem[\protect\citeauthoryear{{Driver}, {Popescu}, {Tuffs}, {Graham},
  {Liske}  \& {Baldry}}{{Driver} et~al.}{2008}]{Driver2008}
{Driver} S.~P.,  {Popescu} C.~C.,  {Tuffs} R.~J.,  {Graham} A.~W.,  {Liske} J.,
    {Baldry} I.,  2008, \mn@doi [\apjl] {10.1086/588582}, \href
  {http://adsabs.harvard.edu/abs/2008ApJ...678L.101D} {678, L101}

\bibitem[\protect\citeauthoryear{{Eisenstein} et~al.,}{{Eisenstein}
  et~al.}{2011}]{Eisenstein2011}
{Eisenstein} D.~J.,  et~al., 2011, \mn@doi [\aj] {10.1088/0004-6256/142/3/72},
  \href {http://adsabs.harvard.edu/abs/2011AJ....142...72E} {142, 72}

\bibitem[\protect\citeauthoryear{{Ellison}, {Nair}, {Patton}, {Scudder},
  {Mendel}  \& {Simard}}{{Ellison} et~al.}{2011}]{Ellison2011}
{Ellison} S.~L.,  {Nair} P.,  {Patton} D.~R.,  {Scudder} J.~M.,  {Mendel}
  J.~T.,   {Simard} L.,  2011, \mn@doi [\mnras]
  {10.1111/j.1365-2966.2011.19195.x}, \href
  {http://adsabs.harvard.edu/abs/2011MNRAS.416.2182E} {416, 2182}

\bibitem[\protect\citeauthoryear{{Elmegreen} \& {Elmegreen}}{{Elmegreen} \&
  {Elmegreen}}{1985}]{Elmegreen1985}
{Elmegreen} B.~G.,  {Elmegreen} D.~M.,  1985, \mn@doi [\apj] {10.1086/162810},
  \href {http://adsabs.harvard.edu/abs/1985ApJ...288..438E} {288, 438}

\bibitem[\protect\citeauthoryear{{Elmegreen}, {Elmegreen}, {Chromey},
  {Hasselbacher}  \& {Bissell}}{{Elmegreen} et~al.}{1996}]{Elmegreen1996}
{Elmegreen} B.~G.,  {Elmegreen} D.~M.,  {Chromey} F.~R.,  {Hasselbacher} D.~A.,
    {Bissell} B.~A.,  1996, \mn@doi [\aj] {10.1086/117957}, \href
  {http://adsabs.harvard.edu/abs/1996AJ....111.2233E} {111, 2233}

\bibitem[\protect\citeauthoryear{{Erwin} \& {Debattista}}{{Erwin} \&
  {Debattista}}{2017}]{Erwin2017}
{Erwin} P.,  {Debattista} V.~P.,  2017, \mn@doi [\mnras]
  {10.1093/mnras/stx620}, \href
  {http://adsabs.harvard.edu/abs/2017MNRAS.468.2058E} {468, 2058}

\bibitem[\protect\citeauthoryear{{Fisher} \& {Drory}}{{Fisher} \&
  {Drory}}{2008}]{Fisher2008}
{Fisher} D.~B.,  {Drory} N.,  2008, \mn@doi [\aj]
  {10.1088/0004-6256/136/2/773}, \href
  {http://adsabs.harvard.edu/abs/2008AJ....136..773F} {136, 773}

\bibitem[\protect\citeauthoryear{{Fisher} \& {Drory}}{{Fisher} \&
  {Drory}}{2010}]{Fisher2010}
{Fisher} D.~B.,  {Drory} N.,  2010, \mn@doi [\apj]
  {10.1088/0004-637X/716/2/942}, \href
  {http://adsabs.harvard.edu/abs/2010ApJ...716..942F} {716, 942}

\bibitem[\protect\citeauthoryear{{Friedli} \& {Benz}}{{Friedli} \&
  {Benz}}{1993}]{Friedli1993}
{Friedli} D.,  {Benz} W.,  1993, \aap, \href
  {http://adsabs.harvard.edu/abs/1993A%26A...268...65F} {268, 65}

\bibitem[\protect\citeauthoryear{{Friedli}, {Benz}  \& {Kennicutt}}{{Friedli}
  et~al.}{1994}]{Friedli1994}
{Friedli} D.,  {Benz} W.,   {Kennicutt} R.,  1994, \mn@doi [\apjl]
  {10.1086/187449}, \href {http://adsabs.harvard.edu/abs/1994ApJ...430L.105F}
  {430, L105}

\bibitem[\protect\citeauthoryear{{Gadotti}}{{Gadotti}}{2008}]{Gadotti2008}
{Gadotti} D.~A.,  2008, \mn@doi [\mnras] {10.1111/j.1365-2966.2007.12723.x},
  \href {http://adsabs.harvard.edu/abs/2008MNRAS.384..420G} {384, 420}

\bibitem[\protect\citeauthoryear{{Gadotti}}{{Gadotti}}{2009}]{Gadotti2009}
{Gadotti} D.~A.,  2009, \mn@doi [\mnras] {10.1111/j.1365-2966.2008.14257.x},
  \href {http://adsabs.harvard.edu/abs/2009MNRAS.393.1531G} {393, 1531}

\bibitem[\protect\citeauthoryear{{Gadotti}}{{Gadotti}}{2010}]{Gadotti2010a}
{Gadotti} D.~A.,  2010, VizieR Online Data Catalog, \href
  {http://adsabs.harvard.edu/abs/2010yCat..73931531G} {739}

\bibitem[\protect\citeauthoryear{{Gadotti}}{{Gadotti}}{2011}]{Gadotti2011}
{Gadotti} D.~A.,  2011, \mn@doi [\mnras] {10.1111/j.1365-2966.2011.18945.x},
  \href {http://adsabs.harvard.edu/abs/2011MNRAS.415.3308G} {415, 3308}

\bibitem[\protect\citeauthoryear{{Gadotti}, {Baes}  \& {Falony}}{{Gadotti}
  et~al.}{2010}]{Gadotti2010b}
{Gadotti} D.~A.,  {Baes} M.,   {Falony} S.,  2010, \mn@doi [\mnras]
  {10.1111/j.1365-2966.2010.16243.x}, \href
  {http://adsabs.harvard.edu/abs/2010MNRAS.403.2053G} {403, 2053}

\bibitem[\protect\citeauthoryear{{Gadotti}, {Seidel},
  {S{\'a}nchez-Bl{\'a}zquez}, {Falc{\'o}n-Barroso}, {Husemann}, {Coelho}  \&
  {P{\'e}rez}}{{Gadotti} et~al.}{2015}]{Gadotti2015}
{Gadotti} D.~A.,  {Seidel} M.~K.,  {S{\'a}nchez-Bl{\'a}zquez} P.,
  {Falc{\'o}n-Barroso} J.,  {Husemann} B.,  {Coelho} P.,   {P{\'e}rez} I.,
  2015, \mn@doi [\aap] {10.1051/0004-6361/201526677}, \href
  {http://adsabs.harvard.edu/abs/2015A%26A...584A..90G} {584, A90}

\bibitem[\protect\citeauthoryear{{Galloway} et~al.,}{{Galloway}
  et~al.}{2015}]{Galloway2015}
{Galloway} M.~A.,  et~al., 2015, \mn@doi [\mnras] {10.1093/mnras/stv235}, \href
  {http://adsabs.harvard.edu/abs/2015MNRAS.448.3442G} {448, 3442}

\bibitem[\protect\citeauthoryear{{Gavazzi} et~al.,}{{Gavazzi}
  et~al.}{2015}]{Gavazzi2015}
{Gavazzi} G.,  et~al., 2015, \mn@doi [\aap] {10.1051/0004-6361/201425351},
  \href {http://adsabs.harvard.edu/abs/2015A%26A...580A.116G} {580, A116}

\bibitem[\protect\citeauthoryear{{Goulding} et~al.,}{{Goulding}
  et~al.}{2017}]{Goulding2017}
{Goulding} A.~D.,  et~al., 2017, \mn@doi [\apj] {10.3847/1538-4357/aa755b},
  \href {http://adsabs.harvard.edu/abs/2017ApJ...843..135G} {843, 135}

\bibitem[\protect\citeauthoryear{{Graham}}{{Graham}}{2016}]{Graham2011}
{Graham} A.~W.,  2016, \mn@doi [Galactic Bulges]
  {10.1007/978-3-319-19378-6_11}, \href
  {http://adsabs.harvard.edu/abs/2016ASSL..418..263G} {418, 263}

\bibitem[\protect\citeauthoryear{{Graham} \& {Worley}}{{Graham} \&
  {Worley}}{2008}]{Graham2008}
{Graham} A.~W.,  {Worley} C.~C.,  2008, \mn@doi [\mnras]
  {10.1111/j.1365-2966.2008.13506.x}, \href
  {http://adsabs.harvard.edu/abs/2008MNRAS.388.1708G} {388, 1708}

\bibitem[\protect\citeauthoryear{{Gunn} et~al.,}{{Gunn}
  et~al.}{1998}]{Gunn1998}
{Gunn} J.~E.,  et~al., 1998, \mn@doi [\aj] {10.1086/300645}, \href
  {http://adsabs.harvard.edu/abs/1998AJ....116.3040G} {116, 3040}

\bibitem[\protect\citeauthoryear{{Hart}, {Bamford}, {Casteels}, {Kruk},
  {Lintott}  \& {Masters}}{{Hart} et~al.}{2017}]{Hart2017}
{Hart} R.~E.,  {Bamford} S.~P.,  {Casteels} K.~R.~V.,  {Kruk} S.~J.,  {Lintott}
  C.~J.,   {Masters} K.~L.,  2017, \mn@doi [\mnras] {10.1093/mnras/stx581},
  \href {http://adsabs.harvard.edu/abs/2017MNRAS.468.1850H} {468, 1850}

\bibitem[\protect\citeauthoryear{{H{\"a}ussler} et~al.,}{{H{\"a}ussler}
  et~al.}{2007}]{Haeussler2007}
{H{\"a}ussler} B.,  et~al., 2007, \mn@doi [\apjs] {10.1086/518836}, \href
  {http://adsabs.harvard.edu/abs/2007ApJS..172..615H} {172, 615}

\bibitem[\protect\citeauthoryear{{H{\"a}u{\ss}ler} et~al.,}{{H{\"a}u{\ss}ler}
  et~al.}{2013}]{Heussler2013}
{H{\"a}u{\ss}ler} B.,  et~al., 2013, \mn@doi [\mnras] {10.1093/mnras/sts633},
  \href {http://adsabs.harvard.edu/abs/2013MNRAS.430..330H} {430, 330}

\bibitem[\protect\citeauthoryear{{Hawarden}, {Mountain}, {Leggett}  \&
  {Puxley}}{{Hawarden} et~al.}{1986}]{Hawarden1986}
{Hawarden} T.~G.,  {Mountain} C.~M.,  {Leggett} S.~K.,   {Puxley} P.~J.~G.,
  1986, \mnras, \href {http://adsabs.harvard.edu/abs/1986MNRAS.221P..41H} {221,
  41P}

\bibitem[\protect\citeauthoryear{{Head} et~al.,}{{Head}
  et~al.}{2014}]{Head2014}
{Head} J.~T.~C.~G.,  et~al., 2014, \mn@doi [\mnras] {10.1093/mnras/stu325},
  \href {http://adsabs.harvard.edu/abs/2014MNRAS.440.1690H} {440, 1690}

\bibitem[\protect\citeauthoryear{{Heller}, {Shlosman}  \&
  {Athanassoula}}{{Heller} et~al.}{2007}]{Heller2007}
{Heller} C.~H.,  {Shlosman} I.,   {Athanassoula} E.,  2007, \mn@doi [\apj]
  {10.1086/523260}, \href {http://adsabs.harvard.edu/abs/2007ApJ...671..226H}
  {671, 226}

\bibitem[\protect\citeauthoryear{{Hickox}, {Mullaney}, {Alexander}, {Chen},
  {Civano}, {Goulding}  \& {Hainline}}{{Hickox} et~al.}{2014}]{Hickox2014}
{Hickox} R.~C.,  {Mullaney} J.~R.,  {Alexander} D.~M.,  {Chen} C.-T.~J.,
  {Civano} F.~M.,  {Goulding} A.~D.,   {Hainline} K.~N.,  2014, \mn@doi [\apj]
  {10.1088/0004-637X/782/1/9}, \href
  {http://adsabs.harvard.edu/abs/2014ApJ...782....9H} {782, 9}

\bibitem[\protect\citeauthoryear{{Hoyle} et~al.,}{{Hoyle}
  et~al.}{2011}]{Hoyle2011}
{Hoyle} B.,  et~al., 2011, \mn@doi [\mnras] {10.1111/j.1365-2966.2011.18979.x},
  \href {http://adsabs.harvard.edu/abs/2011MNRAS.415.3627H} {415, 3627}

\bibitem[\protect\citeauthoryear{{Hubble}}{{Hubble}}{1936}]{Hubble1936}
{Hubble} E.,  1936, \mn@doi [Science] {10.1126/science.84.2188.509}, \href
  {http://adsabs.harvard.edu/abs/1936Sci....84..509M} {84, 509}

\bibitem[\protect\citeauthoryear{{Jacob} et~al.,}{{Jacob}
  et~al.}{2010}]{Jacob2010}
{Jacob} J.~C.,  et~al., 2010, {Montage: An Astronomical Image Mosaicking
  Toolkit} (\mn@eprint {ascl} {1010.036})

\bibitem[\protect\citeauthoryear{{Jarosik} et~al.,}{{Jarosik}
  et~al.}{2011}]{Jarosik2011}
{Jarosik} N.,  et~al., 2011, \mn@doi [\apjs] {10.1088/0067-0049/192/2/14},
  \href {http://adsabs.harvard.edu/abs/2011ApJS..192...14J} {192, 14}

\bibitem[\protect\citeauthoryear{{Johnston}, {Arag{\'o}n-Salamanca}  \&
  {Merrifield}}{{Johnston} et~al.}{2014}]{Johnston2014}
{Johnston} E.~J.,  {Arag{\'o}n-Salamanca} A.,   {Merrifield} M.~R.,  2014,
  \mn@doi [\mnras] {10.1093/mnras/stu582}, \href
  {http://adsabs.harvard.edu/abs/2014MNRAS.441..333J} {441, 333}

\bibitem[\protect\citeauthoryear{{Johnston} et~al.,}{{Johnston}
  et~al.}{2017}]{Johnston2017}
{Johnston} E.~J.,  et~al., 2017, \mn@doi [\mnras] {10.1093/mnras/stw2823},
  \href {http://adsabs.harvard.edu/abs/2017MNRAS.465.2317J} {465, 2317}

\bibitem[\protect\citeauthoryear{{Kauffmann} et~al.,}{{Kauffmann}
  et~al.}{2003a}]{Kauffmann2003a}
{Kauffmann} G.,  et~al., 2003a, \mn@doi [\mnras]
  {10.1046/j.1365-8711.2003.06291.x}, \href
  {http://adsabs.harvard.edu/abs/2003MNRAS.341...33K} {341, 33}

\bibitem[\protect\citeauthoryear{{Kauffmann} et~al.,}{{Kauffmann}
  et~al.}{2003b}]{Kauffmann2003b}
{Kauffmann} G.,  et~al., 2003b, \mn@doi [\mnras]
  {10.1046/j.1365-8711.2003.06292.x}, \href
  {http://adsabs.harvard.edu/abs/2003MNRAS.341...54K} {341, 54}

\bibitem[\protect\citeauthoryear{{Kennedy} et~al.,}{{Kennedy}
  et~al.}{2016}]{Kennedy2016}
{Kennedy} R.,  et~al., 2016, \mn@doi [\mnras] {10.1093/mnras/stw1176}, \href
  {http://adsabs.harvard.edu/abs/2016MNRAS.460.3458K} {460, 3458}

\bibitem[\protect\citeauthoryear{{Kim} et~al.,}{{Kim} et~al.}{2015}]{Kim2015}
{Kim} T.,  et~al., 2015, \mn@doi [\apj] {10.1088/0004-637X/799/1/99}, \href
  {http://adsabs.harvard.edu/abs/2015ApJ...799...99K} {799, 99}

\bibitem[\protect\citeauthoryear{{Kormendy}}{{Kormendy}}{1979}]{Kormendy1979}
{Kormendy} J.,  1979, \mn@doi [\apj] {10.1086/156782}, \href
  {http://adsabs.harvard.edu/abs/1979ApJ...227..714K} {227, 714}

\bibitem[\protect\citeauthoryear{{Kormendy}}{{Kormendy}}{2013}]{Kormendy2013}
{Kormendy} J.,  2013, {Secular Evolution in Disk Galaxies}.
p.~1

\bibitem[\protect\citeauthoryear{{Kormendy} \& {Kennicutt}}{{Kormendy} \&
  {Kennicutt}}{2004}]{Kormendy2004}
{Kormendy} J.,  {Kennicutt} Jr. R.~C.,  2004, \mn@doi [\araa]
  {10.1146/annurev.astro.42.053102.134024}, \href
  {http://adsabs.harvard.edu/abs/2004ARA%26A..42..603K} {42, 603}

\bibitem[\protect\citeauthoryear{{Kormendy}, {Drory}, {Bender}  \&
  {Cornell}}{{Kormendy} et~al.}{2010}]{Kormendy2010}
{Kormendy} J.,  {Drory} N.,  {Bender} R.,   {Cornell} M.~E.,  2010, \mn@doi
  [\apj] {10.1088/0004-637X/723/1/54}, \href
  {http://adsabs.harvard.edu/abs/2010ApJ...723...54K} {723, 54}

\bibitem[\protect\citeauthoryear{{Kraljic}, {Bournaud}  \& {Martig}}{{Kraljic}
  et~al.}{2012}]{Kraljic2012}
{Kraljic} K.,  {Bournaud} F.,   {Martig} M.,  2012, \mn@doi [\apj]
  {10.1088/0004-637X/757/1/60}, \href
  {http://adsabs.harvard.edu/abs/2012ApJ...757...60K} {757, 60}

\bibitem[\protect\citeauthoryear{{Kruk} et~al.,}{{Kruk}
  et~al.}{2017}]{Kruk2017}
{Kruk} S.~J.,  et~al., 2017, \mn@doi [\mnras] {10.1093/mnras/stx1026}, \href
  {http://adsabs.harvard.edu/abs/2017MNRAS.469.3363K} {469, 3363}

\bibitem[\protect\citeauthoryear{{Laurikainen}, {Salo}, {Buta}  \&
  {Vasylyev}}{{Laurikainen} et~al.}{2004}]{Laurikainen2004b}
{Laurikainen} E.,  {Salo} H.,  {Buta} R.,   {Vasylyev} S.,  2004, \mn@doi
  [\mnras] {10.1111/j.1365-2966.2004.08410.x}, \href
  {http://adsabs.harvard.edu/abs/2004MNRAS.355.1251L} {355, 1251}

\bibitem[\protect\citeauthoryear{{Laurikainen}, {Salo}  \&
  {Buta}}{{Laurikainen} et~al.}{2005}]{Laurikainen2005}
{Laurikainen} E.,  {Salo} H.,   {Buta} R.,  2005, \mn@doi [\mnras]
  {10.1111/j.1365-2966.2005.09404.x}, \href
  {http://adsabs.harvard.edu/abs/2005MNRAS.362.1319L} {362, 1319}

\bibitem[\protect\citeauthoryear{{Laurikainen}, {Salo}, {Buta}, {Knapen},
  {Speltincx}  \& {Block}}{{Laurikainen} et~al.}{2006}]{Laurikainen2006}
{Laurikainen} E.,  {Salo} H.,  {Buta} R.,  {Knapen} J.,  {Speltincx} T.,
  {Block} D.,  2006, \mn@doi [\aj] {10.1086/508810}, \href
  {http://adsabs.harvard.edu/abs/2006AJ....132.2634L} {132, 2634}

\bibitem[\protect\citeauthoryear{{Laurikainen}, {Salo}, {Buta}  \&
  {Knapen}}{{Laurikainen} et~al.}{2007}]{Laurikainen2007}
{Laurikainen} E.,  {Salo} H.,  {Buta} R.,   {Knapen} J.~H.,  2007, \mn@doi
  [\mnras] {10.1111/j.1365-2966.2007.12299.x}, \href
  {http://adsabs.harvard.edu/abs/2007MNRAS.381..401L} {381, 401}

\bibitem[\protect\citeauthoryear{{Laurikainen}, {Salo}, {Buta}  \&
  {Knapen}}{{Laurikainen} et~al.}{2009}]{Laurikainen2009}
{Laurikainen} E.,  {Salo} H.,  {Buta} R.,   {Knapen} J.~H.,  2009, \mn@doi
  [\apjl] {10.1088/0004-637X/692/1/L34}, \href
  {http://adsabs.harvard.edu/abs/2009ApJ...692L..34L} {692, L34}

\bibitem[\protect\citeauthoryear{{Laurikainen}, {Salo}, {Buta}  \&
  {Knapen}}{{Laurikainen} et~al.}{2011}]{Laurikainen2011}
{Laurikainen} E.,  {Salo} H.,  {Buta} R.,   {Knapen} J.~H.,  2011, \mn@doi
  [\mnras] {10.1111/j.1365-2966.2011.19283.x}, \href
  {http://adsabs.harvard.edu/abs/2011MNRAS.418.1452L} {418, 1452}

\bibitem[\protect\citeauthoryear{{Laurikainen}, {Salo}, {Athanassoula},
  {Bosma}, {Buta}  \& {Janz}}{{Laurikainen} et~al.}{2013}]{Laurikainen2013}
{Laurikainen} E.,  {Salo} H.,  {Athanassoula} E.,  {Bosma} A.,  {Buta} R.,
  {Janz} J.,  2013, \mn@doi [\mnras] {10.1093/mnras/stt150}, \href
  {http://adsabs.harvard.edu/abs/2013MNRAS.430.3489L} {430, 3489}

\bibitem[\protect\citeauthoryear{{Lintott} et~al.,}{{Lintott}
  et~al.}{2008}]{Lintott2008}
{Lintott} C.~J.,  et~al., 2008, \mn@doi [\mnras]
  {10.1111/j.1365-2966.2008.13689.x}, \href
  {http://adsabs.harvard.edu/abs/2008MNRAS.389.1179L} {389, 1179}

\bibitem[\protect\citeauthoryear{{Lintott} et~al.,}{{Lintott}
  et~al.}{2011}]{Lintott2011}
{Lintott} C.,  et~al., 2011, \mn@doi [\mnras]
  {10.1111/j.1365-2966.2010.17432.x}, \href
  {http://adsabs.harvard.edu/abs/2011MNRAS.410..166L} {410, 166}

\bibitem[\protect\citeauthoryear{{Masters} et~al.,}{{Masters}
  et~al.}{2010}]{Masters2010}
{Masters} K.~L.,  et~al., 2010, \mn@doi [\mnras]
  {10.1111/j.1365-2966.2010.16335.x}, \href
  {http://adsabs.harvard.edu/abs/2010MNRAS.404..792M} {404, 792}

\bibitem[\protect\citeauthoryear{{Masters} et~al.,}{{Masters}
  et~al.}{2011}]{Masters2011}
{Masters} K.~L.,  et~al., 2011, \mn@doi [\mnras]
  {10.1111/j.1365-2966.2010.17834.x}, \href
  {http://adsabs.harvard.edu/abs/2011MNRAS.411.2026M} {411, 2026}

\bibitem[\protect\citeauthoryear{{Masters} et~al.,}{{Masters}
  et~al.}{2012}]{Masters2012}
{Masters} K.~L.,  et~al., 2012, \mn@doi [\mnras]
  {10.1111/j.1365-2966.2012.21377.x}, \href
  {http://adsabs.harvard.edu/abs/2012MNRAS.424.2180M} {424, 2180}

\bibitem[\protect\citeauthoryear{{McDonald} et~al.,}{{McDonald}
  et~al.}{2011}]{McDonald2011}
{McDonald} M.,  et~al., 2011, \mn@doi [\mnras]
  {10.1111/j.1365-2966.2011.18519.x}, \href
  {http://adsabs.harvard.edu/abs/2011MNRAS.414.2055M} {414, 2055}

\bibitem[\protect\citeauthoryear{{Melvin} et~al.,}{{Melvin}
  et~al.}{2014}]{Melvin2014}
{Melvin} T.,  et~al., 2014, \mn@doi [\mnras] {10.1093/mnras/stt2397}, \href
  {http://adsabs.harvard.edu/abs/2014MNRAS.438.2882M} {438, 2882}

\bibitem[\protect\citeauthoryear{{M{\'e}ndez-Abreu}, {Aguerri}, {Corsini}  \&
  {Simonneau}}{{M{\'e}ndez-Abreu} et~al.}{2008}]{Mendez2008}
{M{\'e}ndez-Abreu} J.,  {Aguerri} J.~A.~L.,  {Corsini} E.~M.,   {Simonneau} E.,
   2008, \mn@doi [\aap] {10.1051/0004-6361:20078089}, \href
  {http://adsabs.harvard.edu/abs/2008A%26A...478..353M} {478, 353}

\bibitem[\protect\citeauthoryear{{M{\'e}ndez-Abreu} et~al.,}{{M{\'e}ndez-Abreu}
  et~al.}{2017}]{Califa2017}
{M{\'e}ndez-Abreu} J.,  et~al., 2017, \mn@doi [\aap]
  {10.1051/0004-6361/201629525}, \href
  {http://adsabs.harvard.edu/abs/2017A%26A...598A..32M} {598, A32}

\bibitem[\protect\citeauthoryear{{Minchev} \& {Famaey}}{{Minchev} \&
  {Famaey}}{2010}]{Minchev2010}
{Minchev} I.,  {Famaey} B.,  2010, \mn@doi [\apj]
  {10.1088/0004-637X/722/1/112}, \href
  {http://adsabs.harvard.edu/abs/2010ApJ...722..112M} {722, 112}

\bibitem[\protect\citeauthoryear{{Moore}, {Katz}, {Lake}, {Dressler}  \&
  {Oemler}}{{Moore} et~al.}{1996}]{Moore1996}
{Moore} B.,  {Katz} N.,  {Lake} G.,  {Dressler} A.,   {Oemler} A.,  1996,
  \mn@doi [\nat] {10.1038/379613a0}, \href
  {http://adsabs.harvard.edu/abs/1996Natur.379..613M} {379, 613}

\bibitem[\protect\citeauthoryear{{Nair} \& {Abraham}}{{Nair} \&
  {Abraham}}{2010a}]{Nair2010}
{Nair} P.~B.,  {Abraham} R.~G.,  2010a, \mn@doi [\apjs]
  {10.1088/0067-0049/186/2/427}, \href
  {http://adsabs.harvard.edu/abs/2010ApJS..186..427N} {186, 427}

\bibitem[\protect\citeauthoryear{{Nair} \& {Abraham}}{{Nair} \&
  {Abraham}}{2010b}]{Nair2010b}
{Nair} P.~B.,  {Abraham} R.~G.,  2010b, \mn@doi [\apjl]
  {10.1088/2041-8205/714/2/L260}, \href
  {http://adsabs.harvard.edu/abs/2010ApJ...714L.260N} {714, L260}

\bibitem[\protect\citeauthoryear{{Noguchi}}{{Noguchi}}{1988}]{Noguchi1988}
{Noguchi} M.,  1988, \aap, \href
  {http://adsabs.harvard.edu/abs/1988A%26A...203..259N} {203, 259}

\bibitem[\protect\citeauthoryear{{Pastrav}, {Popescu}, {Tuffs}  \&
  {Sansom}}{{Pastrav} et~al.}{2013a}]{Pastrav2013a}
{Pastrav} B.~A.,  {Popescu} C.~C.,  {Tuffs} R.~J.,   {Sansom} A.~E.,  2013a,
  \mn@doi [\aap] {10.1051/0004-6361/201220962}, \href
  {http://adsabs.harvard.edu/abs/2013A%26A...553A..80P} {553, A80}

\bibitem[\protect\citeauthoryear{{Pastrav}, {Popescu}, {Tuffs}  \&
  {Sansom}}{{Pastrav} et~al.}{2013b}]{Pastrav2013b}
{Pastrav} B.~A.,  {Popescu} C.~C.,  {Tuffs} R.~J.,   {Sansom} A.~E.,  2013b,
  \mn@doi [\aap] {10.1051/0004-6361/201322086}, \href
  {http://adsabs.harvard.edu/abs/2013A%26A...557A.137P} {557, A137}

\bibitem[\protect\citeauthoryear{{Peng}, {Ho}, {Impey}  \& {Rix}}{{Peng}
  et~al.}{2010}]{Peng2010}
{Peng} C.~Y.,  {Ho} L.~C.,  {Impey} C.~D.,   {Rix} H.-W.,  2010, \mn@doi [\aj]
  {10.1088/0004-6256/139/6/2097}, \href
  {http://adsabs.harvard.edu/abs/2010AJ....139.2097P} {139, 2097}

\bibitem[\protect\citeauthoryear{{P{\'e}rez} \&
  {S{\'a}nchez-Bl{\'a}zquez}}{{P{\'e}rez} \&
  {S{\'a}nchez-Bl{\'a}zquez}}{2011}]{Perez2011}
{P{\'e}rez} I.,  {S{\'a}nchez-Bl{\'a}zquez} P.,  2011, \mn@doi [\aap]
  {10.1051/0004-6361/201015002}, \href
  {http://adsabs.harvard.edu/abs/2011A%26A...529A..64P} {529, A64}

\bibitem[\protect\citeauthoryear{{P{\'e}rez} et~al.,}{{P{\'e}rez}
  et~al.}{2017}]{Perez2017}
{P{\'e}rez} I.,  et~al., 2017, \mn@doi [\mnras] {10.1093/mnrasl/slx087}, \href
  {http://adsabs.harvard.edu/abs/2017MNRAS.470L.122P} {470, L122}

\bibitem[\protect\citeauthoryear{{Pierini}, {Gordon}, {Witt}  \&
  {Madsen}}{{Pierini} et~al.}{2004}]{Pierini2004}
{Pierini} D.,  {Gordon} K.~D.,  {Witt} A.~N.,   {Madsen} G.~J.,  2004, \mn@doi
  [\apj] {10.1086/425651}, \href
  {http://adsabs.harvard.edu/abs/2004ApJ...617.1022P} {617, 1022}

\bibitem[\protect\citeauthoryear{{Reese}, {Williams}, {Sellwood}, {Barnes}  \&
  {Powell}}{{Reese} et~al.}{2007}]{Reese2007}
{Reese} A.~S.,  {Williams} T.~B.,  {Sellwood} J.~A.,  {Barnes} E.~I.,
  {Powell} B.~A.,  2007, \mn@doi [\aj] {10.1086/516826}, \href
  {http://adsabs.harvard.edu/abs/2007AJ....133.2846R} {133, 2846}

\bibitem[\protect\citeauthoryear{{Robichaud}, {Williamson}, {Martel}, {Kawata}
  \& {Ellison}}{{Robichaud} et~al.}{2017}]{Robichaud2017}
{Robichaud} F.,  {Williamson} D.,  {Martel} H.,  {Kawata} D.,   {Ellison}
  S.~L.,  2017, \mn@doi [\mnras] {10.1093/mnras/stx1121}, \href
  {http://adsabs.harvard.edu/abs/2017MNRAS.469.3722R} {469, 3722}

\bibitem[\protect\citeauthoryear{{SDSS Collaboration} et~al.,}{{SDSS
  Collaboration} et~al.}{2016}]{Manga2016}
{SDSS Collaboration} et~al., 2016, preprint, \href
  {http://adsabs.harvard.edu/abs/2016arXiv160802013S} {} (\mn@eprint {arXiv}
  {1608.02013})

\bibitem[\protect\citeauthoryear{{Salo} et~al.,}{{Salo}
  et~al.}{2015}]{Salo2015}
{Salo} H.,  et~al., 2015, \mn@doi [\apjs] {10.1088/0067-0049/219/1/4}, \href
  {http://adsabs.harvard.edu/abs/2015ApJS..219....4S} {219, 4}

\bibitem[\protect\citeauthoryear{{S{\'a}nchez-Bl{\'a}zquez}
  et~al.,}{{S{\'a}nchez-Bl{\'a}zquez} et~al.}{2014}]{Blazquez2014}
{S{\'a}nchez-Bl{\'a}zquez} P.,  et~al., 2014, \mn@doi [\aap]
  {10.1051/0004-6361/201423635}, \href
  {http://adsabs.harvard.edu/abs/2014A%26A...570A...6S} {570, A6}

\bibitem[\protect\citeauthoryear{{S{\'a}nchez-Janssen} \&
  {Gadotti}}{{S{\'a}nchez-Janssen} \& {Gadotti}}{2013}]{Sanchez2013}
{S{\'a}nchez-Janssen} R.,  {Gadotti} D.~A.,  2013, \mn@doi [\mnras]
  {10.1093/mnrasl/slt037}, \href
  {http://adsabs.harvard.edu/abs/2013MNRAS.432L..56S} {432, L56}

\bibitem[\protect\citeauthoryear{{S{\'a}nchez} et~al.,}{{S{\'a}nchez}
  et~al.}{2012}]{Califa2012}
{S{\'a}nchez} S.~F.,  et~al., 2012, \mn@doi [\aap]
  {10.1051/0004-6361/201117353}, \href
  {http://adsabs.harvard.edu/abs/2012A%26A...538A...8S} {538, A8}

\bibitem[\protect\citeauthoryear{{Scannapieco}, {Gadotti}, {Jonsson}  \&
  {White}}{{Scannapieco} et~al.}{2010}]{Scannapieco2010}
{Scannapieco} C.,  {Gadotti} D.~A.,  {Jonsson} P.,   {White} S.~D.~M.,  2010,
  \mn@doi [\mnras] {10.1111/j.1745-3933.2010.00900.x}, \href
  {http://adsabs.harvard.edu/abs/2010MNRAS.407L..41S} {407, L41}

\bibitem[\protect\citeauthoryear{{Schawinski} et~al.,}{{Schawinski}
  et~al.}{2014}]{Schawinski2014}
{Schawinski} K.,  et~al., 2014, \mn@doi [\mnras] {10.1093/mnras/stu327}, \href
  {http://adsabs.harvard.edu/abs/2014MNRAS.440..889S} {440, 889}

\bibitem[\protect\citeauthoryear{{Schlegel}, {Finkbeiner}  \&
  {Davis}}{{Schlegel} et~al.}{1998}]{Schlegel1998}
{Schlegel} D.~J.,  {Finkbeiner} D.~P.,   {Davis} M.,  1998, \mn@doi [\apj]
  {10.1086/305772}, \href {http://adsabs.harvard.edu/abs/1998ApJ...500..525S}
  {500, 525}

\bibitem[\protect\citeauthoryear{{Seidel}, {Falc{\'o}n-Barroso},
  {Mart{\'{\i}}nez-Valpuesta}, {D{\'{\i}}az-Garc{\'{\i}}a}, {Laurikainen},
  {Salo}  \& {Knapen}}{{Seidel} et~al.}{2015}]{Seidel2015}
{Seidel} M.~K.,  {Falc{\'o}n-Barroso} J.,  {Mart{\'{\i}}nez-Valpuesta} I.,
  {D{\'{\i}}az-Garc{\'{\i}}a} S.,  {Laurikainen} E.,  {Salo} H.,   {Knapen}
  J.~H. G.,  2015, \mn@doi [\mnras] {10.1093/mnras/stv969}, \href
  {http://adsabs.harvard.edu/abs/2015MNRAS.451..936S} {451, 936}

\bibitem[\protect\citeauthoryear{{Sellwood}}{{Sellwood}}{2014}]{Sellwood2014}
{Sellwood} J.~A.,  2014, \mn@doi [Reviews of Modern Physics]
  {10.1103/RevModPhys.86.1}, \href
  {http://adsabs.harvard.edu/abs/2014RvMP...86....1S} {86, 1}

\bibitem[\protect\citeauthoryear{{Sellwood} \& {Binney}}{{Sellwood} \&
  {Binney}}{2002}]{Sellwood2002}
{Sellwood} J.~A.,  {Binney} J.~J.,  2002, \mn@doi [\mnras]
  {10.1046/j.1365-8711.2002.05806.x}, \href
  {http://adsabs.harvard.edu/abs/2002MNRAS.336..785S} {336, 785}

\bibitem[\protect\citeauthoryear{{Sellwood} \& {Wilkinson}}{{Sellwood} \&
  {Wilkinson}}{1993}]{Sellwood1993}
{Sellwood} J.~A.,  {Wilkinson} A.,  1993, \mn@doi [Reports on Progress in
  Physics] {10.1088/0034-4885/56/2/001}, \href
  {http://adsabs.harvard.edu/abs/1993RPPh...56..173S} {56, 173}

\bibitem[\protect\citeauthoryear{{Sersic}}{{Sersic}}{1968}]{Sersic1968}
{Sersic} J.~L.,  1968, {Atlas de galaxias australes}

\bibitem[\protect\citeauthoryear{{Shen} \& {Sellwood}}{{Shen} \&
  {Sellwood}}{2004}]{Shen2004}
{Shen} J.,  {Sellwood} J.~A.,  2004, \mn@doi [\apj] {10.1086/382124}, \href
  {http://adsabs.harvard.edu/abs/2004ApJ...604..614S} {604, 614}

\bibitem[\protect\citeauthoryear{{Sheth} et~al.,}{{Sheth}
  et~al.}{2008}]{Sheth2008}
{Sheth} K.,  et~al., 2008, \mn@doi [\apj] {10.1086/524980}, \href
  {http://adsabs.harvard.edu/abs/2008ApJ...675.1141S} {675, 1141}

\bibitem[\protect\citeauthoryear{{Sheth} et~al.,}{{Sheth}
  et~al.}{2010}]{Sheth2010}
{Sheth} K.,  et~al., 2010, \mn@doi [\pasp] {10.1086/657638}, \href
  {http://adsabs.harvard.edu/abs/2010PASP..122.1397S} {122, 1397}

\bibitem[\protect\citeauthoryear{{Simard}, {Mendel}, {Patton}, {Ellison}  \&
  {McConnachie}}{{Simard} et~al.}{2011}]{Simard2011}
{Simard} L.,  {Mendel} J.~T.,  {Patton} D.~R.,  {Ellison} S.~L.,
  {McConnachie} A.~W.,  2011, \mn@doi [\apjs] {10.1088/0067-0049/196/1/11},
  \href {http://adsabs.harvard.edu/abs/2011ApJS..196...11S} {196, 11}

\bibitem[\protect\citeauthoryear{{Simmons} et~al.,}{{Simmons}
  et~al.}{2013}]{Simmons2013}
{Simmons} B.~D.,  et~al., 2013, \mn@doi [\mnras] {10.1093/mnras/sts491}, \href
  {http://adsabs.harvard.edu/abs/2013MNRAS.429.2199S} {429, 2199}

\bibitem[\protect\citeauthoryear{{Simmons} et~al.,}{{Simmons}
  et~al.}{2014}]{Simmons2014}
{Simmons} B.~D.,  et~al., 2014, \mn@doi [\mnras] {10.1093/mnras/stu1817}, \href
  {http://adsabs.harvard.edu/abs/2014MNRAS.445.3466S} {445, 3466}

\bibitem[\protect\citeauthoryear{{Skibba} et~al.,}{{Skibba}
  et~al.}{2012}]{Skibba2012}
{Skibba} R.~A.,  et~al., 2012, \mn@doi [\mnras]
  {10.1111/j.1365-2966.2012.20972.x}, \href
  {http://adsabs.harvard.edu/abs/2012MNRAS.423.1485S} {423, 1485}

\bibitem[\protect\citeauthoryear{{Smethurst}, {Lintott}, {Bamford}, {Hart},
  {Kruk}, {Masters}, {Nichol}  \& {Simmons}}{{Smethurst}
  et~al.}{2017}]{Smethurst2017}
{Smethurst} R.~J.,  {Lintott} C.~J.,  {Bamford} S.~P.,  {Hart} R.~E.,  {Kruk}
  S.~J.,  {Masters} K.~L.,  {Nichol} R.~C.,   {Simmons} B.~D.,  2017, \mn@doi
  [\mnras] {10.1093/mnras/stx973}, \href
  {http://adsabs.harvard.edu/abs/2017MNRAS.469.3670S} {469, 3670}

\bibitem[\protect\citeauthoryear{{Spinoso}, {Bonoli}, {Dotti}, {Mayer}, {Madau}
   \& {Bellovary}}{{Spinoso} et~al.}{2017}]{Spinoso2017}
{Spinoso} D.,  {Bonoli} S.,  {Dotti} M.,  {Mayer} L.,  {Madau} P.,
  {Bellovary} J.,  2017, \mn@doi [\mnras] {10.1093/mnras/stw2934}, \href
  {http://adsabs.harvard.edu/abs/2017MNRAS.465.3729S} {465, 3729}

\bibitem[\protect\citeauthoryear{{Stoughton} et~al.,}{{Stoughton}
  et~al.}{2002}]{Stoughton2002}
{Stoughton} C.,  et~al., 2002, \mn@doi [\aj] {10.1086/324741}, \href
  {http://adsabs.harvard.edu/abs/2002AJ....123..485S} {123, 485}

\bibitem[\protect\citeauthoryear{{Tacchella} et~al.,}{{Tacchella}
  et~al.}{2015}]{Tacchella2015}
{Tacchella} S.,  et~al., 2015, \mn@doi [Science] {10.1126/science.1261094},
  \href {http://adsabs.harvard.edu/abs/2015Sci...348..314T} {348, 314}

\bibitem[\protect\citeauthoryear{{Taylor}}{{Taylor}}{2005}]{Topcat}
{Taylor} M.~B.,  2005, in {Shopbell} P.,  {Britton} M.,   {Ebert} R.,  eds,
  Astronomical Society of the Pacific Conference Series Vol. 347, Astronomical
  Data Analysis Software and Systems XIV. p.~29

\bibitem[\protect\citeauthoryear{{Taylor} et~al.,}{{Taylor}
  et~al.}{2011}]{Taylor2011}
{Taylor} E.~N.,  et~al., 2011, \mn@doi [\mnras]
  {10.1111/j.1365-2966.2011.19536.x}, \href
  {http://adsabs.harvard.edu/abs/2011MNRAS.418.1587T} {418, 1587}

\bibitem[\protect\citeauthoryear{{Tuffs}, {Popescu}, {V{\"o}lk}, {Kylafis}  \&
  {Dopita}}{{Tuffs} et~al.}{2004}]{Tuffs2004}
{Tuffs} R.~J.,  {Popescu} C.~C.,  {V{\"o}lk} H.~J.,  {Kylafis} N.~D.,
  {Dopita} M.~A.,  2004, \mn@doi [\aap] {10.1051/0004-6361:20035689}, \href
  {http://adsabs.harvard.edu/abs/2004A%26A...419..821T} {419, 821}

\bibitem[\protect\citeauthoryear{{Unterborn} \& {Ryden}}{{Unterborn} \&
  {Ryden}}{2008}]{Unterborn2008}
{Unterborn} C.~T.,  {Ryden} B.~S.,  2008, \mn@doi [\apj] {10.1086/591898},
  \href {http://adsabs.harvard.edu/abs/2008ApJ...687..976U} {687, 976}

\bibitem[\protect\citeauthoryear{{Vika}, {Bamford}, {H{\"a}u{\ss}ler}, {Rojas},
  {Borch}  \& {Nichol}}{{Vika} et~al.}{2013}]{Vika2013}
{Vika} M.,  {Bamford} S.~P.,  {H{\"a}u{\ss}ler} B.,  {Rojas} A.~L.,  {Borch}
  A.,   {Nichol} R.~C.,  2013, \mn@doi [\mnras] {10.1093/mnras/stt1320}, \href
  {http://adsabs.harvard.edu/abs/2013MNRAS.435..623V} {435, 623}

\bibitem[\protect\citeauthoryear{{Vika}, {Bamford}, {H{\"a}u{\ss}ler}  \&
  {Rojas}}{{Vika} et~al.}{2014}]{Vika2014}
{Vika} M.,  {Bamford} S.~P.,  {H{\"a}u{\ss}ler} B.,   {Rojas} A.~L.,  2014,
  \mn@doi [\mnras] {10.1093/mnras/stu1696}, \href
  {http://adsabs.harvard.edu/abs/2014MNRAS.444.3603V} {444, 3603}

\bibitem[\protect\citeauthoryear{{Wada} \& {Habe}}{{Wada} \&
  {Habe}}{1992}]{Wada1992}
{Wada} K.,  {Habe} A.,  1992, \mn@doi [\mnras] {10.1093/mnras/258.1.82}, \href
  {http://adsabs.harvard.edu/abs/1992MNRAS.258...82W} {258, 82}

\bibitem[\protect\citeauthoryear{{Weinzirl}, {Jogee}, {Khochfar}, {Burkert}  \&
  {Kormendy}}{{Weinzirl} et~al.}{2009}]{Weinzirl2009}
{Weinzirl} T.,  {Jogee} S.,  {Khochfar} S.,  {Burkert} A.,   {Kormendy} J.,
  2009, \mn@doi [\apj] {10.1088/0004-637X/696/1/411}, \href
  {http://adsabs.harvard.edu/abs/2009ApJ...696..411W} {696, 411}

\bibitem[\protect\citeauthoryear{{Willett} et~al.,}{{Willett}
  et~al.}{2013}]{Willett2013}
{Willett} K.~W.,  et~al., 2013, \mn@doi [\mnras] {10.1093/mnras/stt1458}, \href
  {http://adsabs.harvard.edu/abs/2013MNRAS.435.2835W} {435, 2835}

\bibitem[\protect\citeauthoryear{{York} et~al.,}{{York}
  et~al.}{2000}]{York2000}
{York} D.~G.,  et~al., 2000, \mn@doi [\aj] {10.1086/301513}, \href
  {http://adsabs.harvard.edu/abs/2000AJ....120.1579Y} {120, 1579}

\bibitem[\protect\citeauthoryear{{de Souza} et~al.,}{{de Souza}
  et~al.}{2004}]{Souza2004}
{de Souza} R.~E.,  et~al., 2004, \mn@doi [\apjs] {10.1086/421554}, \href
  {http://adsabs.harvard.edu/abs/2004ApJS..153..411D} {153, 411}

\bibitem[\protect\citeauthoryear{{de Vaucouleurs}, {de Vaucouleurs}, {Corwin},
  {Buta}, {Paturel}  \& {Fouqu{\'e}}}{{de Vaucouleurs}
  et~al.}{1991}]{deVauc1993}
{de Vaucouleurs} G.,  {de Vaucouleurs} A.,  {Corwin} Jr. H.~G.,  {Buta} R.~J.,
  {Paturel} G.,   {Fouqu{\'e}} P.,  1991, {Third Reference Catalogue of Bright
  Galaxies. Volume I: Explanations and references. Volume II: Data for galaxies
  between 0$^{h}$ and 12$^{h}$. Volume III: Data for galaxies between 12$^{h}$
  and 24$^{h}$.}

\makeatother
\end{thebibliography}

\appendix

\section{Weak bars}
\label{appendixA}

\begin{figure}
  \includegraphics[width=\columnwidth]{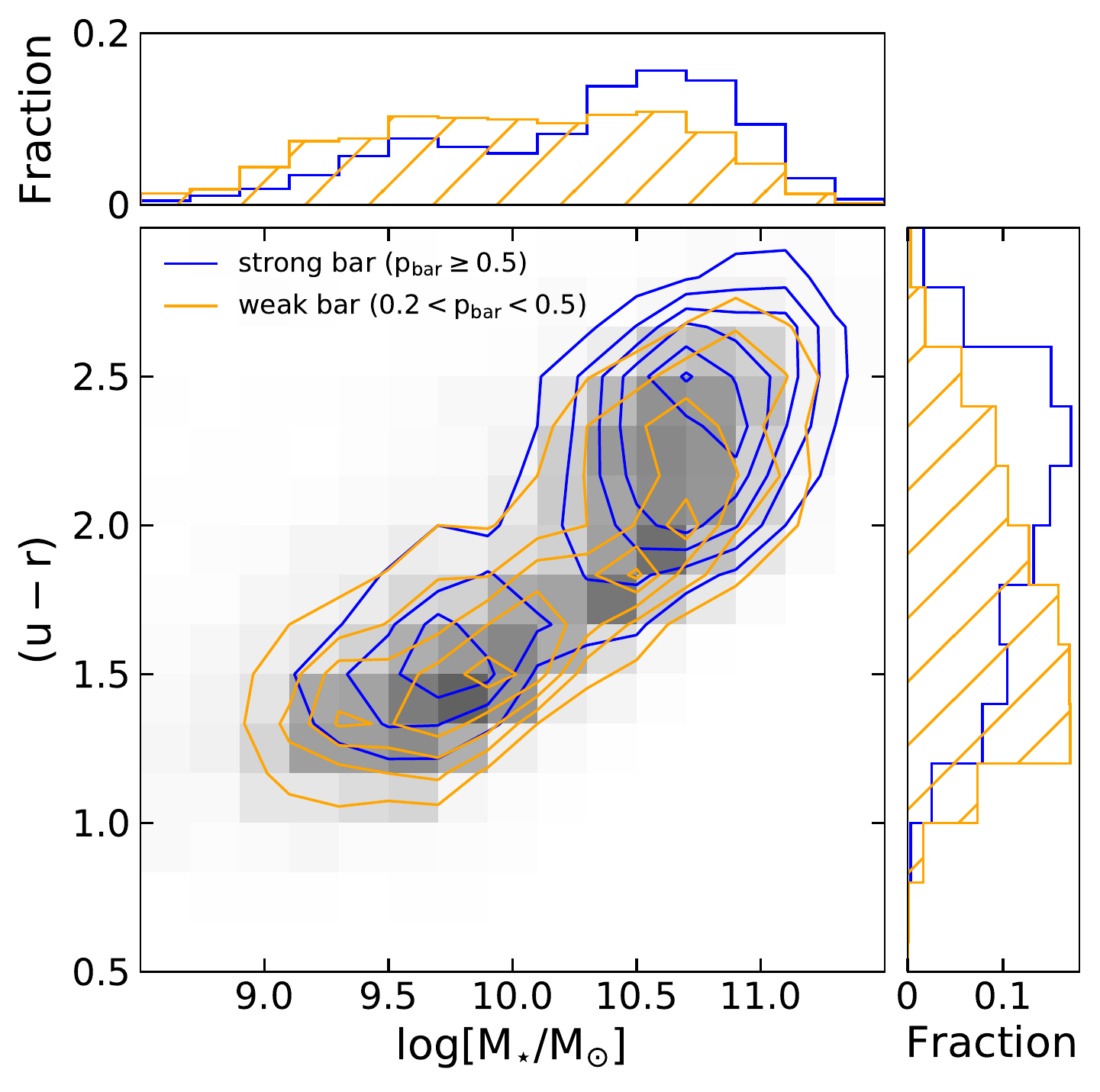}
  \caption{Colour-mass diagram for the weakly and strongly barred galaxies. The histograms show the normalised distributions of the stellar mass and $(u-r)$ colours for the strongly and weakly barred galaxies. }
 \label{weak_strong_CMD}
\end{figure}

\begin{figure*}
 \includegraphics[width=\textwidth]{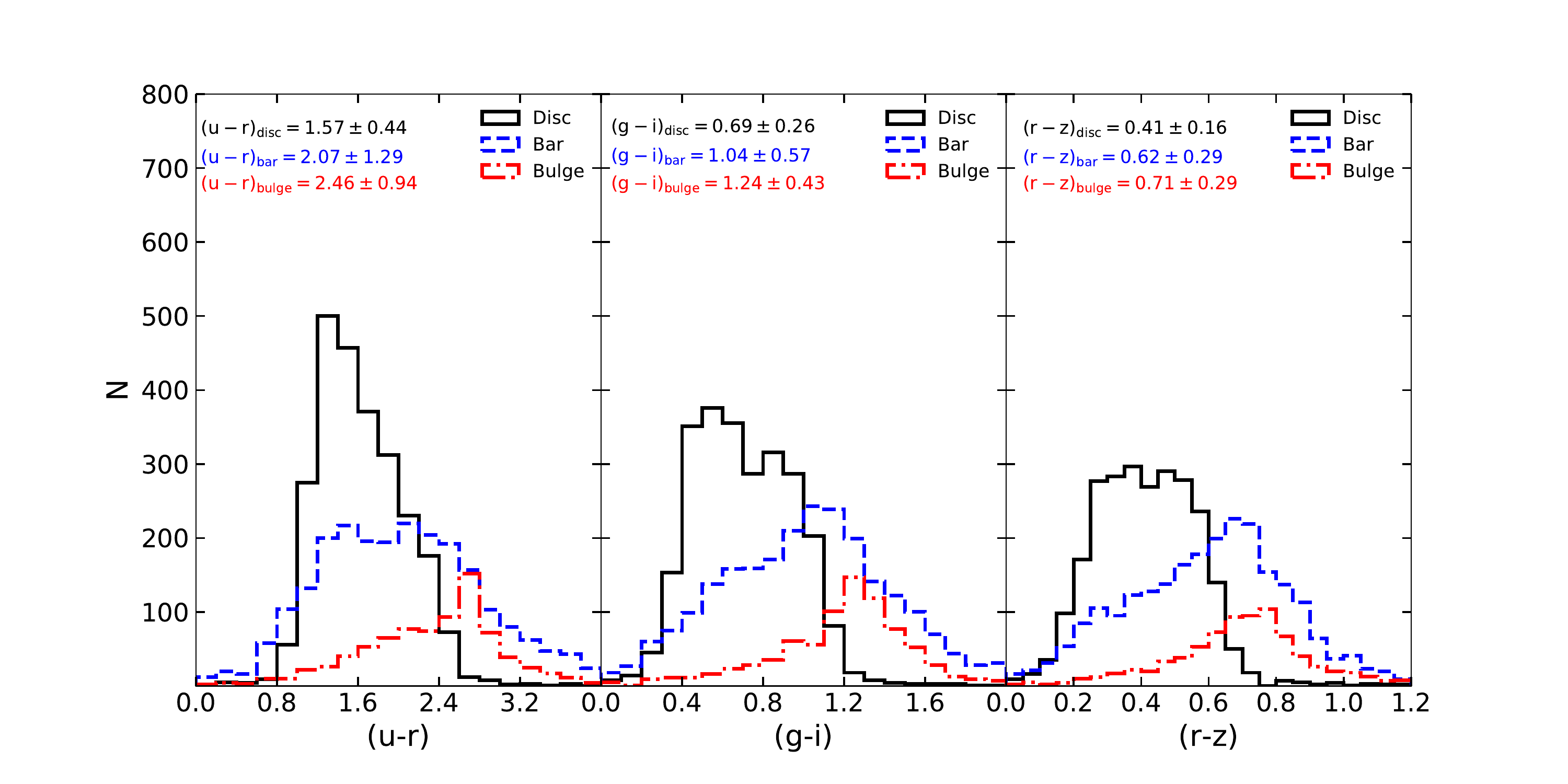}
 \caption{The $(u-r)$, $(g-i)$ and $(r-z)$ colours of the different galaxy components for all the fitted galaxies with weak bars (2,617 galaxies).  This sample contains all the successfully fitted galaxies with weak bars and is not volume-limited. Similarly to galaxies with strong bars, the discs are bluer than the bars, which in turn are slightly bluer than the bulges. The median colours and their corresponding $1\sigma$ spreads are represented for each component.}
 \label{weak_colours}
\end{figure*}

\begin{figure}
  \includegraphics[width=\columnwidth]{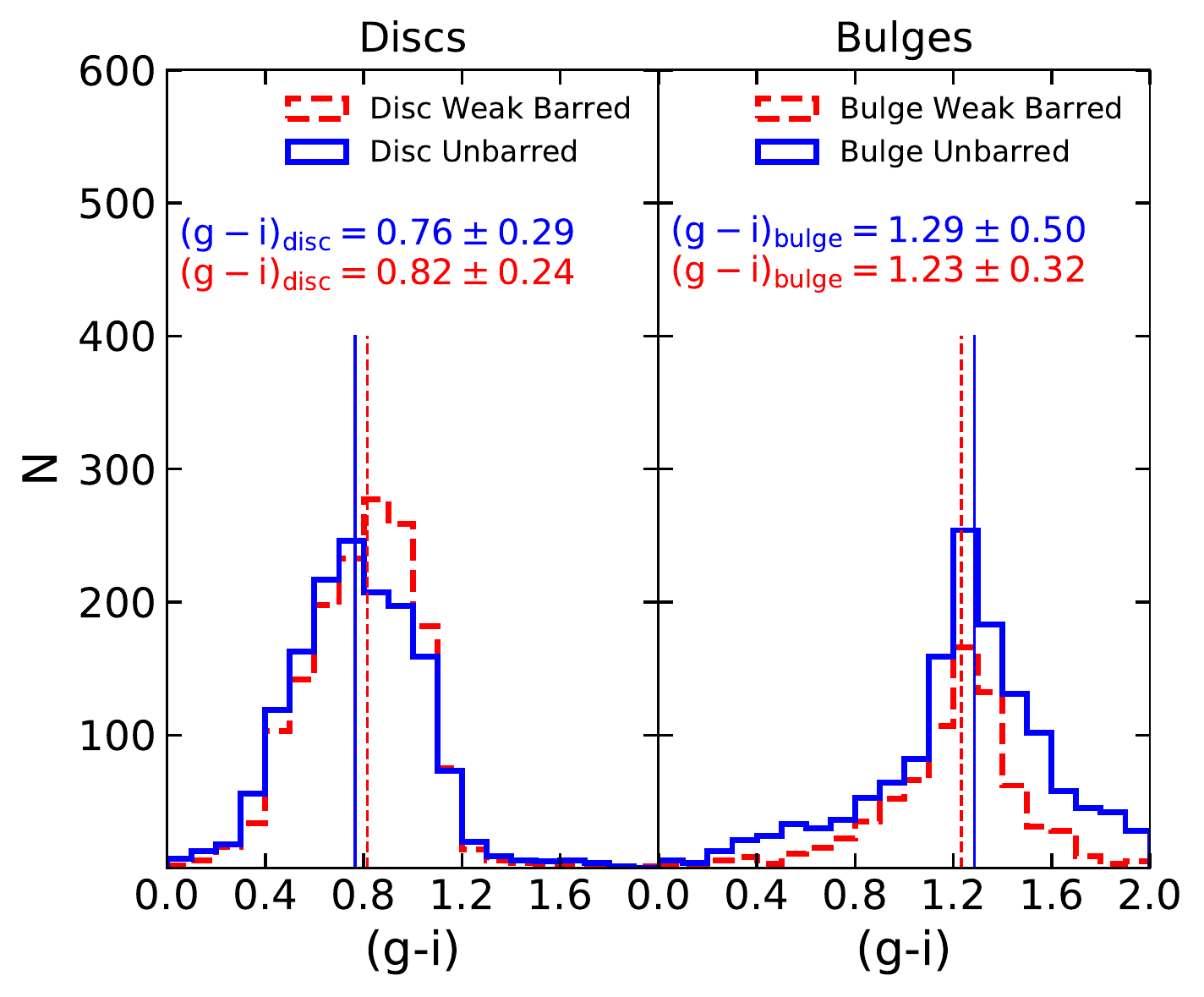}
  \caption{The $(g-i)$ colours of discs (left) and bulges (right) of weakly barred (with red) and unbarred (with blue) galaxies. The two samples are volume-limited and mass-matched. The discs of weakly barred galaxies are redder than the ones of unbarred galaxies, while their bulges have bluer colours when compared to the bulges of unbarred galaxies. Median values for the colours and the $1\sigma$ spread are shown. }
 \label{weak_colour_difference}
\end{figure}

\begin{figure}
 \includegraphics[width=\columnwidth]{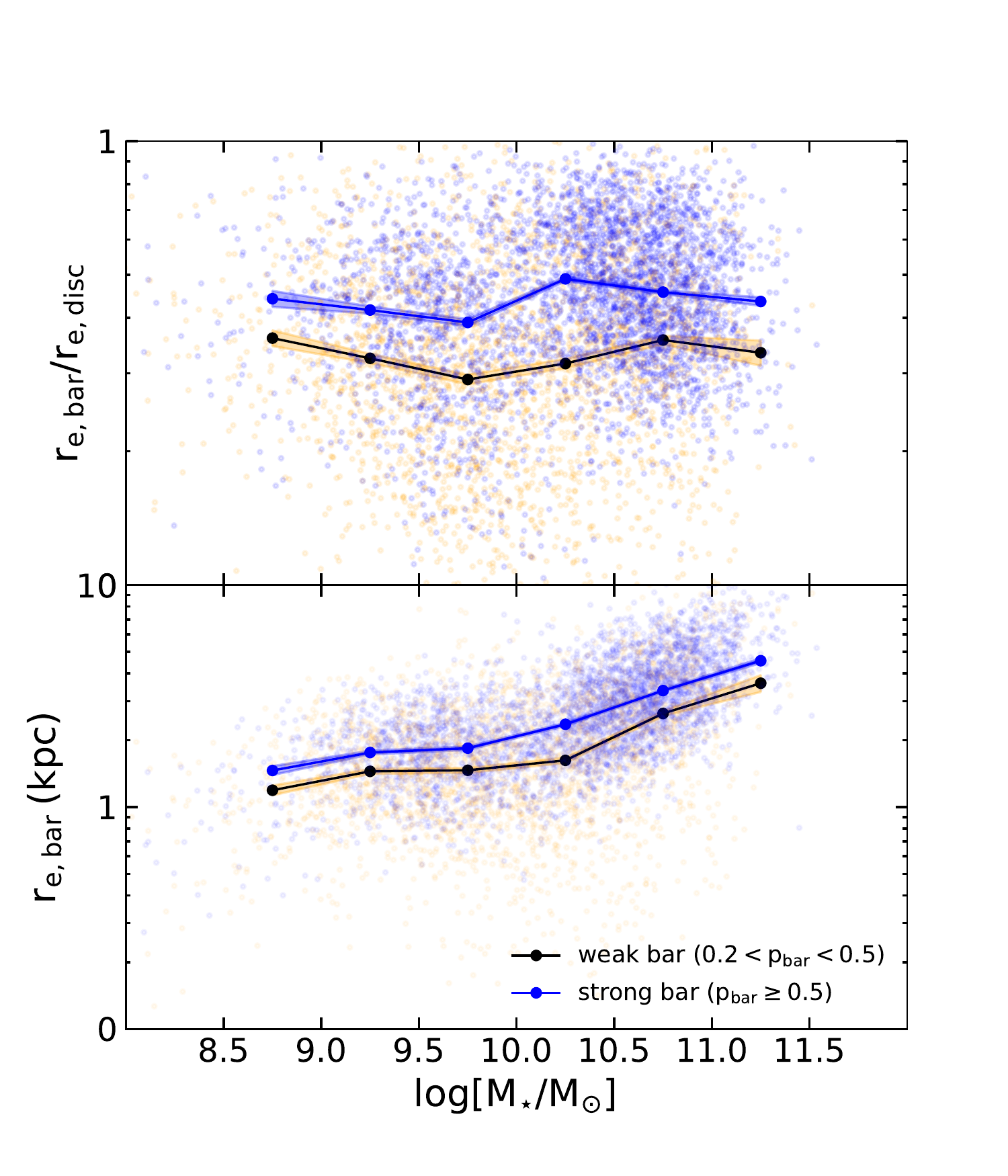}
 \caption{Top panel - The scaled bar effective radius, $r_{\mathrm{e,bar}}/r_{\mathrm{e,disc}}$ for the weak and strong bars in this work as a function of stellar mass. Bottom panel - The projected physical bar effective radius for strong and weak bars. The bar effective radius is a measure of the size of the bars, but does not necessarily correspond to the the length of the bar. Median values in stellar mass bins of $\log{(\frac{M_{\star}}{M_{\odot}}})=0.5$ are plotted and the shaded areas represent the $1\sigma/\sqrt{N}$ error per bin.}
 \label{weak_barlength}
\end{figure}

In this paper we selected barred galaxies with $p_\mathrm{bar}\geq0.5$ and unbarred galaxies with $p_\mathrm{bar}\leq0.2$. What about galaxies with $0.2<p_\mathrm{bar}<0.5$? In this section we explore the possible bias introduced by removing these galaxies from this sample. 

As discussed in Section \ref{data} and shown in \citet{Skibba2012,Masters2012,Willett2013}, galaxies with $0.2<p_\mathrm{bar}<0.5$ correspond mainly to `weak bars', when comparing the volunteers classification with expert classification such as the one in \citet{Nair2010}. Their classification into weak, intermediate and strong bars is based on visual inspection, on the relative size of the bars compared to the disc and the prominence of bars. A bar that dominates the light profile of a galaxy is classified as a strong bar, while weaker bars are smaller in size and contain a smaller percentage of the galaxy's light. As discussed in \citet{Nair2010}, the classification into `weak', `intermediate' and `strong' differs from the more traditional classification into SAB and SB bars of \citet{deVauc1993}: these bar classes correspond to subdivisions of SB bars, rather than to SAB bars. The reason for this is because the data quality of SDSS is lower than the one used by \citet{deVauc1993}. 

Due to the resolution of the SDSS images (median of $1.2\arcsec$ in the $r$-band), image contrast and the presence of other features such as bulges and spiral arms in the vicinity of bars, weak bars are harder to identify in the $gri$ composite images compared to strong bars. The reason for not including weak bars in the analysis of this paper is that a sample of galaxies selected with $0.2<p_\mathrm{bar}<0.5$ is unavoidably contaminated by unbarred galaxies. To assess the degree of this contamination, we fit a single S\'ersic profile to 1,000 galaxies with $0.2<p_\mathrm{bar}<0.5$ and one of the authors (SK) inspected the residuals for the possible presence of a bar. We find that $\sim75\%$ of these galaxies show signatures of a bar feature. In what follows, we repeated the analysis of the barred galaxies, but for weak bars instead of intermediate and strong, and we show the similarity and discrepancies between the two samples. The fits for these galaxies were not individually inspected, and the sample unavoidably contains some unbarred galaxies, therefore the weak sample of galaxies is not expected to be clean or complete.

A selection of galaxies with $0.2<p_\mathrm{bar}<0.5$ (and $N_{\textrm{bar}}\geq10$, $0.005<z<0.06$, $i\lesssim60^{\circ}$) contains 6,013 galaxies with a majority hosting weak bars. We fitted these galaxies in a similar way to the barred galaxies, with disc+bar components (3,236) and disc+bar+bulge components (1,734), with a success rate of automatic fits of $\sim$83\%. Furthermore, we selected galaxies having components with $r_{e}<200$ pixels, $0.12<n<7.8$, disc-bar offsets smaller than $3$ kpc  (as suggested by the analysis in \citealt{Kruk2017}) and $b/a_{\mathrm{bar}}<0.6$, leaving only 2,617 galaxies in the sample, or a final success rate of only $\sim$44\% showing that weak bars are indeed harder to fit.

Figure \ref{weak_strong_CMD} shows the colour-mass diagram of strong and weak bars. When compared to galaxies with strong bars, galaxies with weak bars tend to have lower masses and are bluer in colour. Figure \ref{weak_colours} shows the colours of the components of galaxies with weak bars. The discs of galaxies with weak bars have bluer colours ( $(g-i)_{\mathrm{disc}}=0.69$ compared to $(g-i)_{\mathrm{disc}}=0.90$ for the discs of strongly barred galaxies), which reflects the overall bluer colours of these galaxies, while the bars and bulges have more similar red colours ($(g-i)_{\mathrm{bar}}=1.04$ compared to $(g-i)_{\mathrm{bar}}=1.10$ and $(g-i)_{\mathrm{bulge}}=1.24$ compared to $(g-i)_{\mathrm{bulge}}=1.23$, respectively). There is also a significantly larger spread, an indication of a more diverse population of galaxies.

To compare weakly barred galaxies to unbarred galaxies, we select a volume-limited sample of galaxies with weak bars ($M_\mathrm{r}<-20.15$) and a new volume-limited and mass-matched sample of unbarred galaxies (with 1,580 galaxies in each sample). In Figure \ref{weak_colour_difference} we notice that the disc $(g-i)$ colours of galaxies with weak bars, even though they are on average bluer than the galaxies with strong bars, are still $\Delta(g-i)\sim0.06\pm0.01$ redder compared to the discs of unbarred galaxies. Similarly, the bulges of weakly barred galaxies are $\Delta(g-i)\sim0.06\pm0.01$ bluer compared to the unbarred counterparts, similar to the trends observed for strongly barred galaxies.

What is the difference between the weak and strong bars in our sample?

Apart from the bluer colours of the discs, as well as their lower masses in general, galaxies with weak bars show a similar bimodality in the bar S\'ersic indices as the galaxies with strong bars ($\left< n_{\mathrm{bar}} \right>\sim0.5$ for more massive galaxies, with $M_{\star}\geq10^{10.25} M_{\odot}$ and obvious bulges, and $\left< n_{\mathrm{bar}} \right>\sim1$ for the disc dominated lower mass galaxies, with $M_{\star}<10^{10.25} M_{\odot}$). The median $Bar/T$ is only marginally lower, at $Bar/T\sim0.10$. Figure \ref{weak_barlength} shows the difference in the scaled bar sizes (top panel) between strong and weak bars as well as the projected physical sizes of the bars (bottom panel). Weak bars are on average $\sim1.5$ times shorter than strong bars in our sample, the largest difference being observed at $M_{\star}\sim10^{10.25} M_{\odot}$, in both relative and absolute sizes. 

\bsp	
\label{lastpage}
\end{document}